\documentclass[showpacs,twocolumn,aps,pra,longbibliography,superscriptaddress,notitlepage]{revtex4-2}
\usepackage{qcircuit}
\usepackage[dvips]{graphicx}
\usepackage{amsmath,amssymb,amsthm,mathrsfs,amsfonts,dsfont}
\usepackage{subfigure, epsfig}
\usepackage{braket}
\usepackage{bm}
\usepackage{enumerate}
\usepackage{algorithm}
\usepackage{algpseudocode}
\usepackage{diagbox}
\usepackage{physics}
\usepackage{color}
\usepackage{multirow}
\usepackage[marginal]{footmisc}
\usepackage{comment}
\usepackage{tikz}
\usepackage{gensymb}
\usepackage{mathtools}
\usepackage[]{qcircuit}
\usetikzlibrary{arrows}
\usetikzlibrary{shapes,fadings,snakes}
\usetikzlibrary{decorations.pathmorphing,patterns}
\usetikzlibrary{calc}
\usetikzlibrary{positioning}
\usepackage[colorlinks = true, linkcolor=blue,
            anchorcolor=blue,
            citecolor=blue]{hyperref}

\newcommand{\comments}[1]{}

\newcommand{\bket}[1]{\langle {#1} \rangle}

\usepackage{color}
\definecolor{dred}{rgb}{.8,0.2,.2}
\definecolor{ddred}{rgb}{.8,0.5,.5}
\definecolor{dblue}{rgb}{.2,0.2,.8}
\definecolor{dgreen}{rgb}{.2,0.5,.2}

\usepackage{times}

\newcommand*{\physus}{Department of Physics, State Key Laboratory of Quantum Functional Materials, and Guangdong Basic Research Center of Excellence for Quantum Science, Southern University of Science and Technology, Shenzhen 518055, China}

\begin{document}


\title{Realizing Universal Non-Markovian Noise Suppression}

\author{Hongfeng Liu}
\thanks{These authors contributed equally to this work.}
\affiliation{\physus}

\author{Zizhao Han}
\thanks{These authors contributed equally to this work.}
\affiliation{Center for Quantum Information, IIIS, Tsinghua University, Beijing 100084, China}

\author{Xinfang Nie}
\email{niexinfang@quantumsc.cn}
\affiliation{Quantum Science Center of Guangdong-Hong Kong-Macao Greater Bay Area, Shenzhen 518045, China}
\affiliation{\physus}

\author{Zhenhuan Liu}
\email{qubithuan@gmail.com}
\affiliation{Yau Mathematical Sciences Center, Tsinghua University, Beijing 100084, China}

\author{Dawei Lu}
\email{ludw@sustech.edu.cn}
\affiliation{\physus}
\affiliation{Quantum Science Center of Guangdong-Hong Kong-Macao Greater Bay Area, Shenzhen 518045, China}
\affiliation{{Shenzhen Institute for Quantum Science and Engineering, Southern University of Science and Technology, Shenzhen 518055, China}}

\begin{abstract}
Non-Markovian noise, arising from environmental memory effects, is the most general and challenging form of noise in quantum computing, and is typically difficult to characterize and suppress. 
Here, we analyze and experimentally demonstrate a non-Markovian noise suppression scheme inspired by quantum purification protocols. 
We theoretically prove that, even without noise calibration and assumptions on specific noise models, the scheme can exponentially reduce non-Markovian error rates with respect to the ancillary system size. 
We implement the protocol using nuclear spins, demonstrating that non-Markovian noise can be suppressed for both unitary operations and non-unitary channels.
The observed fidelities and process tomography show close agreement with theoretical predictions, confirming the practicality and effectiveness of the scheme.
\end{abstract}

\maketitle
\begin{figure}[htbp]
\centering
\includegraphics[width=0.48\textwidth]{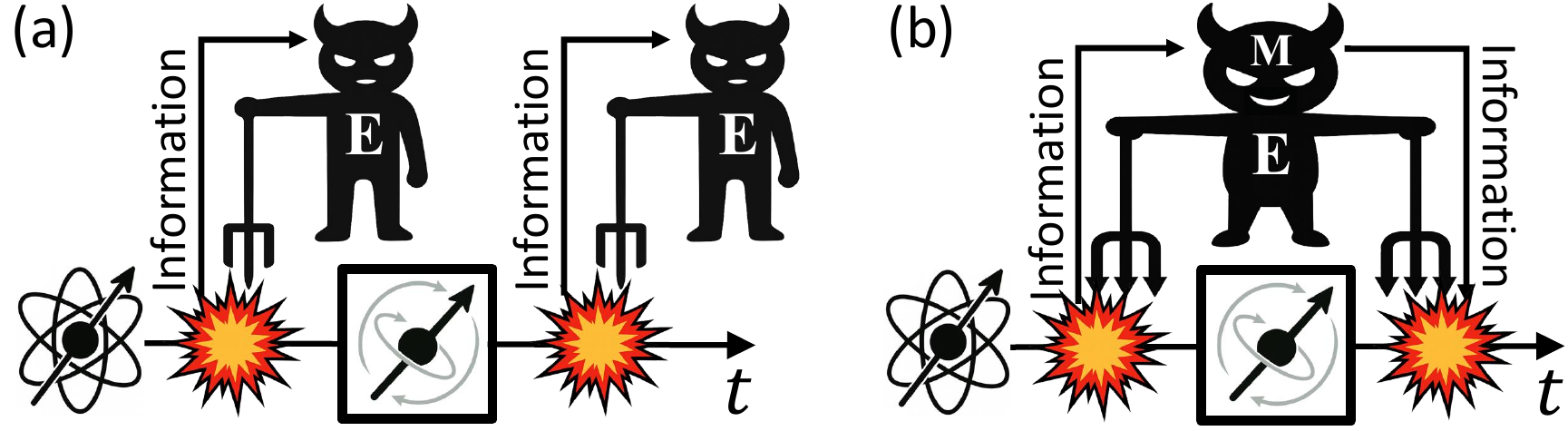}
\caption{An intuitive figure that shows the difference between Markovian and non-Markovian noises, where ``E'' and ``M'' stand for environment and memory, respectively. 
(a) Markovian noise: the interactions between system and environment at different time points are independent (memoryless).
(b) Non-Markovian noise: the interactions at different time points come from the same environment, which keeps a memory.
It can feed information from the past back into the system, leading to history-dependent noises.
}
\label{fig:nm_pic}
\end{figure}

\noindent\emph{\bfseries Introduction.}---Quantum computation is fundamentally limited by noise, which is originated from interacting with environment. 
A common simplifying assumption is that noise is Markovian, meaning information flows from the system to the environment without returning~\cite{breuer2006open,pollock2018operational,LI2018concepts,Taranto2024characterising}, as illustrated in Fig.~\ref{fig:nm_pic}(a). 
In this simplified scenario, noise can be described as the concatenation of independent quantum channels, which can be suppressed by well-established methods such as quantum error correction and mitigation~\cite{terhal2015qec,cai2023qem}.
In practice, however, many platforms exhibit \emph{non-Markovian noise}, where the environment retains memory and feeds the information back into the system~\cite{marcos2011nanostructures,zhang2022predicting,gulacsi2023signitures,Agarwal2024modelling}, as illustrated in Fig.~\ref{fig:nm_pic}(b). 
Such kind of noise is the most general and challenging scenario for quantum devices. 
Compared with Markovian noise, the history dependence of non-Markovian noise introduces an exponentially large number of parameters, making the identification, calibration, and suppression highly challenging~\cite{chiribella2008comb,chiribella2009comb,rivas2014quantum,pollock2018processtensor,milz2021processtensor,figueroa2021nonmarkovianRB,white2025unifying}. 

Multiple paradigms have been developed to combat noise, yet each encounters intrinsic limitations in the non-Markovian regime.
Quantum error correction encodes logical information redundantly across many physical qubits so that errors can be detected and corrected~\cite{terhal2015qec}.
Although it can, in principle, address non-Markovian noise, fault tolerance typically requires much stricter assumptions on the spatial and temporal locality of the noise~\cite{terhal2005fault,aharonov2006fault}, and residual memory effects can still undermine performances of standard codes~\cite{Kam2025detrimental}.
Quantum error mitigation improves computational accuracy by trading off additional sampling overhead~\cite{cai2023qem}, but most techniques either rely on an accurate noise model~\cite{endoPracticalQuantumError2018,guo2022qem,wang2025non} or on tunable noise strength~\cite{temme2017mitigation,Kandala2019ErrorMitigation}, both of which are highly challenging under non-Markovian effects.
Quantum control provides a complementary hardware-level approach, where dynamical decoupling has become a standard technique to suppress environmental noise~\cite{Viola1998DD,Du2009DD}.
Yet, in quantum computation, pulse sequences for dynamical decoupling must be carefully tailored to the target gate and noise characteristics~\cite{Khodjasteh2005,Khodjasteh2009,Zeng2018DD}, and their effectiveness depends sensitively on the degree of non-Markovianity~\cite{Addis2015}.

In this work, we analyze and experimentally validate a calibration-free, model-agnostic protocol for suppressing non-Markovian noise~\cite{liu2024non}, which is inspired by quantum purification techniques~\cite{cirac1999optimal,cotler2019cooling,hugginsVirtualDistillationQuantum2021,koczorExponentialErrorSuppression2021,Childs2025streamingquantum,liu2025virtual}.
Our method is universal and makes minimal assumptions about the noise process and offers strong practical advantages: it does not rely on locality constraints or complex encoding operations, requires no calibrated or tunable noise model, and is applicable to noise affecting idle periods, gate operations, and general channels. 
Moreover, it can be flexibly adjusted to involve different numbers of time points and sizes of ancillary systems to achieve enhanced error suppression performances, which is guaranteed by theoretical analysis. 
We implement the protocol on a five-qubit nuclear magnetic resonance (NMR) quantum processor. 
To assess its performance, we estimate fidelities of output states and perform quantum process tomography in cases with and without the protocol. 
These results consistently reveal a substantial suppression of non-Markovian noise, in close agreement with theoretical predictions.
By avoiding extensive tomographic calibration and large qubit overhead, this approach provides a practical route toward reliable quantum computation in realistic, memory-bearing environments.

\begin{figure}[htbp]
\centering
\includegraphics[width=0.48\textwidth]{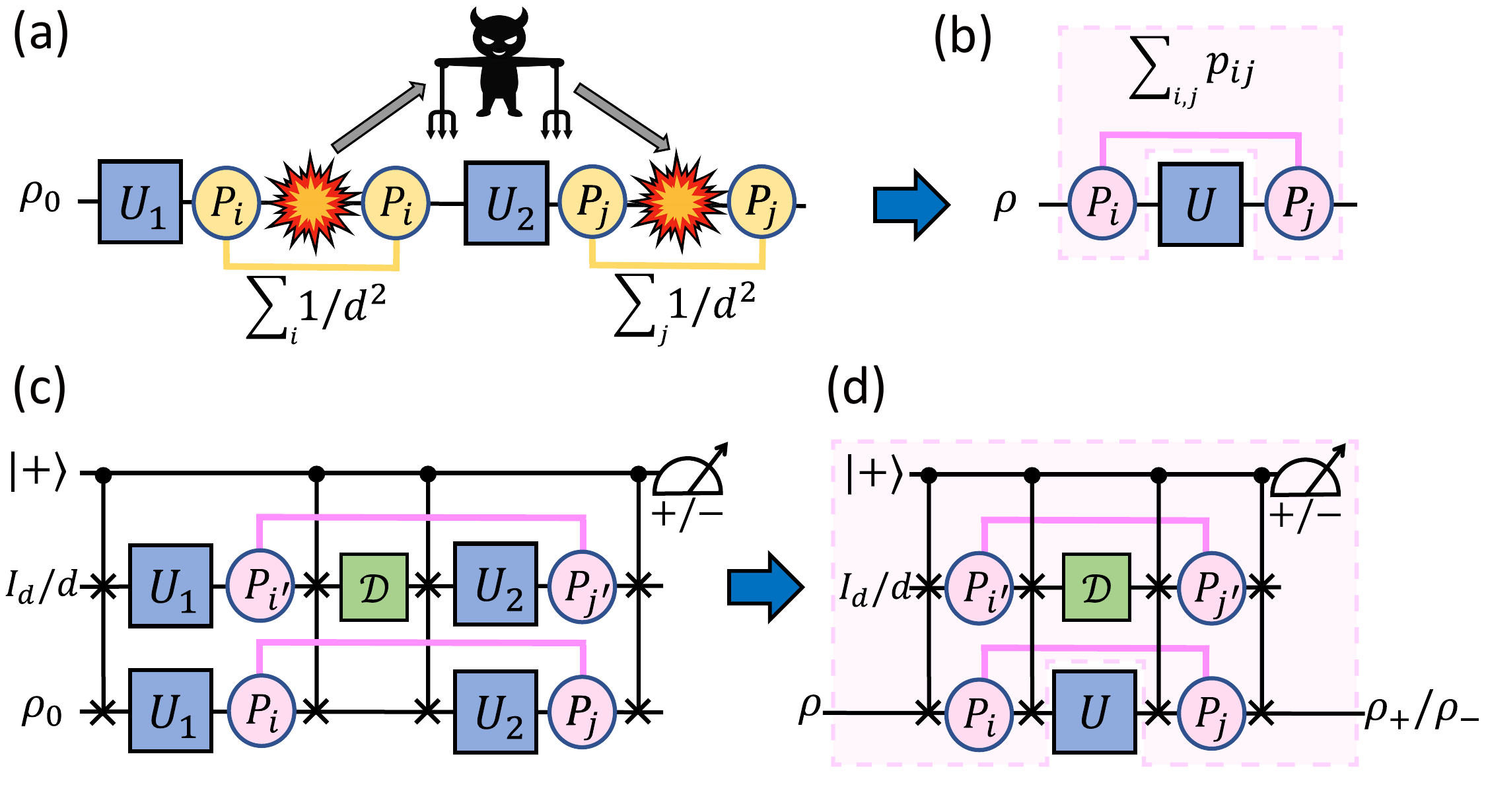}
\caption{Non-Markovian noise suppression protocol, where the yellow circles and lines represent the Pauli twirling and the pink ones represent the effective Pauli noises after Pauli twirling. The summation represents averaging the Pauli operators under a specific probability distribution.  (a) The noisy circuit with non-Markovian noise, represented with two explosion signs and a demon, and the Pauli twirling. (b) Pauli twirling turns a non-Markovian noise into two classically correlated Pauli noises at two time points, with $\rho=U_1 \rho_0 U_1^\dagger$ and $U=U_2$. (c) Purification circuit for non-Markovian noise. (d) The simplified circuit. The pink dashed box denotes the entire non-Markovian noise after purification; compared with (b), it affects the ideal gate $U$ more weakly.}
\label{fig:flowchart}
\end{figure}

\noindent\emph{\bfseries Protocol.}---We first focus on the case in which non-Markovian noise involves two time points in the quantum circuit, as shown in Fig.~\ref{fig:flowchart}(a). 
The system is initialized in state $\rho_0$, and two quantum operations $U_1$ and $U_2$ are applied sequentially.
These two operations are affected by noises, which are temporally correlated through the environment. 
Such non-Markovian noise makes the actual output state deviate from the ideal one, $U_2 U_1 \rho_0U_1^\dagger U_2^\dagger$.

Our suppression protocol is consisted of two main steps, Pauli twirling and purification.
Since the non-Markovian noise manifests complex quantum memory effect, we first introduce the Pauli twirling method to simplify it and benefit further manipulation~\cite{wallmanNoiseTailoringScalable2016,Cai2020Mitigating}.
To implement the Pauli twirling on non-Markovian noise, one can sandwich the noises at two time points with two pairs of independent random Pauli operations, as shown in Fig.~\ref{fig:flowchart}(a). 
In a practical weak-noise scenario, the noisy gate can be approximately represented as the ideal gate followed by the noise~\cite{Wallman2018randomized}.
Therefore, to sandwich the noise with the same random Pauli operators, we can replace the target gate $U$ by $P_i U$ for some random Pauli operator $P_i$, and apply the same $P_i$ after the noisy gate.
After Pauli twirling, the original non-Markovian noise is converted into random Pauli errors occurring at two time points, which are classically correlated with some joint probability distribution $\{p_{ij}\}_{i,j}$~\cite{liu2024non,winickConceptsConditionsError2022,figueroa-romeroOperationalMarkovianizationRandomized2024}.
As shown in Fig.~\ref{fig:flowchart}(b), the output state after Pauli twirling is
\begin{equation}
\label{eq:pauli_twirling}
\sum_{ij}p_{ij}(P_j U P_i) \rho (P_i U^\dagger P_j),
\end{equation}
where $\rho=U_1\rho_0U_1^\dagger$ and $U=U_2$.
Here, both \( P_i \) and \( P_j \) denote Pauli operators that are consistent with the dimension of the system, where \( P_0 \) is defined as the identity operator. We leave the proof and extension to multiple time points to Appendix~\ref{app:pauli_twirling}.

Quantum purification aims to consume multiple copies of the noisy state or process to get a purified one with a lower noise rate~\cite{cotler2019cooling,hugginsVirtualDistillationQuantum2021,koczorExponentialErrorSuppression2021,liu2025virtual}.
The advantages of purification-based error suppression protocols include that they can work for unknown quantum states or processes, and the error rate can be suppressed exponentially with the number of copies.
We thus employ the same logic and devise a purification circuit for the non-Markovian noise processed by Pauli twirling.
As shown in Fig.~\ref{fig:flowchart}(c), our setup involves three key components: a main register initialized in the input state $\rho_0$ of the original circuit; an ancillary register with the same size initialized in a maximally mixed state; and a control qubit prepared in the $\ket{+}$ state.
After that, we use two pairs of CSWAP gates to sandwich two sequential noisy gates.
The CSWAP gate swaps two states on the main and ancillary registers depending on the state of the control qubit.
A completely depolarizing channel labeled by $\mathcal{D}$ is applied on the ancillary register at the middle of the two pairs of CSWAP gates.
At the end of the circuit, we discard the ancillary register and perform Pauli-$X$ basis measurements on the control qubit, which is used to post-process the output state of the main register. 

\begin{figure*}[htbp]
\centering
\includegraphics[width=1\linewidth]{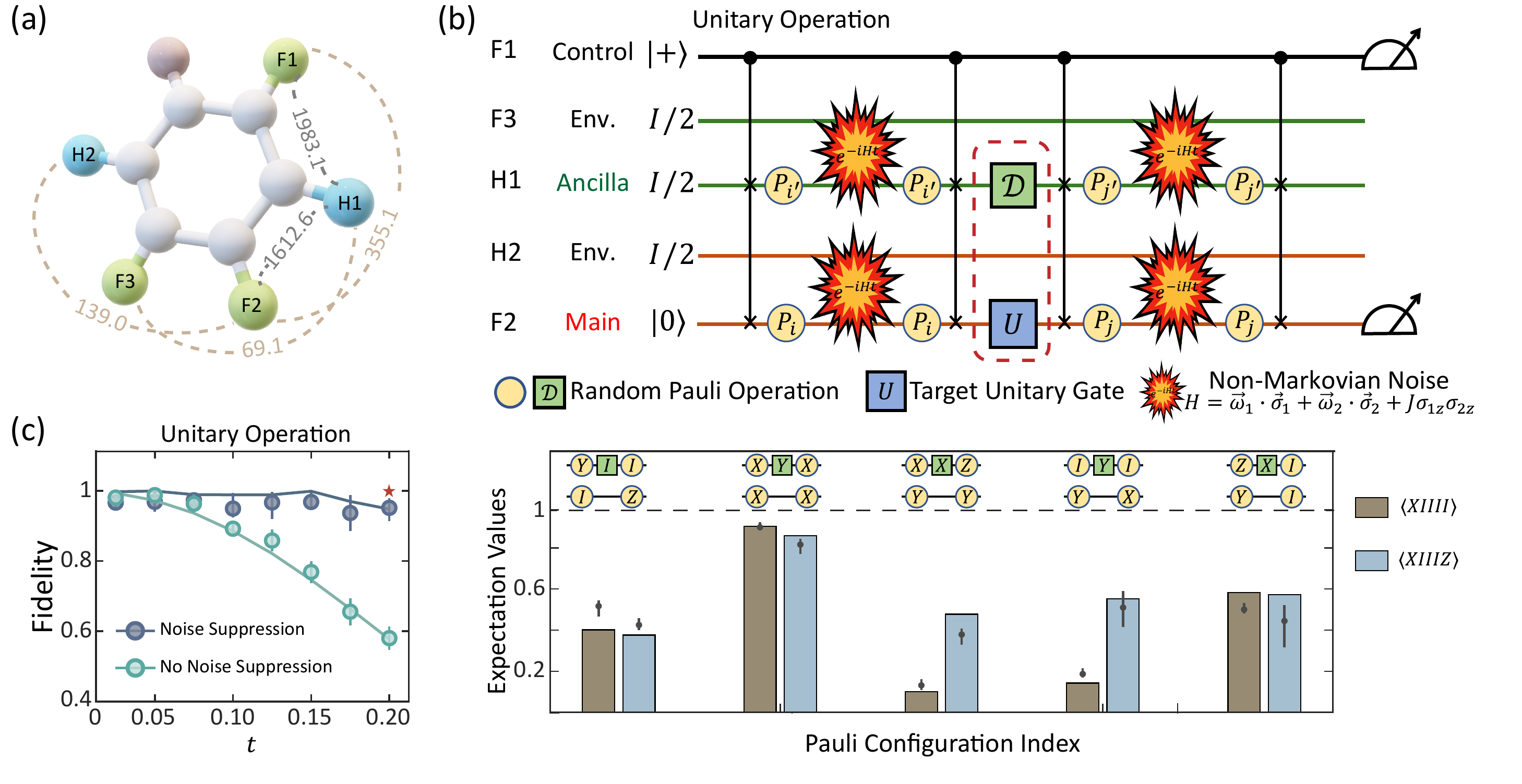}
\caption{(a) Molecular structure of the five-qubit NMR sample, comprising three $^{19}$F and two $^{1}$H spins as qubits. The dashed lines indicate the strong couplings, with the corresponding coupling strengths (in Hz) labeled alongside.
(b) Experimental circuit for non-Markovian noise suppression in the case of unitary gates.
The yellow circles $P_{i,j}$, $P_{i',j'}$ and the depolarizing operations $\mathcal{D}$ denote uniformly random Pauli operations, used for Pauli twirling and resetting the ancillary register to the maximally mixed state, respectively. 
The non-Markovian noise is generated by the joint Hamiltonian evolution governed by the Hamiltonian $H=\vec{\omega}_1 \cdot \vec{\sigma}_1+\vec{\omega}_2 \cdot \vec{\sigma}_2 +J\sigma_{1z}\sigma_{2z}$, where $\vec{\sigma}_1$ and $\vec{\sigma}_2$ are Pauli vectors for the upper and lower qubits, respectively. 
Here $\vec{\omega}_1=(\omega_{1x},\omega_{1y},\omega_{1z})=(2.0,1.3,1.0)=-\vec{\omega}_2$ and $J=1$. 
(c) Fidelity results for the unitary operation case.
For each evolution time $t \in [0,0.2]$, 10 configurations of random Pauli operators are sampled, and the experiment is repeated once for each configuration. 
Blue and green circles represent the experimental fidelities with and without the noise suppression protocol, respectively, while solid lines denote the simulated results.
At $t=0.2$ (red star), the expectation values obtained from five random configurations are shown on the right. 
The bars denote the simulated expectation values, while the gray solid circles and error bars indicate the corresponding experimental results. 
The Pauli gate combinations used in the experiment are illustrated above each group.
}
\label{nMfig1:exp}
\end{figure*}

To understand why the circuit can suppress the non-Markovian noise, one can notice that $U\otimes U$ commutes with the CSWAP gate.
Therefore, the two $U_1$ gates and the two $U_2$ gates can be moved in front of the first and the third CSWAP gates, which transforms Fig.~\ref{fig:flowchart}(c) into a simplified circuit shown in Fig.~\ref{fig:flowchart}(d).
Compared with the original circuit shown in Fig.~\ref{fig:flowchart}(b), the non-Markovian noise does not directly affect the ideal gate $U$.
Instead, it is purified with the purification circuit and then affects $U$.
Specifically, suppose the probabilities of measuring the control qubit in $\ket{+}$ and $\ket{-}$ are $p_+$ and $p_-$, respectively, and the corresponding conditional states of the main register are $\rho_+$ and $\rho_-$. Assuming Pauli distributions of the two Pauli twirled non-Markovian noises on the main and ancillary registers are both $\{p_{ij}\}_{i,j}$, one can effectively obtain the state 
\begin{equation}\label{eq:virtual_purification}
\rho_{\mathrm{eff}} \coloneqq\frac{p_+\rho_+ - p_-\rho_-}{p_+-p_-}=\frac{\sum_{ij}p_{ij}^2(P_j U P_i) \rho (P_i U^\dagger P_j)}{\sum_{ij}p_{ij}^2}
\end{equation}
through post-processing, which is proved in Appendix~\ref{app:proof_vp}.
In the weak noise regime where $p_{00} > p_{ij}$ with $i\neq 0$ or $j\neq 0$ and $P_0=I$ represents the noiseless component, the error rate of $\rho_{\mathrm{eff}}$ is lower than the one in Eq.~\eqref{eq:pauli_twirling} as $p_{00}^2/\sum_{ij}p_{ij}^2\ge p_{00}$.
To extract information from $\rho_{\mathrm{eff}}$, such as $\Tr(O\rho_{\mathrm{eff}})$ with $O$ being some observable, one can measure both the control qubit and the main register,
\begin{equation}
\Tr(O\rho_{\mathrm{eff}})=\frac{\expval{X\otimes O}}{\expval{X}},
\label{TrO}
\end{equation}
where $\expval{X}$ is the expectation value of Pauli $X$ operator on the control qubit and $\expval{X\otimes O}$ is the expectation value of $X\otimes O$ on the control qubit and main register.
It is worth mentioning that all these conclusions can be easily generalized to cases where Pauli distributions of non-Markovian noises acting on the main and ancillary registers are not identical, shown in Appendix~\ref{app:non_identical}.

With these analyses, we can summarize the advantages of the non-Markovian noise suppression protocol.
First, both Pauli twirling and purification do not utilize the knowledge of the target noise, avoiding the need for calibrating complex non-Markovian noise.
Besides, the only requirement of the noise is that the noise-free probability is larger than probabilities of all other noise components, i.e., $p_{00}>p_{ij}$ in Eq.~\eqref{eq:virtual_purification}, without assuming the spatial and temporal locality.
According to the commutation relation between the $U\otimes U$ and the CSWAP, our protocol can be applied to both idle and gate noises.
Moreover, the performances of our protocol, including the error suppression rate and sample complexity, can be theoretically derived and shown in Appendix~\ref{app:pauli_twirling}, \ref{app:proof_vp}, and \ref{app:complexity}.
In Appendix~\ref{app:generalization}, we show that our protocol can be flexibly adjusted to multi-time and multi-copy versions for handling different practical scenarios and achieving better error suppression performances.
Specifically, the error suppression rate grows exponentially with the copy number of the non-Markovian noise involved in the protocol, similar to the state and channel purification cases~\cite{hugginsVirtualDistillationQuantum2021,koczorExponentialErrorSuppression2021,liu2025virtual}.

\begin{figure*}[t]
\centering
\includegraphics[width=1\linewidth]{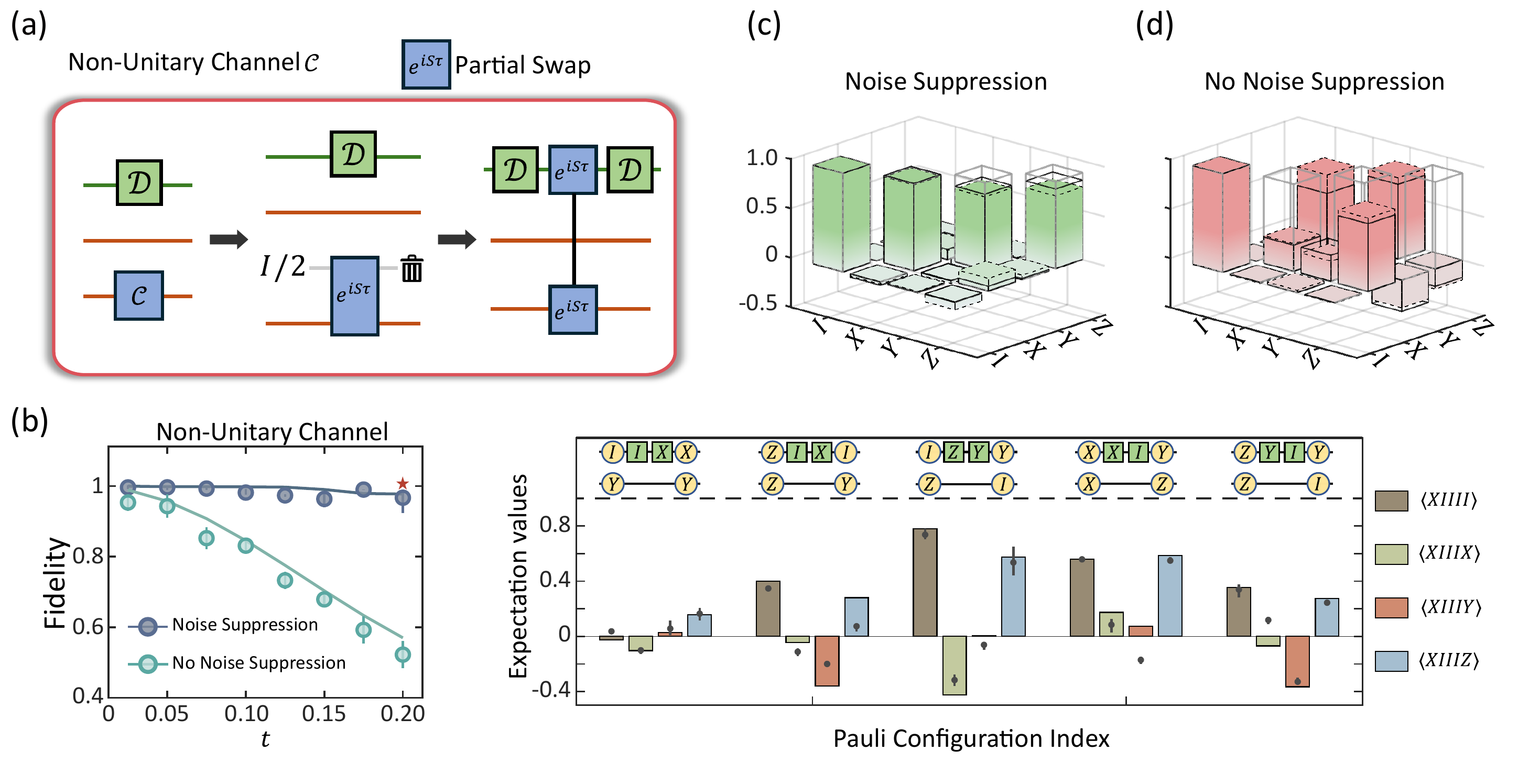}
\caption{(a) Experimental circuit for the non-unitary channel $\mathcal{C}$. Shown here is the segment that replaces the part enclosed by the red dashed box in the unitary operation case of Fig.~\ref{nMfig1:exp}(b).
The left circuit shows the target evolution, while the right depicts the experimental implementation, with $\tau=0.35$.
(b) Fidelity results for the non-unitary channel case.
For evolution times $t \in [0,0.2]$, 10 random Pauli configurations are sampled, with one experimental run performed for each configuration. 
Experimental fidelities with and without the suppression protocol are shown as blue and green circles, respectively, while the solid lines represent numerical simulations.
At $t=0.2$ (red star), the expectation values obtained from five random configurations are shown on the right. 
The bars denote the simulated expectation values, while the gray solid circles and error bars indicate the corresponding experimental measurements. 
The Pauli gate combinations used in the experiment are illustrated above each group.
(c–d) Process matrices of the non-unitary channel. Gray solid bars correspond to the ideal partial SWAP channel, while black solid bars and colored bars with dashed outlines represent the simulated and experimental results at $t=0.2$, respectively. In (c), green bars correspond to the result with the noise suppression protocol, and in (d), red bars correspond to the result without it.}
\label{nMfig2:exp}
\end{figure*}

However, the practical realization of our protocol also faces challenges.
The primary difficulty lies in the implementation of a high-fidelity CSWAP gate. In a qubit-based system, a CSWAP can be decomposed into a product of simple CSWAP gates, each involving one control qubit and two target qubits. In digital quantum simulators, such a three-qubit gate is typically decomposed into a sequence of multiple two-qubit entangling gates~\cite{smolin1996cswap,yu2013cswap}, which substantially increases the noise level.
This makes it crucial to explore physical platforms that can efficiently and robustly implement the CSWAP gate.

\noindent\emph{\bfseries Experimental setup.}---To experimentally implement a proof-of-principle demonstration of our non-Markovian noise suppression protocol, we employ a five-qubit NMR quantum processor. 
The sample consists of two $^1$H nuclear spins and three $^{19}$F nuclear spins in a 1-bromo-2,4,5-trifluorobenzene molecule, dissolved in the liquid crystal solvent MBBA~\cite{Niexf2024,Lizhaokai2014,CHEN202123}.
The molecular structure of the five-qubit NMR sample is shown in Fig.~\ref{nMfig1:exp}(a), where F1, F3, H1, H2, and F2 are designated as qubits. 
The choice of this sample is motivated by its strong spin-spin couplings, which facilitate multi-body operations, such as the CSWAP gate. 
All experiments were conducted at 299 K on a Bruker 600 MHz NMR spectrometer. 
We further improve the experimental control accuracy by designing a shaped pulse using gradient-based optimization~\cite{khaneja2005optimal}.

We aim to experimentally implement the simplified circuit in Fig.~\ref{fig:flowchart}(d), which essentially purifies the non-Markovian noise before it affects the ideal gate.
Therefore, a natural corollary is that the ideal gate is not necessarily unitary.
To evaluate the effectiveness and generality of our suppression protocol, we thus consider both the standard case with a unitary target gate and a more general case in which the target process is a non-unitary channel, such as engineered dissipation.
In both cases, we compare the noise rates with and without the suppression protocol. 
The experimental procedure consists of three steps: (i) initial state preparation, (ii) evolution, and (iii) measurement and post-processing.

Initial state preparation: The five-qubit system is first prepared in a pseudo-pure state $\ket{00000}$ from thermal equilibrium using the line-selective approach~\cite{PENG2001509,Niexf2024}. 
Then, the system is initialized in $\ketbra{+} \otimes \left( I/2 \right)^{\otimes 3} \otimes \ketbra{0}$ via single qubit rotations combined with gradient-field pulses, as shown in Fig.~\ref{nMfig1:exp}(b). 
Here and in what follows, $I$ denotes the 2-dimensional identity operator.
The details are shown in Appendix~\ref{app:exp}.
The first spin serves as a control qubit, the second and fourth spins act as environment qubits $I/2$. 
The third and fifth spins serve as the ancillary and main registers, respectively.

Evolution: 
The experimental circuits are depicted in Fig.~\ref{nMfig1:exp}(b).
Non-Markovian noise is engineered by coupling an environment qubit to the main register with joint Hamiltonian evolution $e^{-iHt}$ at two time points, as illustrated by the fourth and fifth spins (orange lines in the circuit). 
The evolution time $t$ is tunable to adjust the noise strength.
The completely depolarizing channel $\mathcal{D}$ is achieved by applying a uniformly random Pauli gate.
In addition, a copy of the noise is applied to the ancillary register, represented by the second and third spins (green lines in the circuit). 
In the non-unitary case shown in Fig.~\ref{nMfig2:exp}(a), which corresponds to the section enclosed by the red dashed box in Fig.~\ref{nMfig1:exp}(b), we aim to implement a partial SWAP channel on the main register defined as $\mathcal{C}(\rho)=\Tr_{2}[e^{iS\tau}(\rho\otimes I/2)e^{-iS\tau}]$, where $S$ is the SWAP operator, $\Tr_2$ represents partial trace over the second qubit.
Although realizing such a non-unitary channel would normally require an additional ancillary qubit, we avoid this by reusing the third qubit. 
We reset the third qubit to a maximally mixed state after the first pair of CSWAP gates and use it as an ancilla to implement the partial SWAP gate.
Afterwards, we reset it again to a maximally mixed state for the remaining circuit. 

Measurement and post-processing:
At the end of the circuit, we measure the control qubit and the main register and perform post-processing to obtain the noise-suppressed effective outcomes. 
The specific measurement sets and reconstruction procedures differ between the unitary and non-unitary cases and are described in detail below.

\noindent\emph{\bfseries Results.}---We first demonstrate the effectiveness of our protocol in the unitary operation case, where $U$ is set to be a Hadamard gate. 
To estimate the fidelity between the outputs of the actual and ideal circuits, a $U^\dagger$ operation is applied to the main register at the end of the circuit.
Then, the fidelity can be obtained from the expectation value $\expval{Z}$ in the output state.
According to Eq.~\eqref{TrO}, the noise-suppressed effective expectation value is obtained by measuring $\langle XIIII \rangle$ and $\langle XIIIZ \rangle$ in each experimental configuration and then averaging the results to calculate
\begin{equation}
\bket{Z_{\text{eff}}} = \frac{\bket{\overline{X III Z}}}{\bket{\overline{X IIII}}}.
\label{O_eff}
\end{equation}
Here and in what follows, $X$, $Y$, $Z$ share the same meaning as the Pauli operators $\sigma_x$, $\sigma_y$, $\sigma_z$. For reference, an experiment without the suppression protocol is also performed, where the output is denoted as $\bket{Z_\text{noise}}$. 
The comparison between cases with and without the suppression protocol is presented in Fig.~\ref{nMfig1:exp}(c), where the fidelity $F$ is calculated by $F=1-\frac{|1-\bket{Z}|}{2}$. 
As the evolution time $t$ increases, indicating stronger non-Markovian noise, the reference fidelity drops rapidly. 
In contrast, the fidelity under our suppression protocol remains robust, staying markedly higher even at large $t$. On the right of Fig.~\ref{nMfig1:exp}(c), we further display the experimental results from five random configurations at $t=0.2$, while the complete set of results is provided in the Appendix~\ref{app:pauliresult}.

For the non-unitary case, we perform quantum state tomography on the effective output state of the main register. In addition to $\langle X IIII \rangle$ and $\langle XIIIZ \rangle$, we also measure the expectation values $\langle X IIIX \rangle$ and $\langle X IIIY \rangle$. 
The effective Pauli expectation values are estimated following the procedure defined in Eq.~\eqref{O_eff}, which together yield the effective noise-suppressed output state
$\rho_{\text{eff}} = \frac{1}{2}\left( I + \langle X_{\text{eff}} \rangle X + \langle Y_{\text{eff}} \rangle Y + \langle Z_{\text{eff}} \rangle Z \right)$.
A corresponding reference experiment without the suppression protocol is also conducted under the same partial SWAP channel with the tomographic reconstructed output density matrix $\rho_{\text{noise}}$. 
The fidelity $F$ in the non-unitary case is calculated as $F=(\text{Tr}(\sqrt{\sqrt{\rho}\rho_\text{ideal}\sqrt{\rho}}))^2$, where $\rho_\text{ideal}=\mathcal{C}(\ketbra{0}{0})$ denotes the ideal output state of the channel $\mathcal{C}$ and $\rho$ is either $\rho_{\text{eff}}$ or $\rho_{\text{noise}}$. 
The fidelity comparison is shown in Fig.~\ref{nMfig2:exp}(b). 
Similar to the unitary case, as the evolution time $t$ increases, the fidelity obtained with the suppression protocol exhibits increasingly pronounced advantages.
At $t=0.2$, the measurement results from five randomly selected configurations are displayed on the right: simulated expectation values are shown as bars, while the corresponding experimental data with error bars are overlaid. 
The Pauli gate combinations used in each configuration are indicated above the respective groups.
More results are provided in the Appendix~\ref{app:pauliresult}.

We further perform quantum process tomography on the non-unitary channel, both with and without the suppression protocol. 
The reconstructed process matrices are compared with those of the ideal partial SWAP channel. 
Representative experimental results for $t=0.2$ are shown in Figs.~\ref{nMfig2:exp}(c) and (d). The close agreement between the ideal partial SWAP process matrix and the experimental process matrix obtained with noise suppression highlights the effectiveness of the suppression scheme.

\emph{\bfseries Conclusion.}---Given that non-Markovian noise is the most general and challenging form of noise in quantum computing, devising suppression schemes is particularly significant. 
We experimentally demonstrate on an NMR platform a calibration-free, model-agnostic protocol for suppressing non-Markovian noise, showing both effectiveness and agreement with theoretical analysis. 
The protocol is flexible to be extended to multiple time points or to larger ancillary systems to achieve stronger suppression. As our experiments are proof-of-principle, demonstrating the protocol under more realistic noise conditions and across a broader range of practical quantum information processing tasks will be a promising direction for future research.

\emph{\bfseries Acknowledgments.}---H.L., X.N., and D.L. are supported by the National Natural Science Foundation of China (Grants No. 12574543, 12575020), Guangdong Provincial Quantum Science Strategic Initiative (GDZX2203001, GDZX2203001, GDZX2403004), and GuangDong Basic and Applied Basic Research Foundation (2025A1515011599). Z.H. is supported by the National Natural Science Foundation of China (Grants No. T2225008), the Innovation Program for Quantum Science and Technology (No. 2021ZD0302203), Tsinghua University Dushi Program, and the Shanghai Qi Zhi Institute Innovation Program SQZ202318.
Z.L. is supported in part by a startup funding from YMSC, Dushi Program, and NSFC under Grant No.~12475023.

%

\appendix

\section{Proof for Pauli Twirling}\label{app:pauli_twirling}

In this section, we prove the effect of Pauli twirling on non-Markovian noise. A noisy process can be described as a joint channel acting on both the system and the environment. For Markovian noise involving two time points, it can be represented as two such joint channels, with some operations performed on the system in between, and the two noise channels share the same environment that retains memory. If we denote the initial states of the system and the environment as \(\rho\) and \(\sigma\), respectively, and the two noise channels as \(\mathcal{E}_1\) and \(\mathcal{E}_2\), then the state of the system after this noisy process can be expressed as:

\begin{equation} \rho_f=\mathrm{Tr}_{\mathrm{E}}[\mathcal{E}_2\circ(\mathcal{U}\otimes\mathcal{I})\circ\mathcal{E}_1(\rho\otimes\sigma)]
\end{equation}
Here \(\mathcal{U}\) and \(\mathcal{I}\) denote the intended quantum gate acting on the system and the identity channel on the environment, respectively, and \(\circ\) represents the sequential application of channels.

Our main result regarding Pauli twirling can be summarized in the following lemma:

\textit{Lemma 1}---Applying Pauli twirling to a two-time-point non-Markovian noise transforms the noise into random Pauli errors at the two points, with a correlated probability distribution:

\begin{equation} \mathrm{Tr_E}[\mathrm{T}_{\mathrm{S}}[\mathcal{E}_2]\circ(\mathcal{U}\otimes\mathcal{I})\circ\mathrm{T}_{\mathrm{S}}[\mathcal{E}_1](\rho\otimes\sigma)]=\sum_{ij}p_{ij}\mathcal{P}_j\circ \mathcal{U}\circ \mathcal{P}_i(\rho)
\end{equation}
Here, \(\mathcal{P}\) denotes a Pauli channel, and \(p_{ij}\) is the joint probability that Pauli errors \(\mathcal{P}_i\) and \(\mathcal{P}_j\) occur at the two respective time points. $\mathrm{T}_{\mathrm{S}}[\cdot]$ denotes performing Pauli twirling to a process on the system part.

\begin{proof}
As a starting point, we first establish the following basic result:

\textit{Lemma 1.1}---Let $\rho$ be a density matrix, and let $P_i$ and $P_j$ be arbitrary $n$-qubit Pauli operators. Then:

\begin{equation}
  \frac{1}{4^n}\sum_{k=0}^{4^n-1}P_kP_i(P_k\rho P_k)P_jP_k=\delta_{ij} P_i \rho P_i
\end{equation}

\begin{proof} 
First, when a Pauli operator is conjugated by another Pauli operator, it may acquire a sign change: $P_k P_i P_k = (-1)^{m_{ik}} P_i$, $m_{ik}$ is the number of sites at which the single-qubit Pauli operators composing $P_i$ and $P_k$ anticommute.

For $i = j$, a direct calculation yields:

\begin{equation}
  \frac{1}{4^n}\sum_{k=0}^{4^n-1}P_kP_iP_k\rho P_kP_iP_k=P_i\rho P_i
\end{equation}

For $i\neq j$, we have:
\begin{equation}
    \begin{split}
        &\frac{1}{4^n}\sum_{k=0}^{4^n-1}P_kP_iP_k\rho P_kP_jP_k\\
        =&\frac{1}{4^n}\sum_{k=0}^{4^n-1}(-1)^{m_{ik}}P_i\rho (-1)^{m_{jk}}P_j\\
        =&\frac{1}{4^n}\sum_{k=0}^{4^n-1}(-1)^{m_{ik}+m_{jk}}P_i\rho P_j
    \end{split}
\end{equation}
To compute $m$, we can evaluate it site by site: $m_{ik} = \sum_{s=0}^{n-1} m_{ik}^{(s)}$, where $m_{ik}^{(s)}$ indicates whether the single-qubit Pauli operators of $P_i$ and $P_k$ at site $s$ commute or anticommute, taking the value $0$ if they commute and $1$ if they anticommute. Thus:

\begin{equation}
\label{eq:a4}
    \begin{split}
        &\frac{1}{4^n}\sum_{k=0}^{4^n-1}P_kP_iP_k\rho P_kP_jP_k\\
        =&\frac{1}{4^n}\sum_{k=0}^{4^n-1}(-1)^{\sum_{s=0}^{n-1}(m_{ik}^{(s)}+m_{jk}^{(s)})}P_i\rho P_j\\
        =&\frac{1}{4^n}P_i\rho P_j\sum_{k=0}^{4^n-1}\prod_{s=0}^{n-1}(-1)^{m_{ik}^{(s)}+m_{jk}^{(s)}}\\
        =&\frac{1}{4^n}P_i\rho P_j\prod_{s=0}^{n-1}\left[\sum_{P_k^{(s)}\in\{I,X,Y,Z\}}(-1)^{m_{ik}^{(s)}+m_{jk}^{(s)}}\right]
    \end{split}
\end{equation}
Here, $P_k^{(s)}$ denotes the single-qubit operator at the $s$-th site of the Pauli operator $P_k$. Since $P_i \neq P_j$, there exists at least one $s$ such that $P_i^{(s)} \neq P_j^{(s)}$. Find one such $s$, the relation between $P_i^{(s)}$ and $P_j^{(s)}$ can be further split into two cases. 

First case, one of $P_i^{(s)}$ and $P_j^{(s)}$ is the identity. Without loss of generality, assume $P_i^{(s)}=I$ and $P_j^{(s)}=X$. Then, as $P_k^{(s)}$ takes the values $I,X,Y,Z$ in turn, the corresponding pairs $(m_{ik}^{(s)},m_{jk}^{(s)})$ are $(0,0),(0,0),(0,1),(0,1)$, respectively, and thus the sum inside the brackets $[\ ]$ in Eq.~\eqref{eq:a4} evaluates to $0$.

Second case, both $P_i^{(s)}$ and $P_j^{(s)}$ are not the identity $I$. Without loss of generality, assume $P_i^{(s)}=X$ and $P_j^{(s)}=Y$. Then, as $P_k^{(s)}$ takes the values $I,X,Y,Z$ in turn, the corresponding pairs $(m_{ik}^{(s)},m_{jk}^{(s)})$ are $(0,0),(0,1),(1,0),(1,1)$, respectively, and the sum inside the brackets in Eq.~\eqref{eq:a4} still evaluates to $0$.

Therefore, when $P_i \neq P_j$, at least one factor in the product in Eq.~\eqref{eq:a4} vanishes, making the entire expression zero. This completes the proof of Lemma 1.1.
\end{proof}

We now return to Lemma 1. The initial system and environment are uncorrelated, with the joint initial state given by $\rho \otimes \sigma$. The first noise segment $\mathcal{E}_1$, as a joint channel acting on the system and the environment, can be expanded in terms of a $\chi$-matrix representation: 
\begin{equation}
\mathcal{E}_1(\rho\otimes\sigma)=\sum_{ijkl}\chi_{ijkl}(P_i^{\mathrm{S}}\otimes P_j^{\mathrm{E}})(\rho\otimes\sigma)(P_k^{\mathrm{S}}\otimes P_l^{\mathrm{E}})
\end{equation}
The superscripts $\mathrm{S}$ and $\mathrm{E}$ indicate that the Pauli operators act on the system and on the environment, respectively. We focus on the system part, thereby hiding the summation over the environment degrees of freedom, and define $\sigma_{ik} = \sum_{j l} \chi_{ijkl} \, P_j^{\mathrm{E}} \, \sigma \, P_l^{\mathrm{E}}$. Thus, the action of the channel $\mathcal{E}_1$ can be written as:

\begin{equation}
\label{eq:E1_act}
\mathcal{E}_1(\rho\otimes\sigma)=\sum_{ik}(P_i^{\mathrm{S}}\rho P_k^{\mathrm{S}})\otimes \sigma_{ik}
\end{equation}
According to Lemma 1.1, performing Pauli twirling on the system part of this process yields:

\begin{equation}
\label{eq:E1_tw_act}
\mathrm{T}_{\mathrm{S}}[\mathcal{E}_1](\rho\otimes\sigma)=\sum_{i}(P_i^{\mathrm{S}}\rho P_i^{\mathrm{S}})\otimes \sigma_{ii}
\end{equation}
Applying the ideal quantum gate $U$ afterwards, we obtain:
\begin{equation}
(\mathcal{U}\otimes \mathcal{I})\circ\mathrm{T}_{\mathrm{S}}[\mathcal{E}_1](\rho\otimes\sigma)=\sum_{i}(UP_i^{\mathrm{S}}\rho P_i^{\mathrm{S}}U^\dag)\otimes \sigma_{ii}
\end{equation}

Next, the system and environment undergo a second joint noise channel $\mathcal{E}_2$. Owing to linearity, we can consider the action of $\mathcal{E}_2$ on each term in the above summation separately:

\begin{equation}
\mathcal{E}_2\circ(\mathcal{U}\otimes \mathcal{I})\circ\mathrm{T}_{\mathrm{S}}[\mathcal{E}_1](\rho\otimes\sigma)=\sum_{i}\mathcal{E}_2((UP_i^{\mathrm{S}}\rho P_i^{\mathrm{S}}U^\dag)\otimes \sigma_{ii})
\end{equation}
The action on each term has exactly the same form as the effect of $\mathcal{E}_1$ in Eq.~\eqref{eq:E1_act}. By introducing two new indices $i'$ and $j'$, we then have:

\begin{equation}
\begin{split}
&\mathcal{E}_2\circ(\mathcal{U}\otimes \mathcal{I})\circ\mathrm{T}_{\mathrm{S}}[\mathcal{E}_1](\rho\otimes\sigma)\\
=&\sum_{ii'k'}(P_{i'}^{\mathrm{S}} UP_i^{\mathrm{S}}\rho P_i^{\mathrm{S}}U^\dag P_{k'}^{\mathrm{S}}) \otimes \sigma_{ii'k'}
\end{split}
\end{equation}
Here, we do not need the explicit form of $\sigma_{ii'k'}$; it is only important to emphasize that, in each summation term, it appears in the form of a tensor product with the system part.

Finally, we perform Pauli twirling on the system part of the second noise $\mathcal{E}_2$ that we just introduced. By the same derivation as from Eq.~\eqref{eq:E1_act} to Eq.~\eqref{eq:E1_tw_act}, we obtain:

\begin{equation}
\begin{split}
\label{eq:E2_tw_act}
&\mathrm{T}_{\mathrm{S}}[\mathcal{E}_2]\circ(\mathcal{U}\otimes \mathcal{I})\circ\mathrm{T}_{\mathrm{S}}[\mathcal{E}_1](\rho\otimes\sigma)\\
=&\sum_{ii'}(P_{i'}^{\mathrm{S}} UP_i^{\mathrm{S}}\rho P_i^{\mathrm{S}}U^\dag P_{i'}^{\mathrm{S}}) \otimes \sigma_{ii'i'}
\end{split}
\end{equation}

Tracing out the environment yields the system's final state:

\begin{equation}
\begin{split}
&\mathrm{Tr}_{\mathrm{E}} [\mathrm{T}_{\mathrm{S}}[\mathcal{E}_2]\circ(\mathcal{U}\otimes \mathcal{I})\circ\mathrm{T}_{\mathrm{S}}[\mathcal{E}_1](\rho\otimes\sigma)]\\
=&\sum_{ii'}p_{ii'}P_{i'}^{\mathrm{S}} UP_i^{\mathrm{S}}\rho P_i^{\mathrm{S}}U^\dag P_{i'}^{\mathrm{S}} \\
=&\sum_{ii'}p_{ii'}\mathcal{P}_{i'}^{\mathrm{S}}\circ \mathcal{U}\circ \mathcal{P}_{i}^{\mathrm{S}} (\rho)
\end{split}
\end{equation}
Here $p_{ii'}$ is a joint probability distribution, with $\sum_{ii'} p_{ii'} = 1$, $\mathcal{P}_i$ denotes the Pauli channel corresponding to operator $P_i$. The probability $p_{ii'}$ can be written in terms of the parameters in Eq.~\eqref{eq:E2_tw_act} as
$$
p_{ii'} = \frac{\,\operatorname{Tr}\big[\sigma_{ii'i'}\big]}{\sum_{ii'} \,\operatorname{Tr}\big[\sigma_{ii'i'}\big]}.
$$
This completes the proof of Lemma 1.
\end{proof}

It is worth noting that the result of Lemma 1 can be generalized to non-Markovian noise with an arbitrary number of time points involved, following a similar derivation. Assume non-Markovian noise at $n$ time points, $\mathcal{E}_1,\ldots,\mathcal{E}_n$, interleaved with $n-1$ quantum gates $\mathcal{U}_1,\ldots,\mathcal{U}_{n-1}$. After applying Pauli twirling to the $n$ noise points, one obtains $n$ correlated random Pauli errors:

\begin{equation} 
\begin{split}
&\mathrm{Tr_E}\left[\mathrm{T}[\mathcal{E}_n]\circ(\mathcal{U}_{n-1}\otimes\mathcal{I})\cdots\mathrm{T}[\mathcal{E}_2]\circ(\mathcal{U}_1\otimes\mathcal{I})\circ\mathrm{T}[\mathcal{E}_1](\rho\otimes\sigma)\right]\\
&=\sum_{i_1,...i_n}p_{i_1,...i_n}\mathcal{P}_{i_n}\circ\mathcal{U}_{n-1}\cdots\mathcal{P}_{i_2}\circ \mathcal{U}_1\circ \mathcal{P}_{i_1}(\rho)
\end{split}
\end{equation}

\section{Proof for purification}

\label{app:proof_vp}

In this section, we will prove that for a twirled non-Markovian noise involving two time points, when the noise rate is not too large, the purification circuit in Fig.~\ref{fig:flowchart}(c) and (d) of the main text can effectively yield a final state with less noise.

Since Figs.~\ref{fig:flowchart} (c) and (d) are equivalent, we proceed directly from Fig.~\ref{fig:flowchart}(d). The initial state of the main register is $\rho$, so the joint initial state is:$\ketbra{+}{+}\otimes I_d/d\otimes \rho$. Executing the circuit in Fig.~\ref{fig:flowchart}(d) step by step, we have:

\begin{widetext}
\begin{equation}\label{eq:rho_f}
\begin{aligned}
&\ketbra{+}{+}\otimes I_d/d\otimes \rho\\
\xrightarrow{\mathrm{CSWAP}\ \ }& \frac{1}{2d}(\ketbra{0}{0}\otimes I_d\otimes \rho +\ketbra{0}{1}\otimes (I_d\otimes\rho)S
+\ketbra{1}{0}\otimes S(I_d\otimes \rho)+\ketbra{1}{1}\otimes S(I_d\otimes \rho)S)\\
\xrightarrow{\mathrm{Noise}\ \ \ \ \ \ }& \mathbb{E}_{ii'}\bigg[\frac{1}{2d}(\ketbra{0}{0}\otimes I_d\otimes P_i\rho P_i +\ketbra{0}{1}\otimes (P_{i'}P_i\otimes P_i\rho P_{i'})S
+\ketbra{1}{0}\otimes S(P_i P_{i'}\otimes P_{i'}\rho P_i)\\
&+\ketbra{1}{1}\otimes (P_{i'}\rho P_{i'})\otimes I_d)\bigg]\\
\xrightarrow{\mathrm{CSWAP}\ \ } & \mathbb{E}_{ii'}\bigg[\frac{1}{2d}(\ketbra{0}{0}\otimes I_d\otimes P_i\rho P_i +\ketbra{0}{1}\otimes P_{i'}P_i\otimes P_i\rho P_{i'}
+\ketbra{1}{0}\otimes P_i P_{i'}\otimes P_{i'}\rho P_i+\ketbra{1}{1}\otimes  I_d\otimes P_{i'}\rho P_{i'})\bigg]\\
\xrightarrow{\mathcal{D}\ \mathrm{and}\ \mathcal{U}\ \ } &\mathbb{E}_{ii'}\bigg[ \frac{1}{2d}(\ketbra{0}{0}\otimes I_d\otimes \mathcal{U}(P_i\rho P_i) +\delta_{ii'}\ketbra{0}{1}\otimes I_d\otimes \mathcal{U}(P_i\rho P_{i'})
+\delta_{ii'}\ketbra{1}{0}\otimes I_d\otimes \mathcal{U}(P_{i'}\rho P_i)\\
&+\ketbra{1}{1}\otimes  I_d\otimes \mathcal{U}(P_{i'}\rho P_{i'}))\bigg]\\
\xrightarrow{\mathrm{CSWAP}\ \ }&\mathbb{E}_{ii'}\bigg[\frac{1}{2d}(\ketbra{0}{0}\otimes I_d\otimes \mathcal{U}(P_i\rho P_i)+\delta_{ii'}\ketbra{0}{1}\otimes S(I_d\otimes \mathcal{U}(P_i\rho P_{i'}))
+\delta_{ii'}\ketbra{1}{0}\otimes (I_d\otimes \mathcal{U}(P_{i'}\rho P_i))S\\
&+\ketbra{1}{1}\otimes  S(I_d\otimes \mathcal{U}(P_{i'}\rho P_{i'}))S)\bigg]\\
\xrightarrow{\mathrm{Noise}\ \ \ \ \ \ }&\mathbb{E}_{ii'jj'}\bigg[\frac{1}{2d}(\ketbra{0}{0}\otimes I_d\otimes P_j\mathcal{U}(P_i\rho P_i) P_j +\delta_{ii'}\ketbra{0}{1}\otimes S(P_{j'}P_j\otimes P_j\mathcal{U}(P_i\rho P_{i'}) P_{j'})\\
&+\delta_{ii'}\ketbra{1}{0}\otimes (P_j P_{j'}\otimes P_{j'}\mathcal{U}(P_{i'}\rho P_i) P_j)S+\ketbra{1}{1}\otimes (P_{j'}\mathcal{U}(P_{i'}\rho P_{i'}) P_{j'}\otimes I_d))\bigg]\\
\xrightarrow{\mathrm{CSWAP}\ \ }&\mathbb{E}_{ii'jj'}\bigg[\frac{1}{2d}(\ketbra{0}{0}\otimes I_d\otimes P_j\mathcal{U}(P_i\rho P_i) P_j +\delta_{ii'}\ketbra{0}{1}\otimes P_{j'}P_j\otimes P_j\mathcal{U}(P_i\rho P_{i'}) P_{j'}\\
&+\delta_{ii'}\ketbra{1}{0}\otimes P_j P_{j'}\otimes P_{j'}\mathcal{U}(P_{i'}\rho P_i) P_j+\ketbra{1}{1}\otimes I_d\otimes P_{j'}\mathcal{U}(P_{i'}\rho P_{i'}) P_{j'})\bigg]
\end{aligned}
\end{equation}
\end{widetext}
Here, $\mathbb{E}$ denotes the expectation taken over the Pauli operators in the twirled non-Markovian noise, $S$ is the swap operator between the main and the ancillary registers, \(\mathcal{U}\) is the channel corresponding to the ideal gate. In the fourth step, we regard the completely depolarizing channel $\mathcal{D}$ as the action of a uniformly random Pauli operator on the ancillary register. We denote the final state obtained from the above expression as $\rho_f$.

Next, we calculate the post-selected states obtained by measuring the control qubit:

\begin{equation}
\begin{split}
p_+\rho_+=&\mathrm{Tr}_{1,2}[(\ketbra{+}{+}\otimes I_d\otimes I_d)\rho_f]\\
=&\mathbb{E}_{ii'jj'}\bigg[\frac{1}{4}(P_j\mathcal{U}(P_i\rho P_i) P_j+2\delta_{ii'}\delta_{jj'}P_j\mathcal{U}(P_i\rho P_i) P_j\\
&+P_{j'}\mathcal{U}(P_{i'}\rho P_{i'}) P_{j'})\bigg]
\end{split}
\end{equation}

\begin{equation}
\begin{split}
p_-\rho_-=&\mathrm{Tr}_{1,2}[(\ketbra{-}{-}\otimes I_d\otimes I_d)\rho_f]\\
=&\mathbb{E}_{ii'jj'}\bigg[\frac{1}{4}(P_j\mathcal{U}(P_i\rho P_i) P_j-2\delta_{ii'}\delta_{jj'}P_j\mathcal{U}(P_i\rho P_i) P_j\\
&+P_{j'}\mathcal{U}(P_{i'}\rho P_{i'}) P_{j'})\bigg]
\end{split}
\end{equation}
Here, $\mathrm{Tr}_{1,2}$ denotes taking the trace over the control qubit and the ancillary register. Take the difference between the two terms above:

\begin{equation}
\begin{split}
\label{eq:vp_sub}
p_+\rho_+-p_-\rho_-=&\mathbb{E}_{ii'jj'}\bigg[\delta_{ii'}\delta_{jj'}P_j\mathcal{U}(P_i\rho P_i) P_j\bigg]\\
=&\sum_{ii'jj'}p_{ij}p_{i'j'}\delta_{ii'}\delta_{jj'}P_j\mathcal{U}(P_i\rho P_i) P_j\\
=&\sum_{ij}(p_{ij})^2P_j\mathcal{U}(P_i\rho P_i) P_j
\end{split}
\end{equation}
Here, $p_{ij}$ and $p_{i'j'}$ are the joint distributions of Pauli errors in the twirled non-Markovian noise acting on the main and the ancillary register, respectively. They are assumed to be the same distribution here.

After normalization, we obtain the resulting state of the purification:

\begin{equation}
\begin{split}
\rho_{\mathrm{eff}}=&\frac{p_+\rho_+-p_-\rho_-}{p_+-p_-}\\
=&\frac{\sum_{ij}(p_{ij})^2P_j\mathcal{U}(P_i\rho P_i) P_j}{\sum_{ij}(p_{ij})^2}
\end{split}
\end{equation}

As a comparison, we consider the case without purification, where only Pauli twirling is performed on the non-Markovian noise on the system, and the final state of the system is:

\begin{equation}
\begin{split}
\rho_{\mathrm{tw}}=\sum_{ij}p_{ij}P_j\mathcal{U}(P_i\rho P_i) P_j
\end{split}
\end{equation}

It is easy to see that purification effectively squares the weights $p_{ij}$ corresponding to the error terms $P_i$ and $P_j$ in the final state without purification, followed by normalization. This transformation increases the proportion of the originally largest term while reducing the proportions of all other terms. Therefore, if the term where both Pauli operators are the identity (i.e., the error-free term) is the most significant in the original final state, purification will further amplify the proportion of this error-free term in the mixture, thereby achieving error suppression:

\begin{equation}
    \begin{split}
        \frac{p_{00}^2}{\sum_{ij}p_{ij}^2}=&\frac{1}{1+\sum_{ij}'(p_{ij}/p_{00})^2}\\
        >&\frac{1}{1+\sum_{ij}'p_{ij}/p_{00}}\\
        =&\frac{p_{00}}{p_{00}+\sum'_{ij}p_{ij}}\\
        =&\frac{p_{00}}{\sum_{ij}p_{ij}}\\
        =& p_{00}
    \end{split}
\end{equation}
Where \(\sum'_{ij}\) denotes the sum over all terms except \(ij = 00\). The inequality in the second line follows from the fact that when \( p_{00} > p_{ij} \), we have $(p_{ij}/p_{00})^2<p_{ij}/p_{00}$.

Notably, throughout the derivation in this section, we did not rely on the unitary nature of the channel $\mathcal{U}$. Consequently, the derivation remains valid if $\mathcal{U}$ is replaced by a non-unitary channel. This shows that our non-Markovian noise-suppression protocol also works when the ideal process is a non-unitary channel.

\section{Sample Complexity}\label{app:complexity}

To measure an observable $O$ on the main register after our error-suppression protocol, we need to estimate

\begin{equation}
\langle O_{\mathrm{eff}}\rangle=\mathrm{Tr}[O\rho_{\mathrm{eff}}]
=\frac{\operatorname{Tr}\!\left[\,O\,(p_{+}\rho_{+}-p_{-}\rho_{-})\right]}{p_{+}-p_{-}}
\end{equation}
to a specified accuracy. We need to estimate it by measuring the final state in Eq.~\eqref{eq:rho_f}. By definition, it is easy to see that $p_+ - p_-$ is equal to the expectation value of measuring the Pauli-$X$ on the control qubit. And $\operatorname{Tr}\!\left[\,O\,(p_{+}\rho_{+}-p_{-}\rho_{-})\right]=p_+\langle O\rangle_{+}-p_-\langle O\rangle_{-}=p_+\langle X\otimes O\rangle_{+}+p_-\langle X\otimes O\rangle_{-}=\langle X\otimes O\rangle$, is equal to simultaneously measuring $X$ on the control qubit and $O$ on the main register and taking the product of the outcomes. Here $\langle O\rangle_{+}$ and $\langle O\rangle_{-}$ denote the conditional expectation values of observable $O$ given that the measurement of Pauli-$X$ on the control qubit yields $+$ and $-$, respectively. Throughout this section, we always ignore (i.e., trace out) the ancillary register, since we do not perform any measurements on it.

Therefore, the estimator for the desired expectation value of $O$ on the main register after purification is given by the ratio of two measured expectation values:

\begin{equation}    \expval{O_{\mathrm{eff}}}=\frac{\expval{X\otimes O}}{\expval{X\otimes I}}
\end{equation}
Since in practical experiments we can perform only a finite number of measurements, the estimator we actually use can be written as $\langle O_{\mathrm{eff}}\rangle\approx x/y$, where $x$ and $y$ are the averages (over a finite number of shots) of measurements of $X\!\otimes\! O$ and $X\!\otimes\! I$, respectively.

To assess the complexity, we examine the variance of this estimator when $K$ measurement shots are performed. When $K$ is sufficiently large—so that $\mathrm{Var}(y)$ is small—the variance of the ratio can be approximated as \cite{seltman2012approximations,liu2025virtual}:

\begin{equation}
\label{eq:var_x_y}
    \mathrm{Var}\left( \frac{x}{y}\right)\approx \frac{\mu_x^2}{\mu_y^2}\left(\frac{\mathrm{Var}(x)}{\mu_x^2}-2\frac{\mathrm{Cov}(x,y)}{\mu_x\mu_y}+\frac{\mathrm{Var}(y)}{\mu_y^2}\right)
\end{equation}
We will evaluate each term using the final state in Eq.~\eqref{eq:rho_f}. $\mu_x$ and $\mu_y$ are the expectation values of $x$ and $y$, respectively.

\begin{widetext}
\begin{equation}
\begin{split}
    \mu_y=&\expval{X\otimes I}\\
    =&\sum_{iji'j'}p_{ij}p_{i'j'}\frac{1}{2}\bigg[\bra{0}X\ket{0}\mathrm{Tr}\left(P_jUP_i\rho P_i U^\dag P_j\right) + \delta_{ii'}\delta_{jj'}\bra{1}X\ket{0}\mathrm{Tr}(P_jUP_i \rho P_{i'}U^\dag P_{j'})\\
    & +\delta_{ii'}\delta_{jj'} \bra{0}X\ket{1} \mathrm{Tr}(P_{j'}UP_{i'} \rho P_{i}U^\dag P_{j})+\bra{1}X\ket{1}\mathrm{Tr}\left(P_{j'}UP_{i'}\rho P_{i'} U^\dag P_{j'}\right) \bigg]\\
    =&\sum_{ij}p_{ij}^2
    \label{eq:Z_I}
\end{split}
\end{equation}

\begin{equation}
\begin{split}
    \mu_x=&\expval{X\otimes O}\\
    =&\sum_{iji'j'}p_{ij}p_{i'j'}\frac{1}{2}\bigg[\bra{0}X\ket{0}\mathrm{Tr}\left(OP_jUP_i\rho P_i U^\dag P_j\right) + \delta_{ii'}\delta_{jj'}\bra{1}X\ket{0}\mathrm{Tr}(OP_jUP_i \rho P_{i'}U^\dag P_{j'})\\
    & +\delta_{ii'}\delta_{jj'} \bra{0}X\ket{1} \mathrm{Tr}(OP_{j'}UP_{i'} \rho P_{i}U^\dag P_{j})+\bra{1}X\ket{1}\mathrm{Tr}\left(OP_{j'}UP_{i'}\rho P_{i'} U^\dag P_{j'}\right) \bigg]\\
    =&\sum_{ij}(p_{ij})^2\mathrm{Tr}\left[OP_jUP_i\rho P_iU^\dag P_j\right]\\
    =&(\sum_{ij}p_{ij}^2)\mathrm{Tr}\left[O\rho_{\mathrm{eff}}\right]\\
    =&\mu_y \mathrm{Tr}\left[O\rho_{\mathrm{eff}}\right]
    \label{eq:Z_O}
\end{split}
\end{equation}
\end{widetext}
$\mathrm{Var}(x)$ and $\mathrm{Var}(y)$ denote, respectively, the variances of the estimators $x$ and $y$ when averaged over $K$ measurement shots. They can be written as

\begin{equation}
\label{eq:var_x}
\mathrm{Var}(x)=\frac{\big\langle (X\!\otimes\! O)^{2}\big\rangle-\big\langle X\!\otimes\! O\big\rangle^{2}}{K}
=\frac{\big\langle I\!\otimes\! O^{2}\big\rangle-\mu_x^{2}}{K},
\end{equation}

\begin{equation}
\label{eq:var_y}
\mathrm{Var}(y)=\frac{\big\langle (X\!\otimes\! I)^{2}\big\rangle-\big\langle X\!\otimes\! I\big\rangle^{2}}{K}
=\frac{1-\mu_y^{2}}{K},
\end{equation}

The calculation of $\langle I\!\otimes\! O^{2}\rangle$ is as follows:

\begin{widetext}
\begin{equation}
\begin{split}
    \expval{I\otimes O^2}    =&\sum_{iji'j'}p_{ij}p_{i'j'}\frac{1}{2}\bigg[\bra{0}I\ket{0}\mathrm{Tr}\left(O^2P_jUP_i\rho P_i U^\dag P_j\right) + \delta_{ii'}\delta_{jj'}\bra{1}I\ket{0}\mathrm{Tr}(O^2P_jUP_i \rho P_{i'}U^\dag P_{j'})\\
    & +\delta_{ii'}\delta_{jj'} \bra{0}I\ket{1} \mathrm{Tr}(O^2P_{j'}UP_{i'} \rho P_{i}U^\dag P_{j})+\bra{1}I\ket{1}\mathrm{Tr}\left(O^2P_{j'}UP_{i'}\rho P_{i'} U^\dag P_{j'}\right) \bigg]\\
    =&\sum_{ij}p_{ij}\mathrm{Tr}\left[O^2P_jUP_i\rho P_iU^\dag P_j\right]\\
    =&\mathrm{Tr}\left[O^2\rho_{\mathrm{tw}}\right]
\label{eq:I_O2}
\end{split}
\end{equation}
\end{widetext}

Finally, $\mathrm{Cov}(x,y)$ is the covariance between the estimators $x$ and $y$; its calculation is as follows:

\begin{equation}
    \begin{split}
        \mathrm{Cov}(x,y)=&\frac{\expval{(X\otimes O)(X\otimes I)}-\expval{X\otimes O}\expval{X\otimes I}}{K}\\
        =&\frac{\expval{I\otimes O}-\expval{X\otimes O}\expval{X\otimes I}}{K}
    \end{split}
\end{equation}
where, by analogy with Eq.~\eqref{eq:I_O2}, we obtain $\expval{I\otimes O}=\mathrm{Tr}\left[O\rho_{\mathrm{tw}}\right]$. Together with Eqs.~\eqref{eq:Z_I} and~\eqref{eq:Z_O}, we have:

\begin{equation}
    \begin{split}
    \label{eq:cov_x_y}
        \mathrm{Cov}(x,y)=&\frac{\mathrm{Tr}\left[O\rho_{\mathrm{tw}}\right]-\mu_y^2\mathrm{Tr}\left[O\rho_{\mathrm{eff}}\right]}{K}
    \end{split}
\end{equation}

By substituting Eqs.~\eqref{eq:Z_I},~\eqref{eq:Z_O},~\eqref{eq:var_x},~\eqref{eq:var_y},~\eqref{eq:I_O2}, and~\eqref{eq:cov_x_y} into Eq.~\eqref{eq:var_x_y}, we obtain the variance of the estimator:

\begin{equation}
\begin{split}
    &\mathrm{Var}\left(\frac{x}{y}\right)\\
    \approx &\frac{\mathrm{Tr}\left[O^2\rho_{\mathrm{tw}}\right]-2\mathrm{Tr}\left[O\rho_{\mathrm{tw}}\right]\mathrm{Tr}\left[O\rho_{\mathrm{eff}}\right]+\mathrm{Tr}\left[O\rho_{\mathrm{eff}}\right]^2}{K\mu_y^2}
\end{split}
\end{equation}
For an observable $O$ whose spectral norm is bounded by a constant, all terms in the numerator are of constant order, and the overall variance scales as $\mathrm{Var}(x/y)\sim \mathcal{O}(\frac{1}{K\mu_y^2})$. Therefore, if we require the estimation error to be of order $\epsilon$ (so that $\epsilon^{2}\sim \mathrm{Var}(x/y)$), the required number of repeated measurements is $
K \sim \mathcal{O}\!\left(\frac{1}{\epsilon^{2}\mu_y^{2}}\right),
$.

\section{Generalization}\label{app:generalization}
We will discuss how to generalize our protocol to versions with multiple time points and multiple ancillary registers, and also to the case where the noises on the ancillary register and the main register are not identical. 

\subsection{Multiple time points}

\begin{figure}[htbp]
\label{fig:vp_n_point}
\centering
\includegraphics[width=1.0\linewidth]{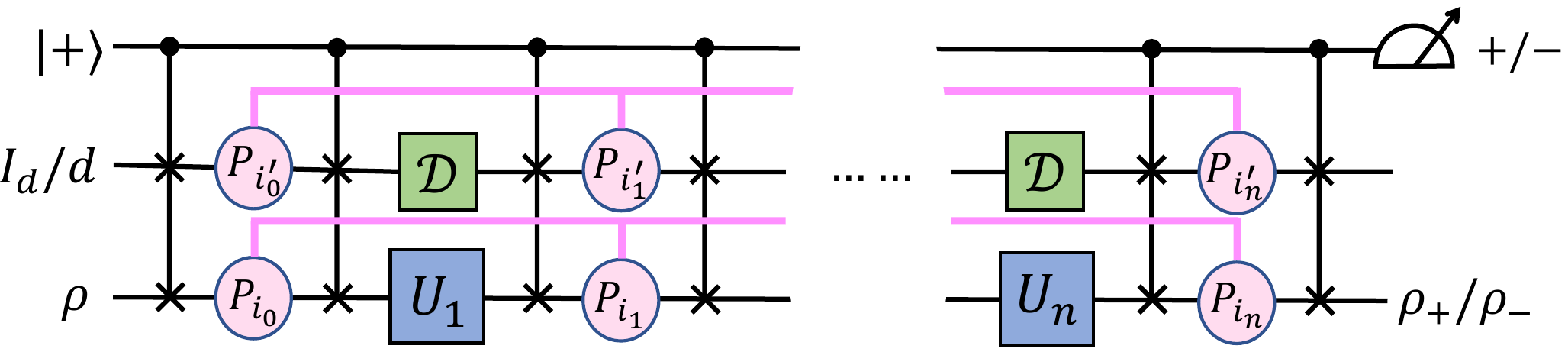}
\caption{The generalized version of our scheme. The non-Markovian noise consists of $n{+}1$ time points, separated by $n$ target quantum gates $U_i$. The joint probability distributions of Pauli errors on the main and ancillary registers are assumed to be identical.}
\end{figure}

In the most general case, non-Markovian noise can involve more than two time points, and our protocol can be extended to this setting as well. As shown in Fig.~\ref{fig:vp_n_point}, the noise at each time point is Pauli-twirled. An ancillary register and a control qubit are employed. Before and after each noise event, apply a pair of CSWAP operations to perform purification. The measurement of the control qubit and the post-processing of the output state are the same as in the two-point case.

Just like in the two-point case, performing Pauli twirling at each noise point converts a general non-Markovian noise into multi-time Pauli errors with correlated probabilities:

\begin{equation}
\begin{split}
    &\mathrm{Tr}_\mathrm{E}[\mathrm{T}_\mathrm{S}[\mathcal{E}_n]\circ \mathcal{U}_n\circ \mathrm{T}_\mathrm{S}[\mathcal{E}_{n-1}]\circ\cdots\circ \mathcal{U}_1 \circ\mathrm{T}_\mathrm{S}[\mathcal{E}_0](\rho\otimes\sigma)]\\
    =&\sum_{i_0,...,i_n}p_{i_0,...,i_n}\mathcal{P}_{i_n}\circ \mathcal{U}_n\circ \mathcal{P}_{i_{n-1}}\circ\cdots\circ \mathcal{U}_1\circ \mathcal{P}_{i_0}(\rho)
\end{split}
\end{equation}
\(\mathcal{U}\) and \(\mathcal{P}\) represent the unitary operations in the original circuit and the random Pauli operations, respectively. After Pauli twirling, purification squares the probability of each component and then re-normalizes, resulting in the effective final state

\begin{equation} \rho_{\mathrm{eff}}=\frac{\sum_{i_0,...,i_n}p_{i_0,...,i_n}^2 \mathcal{P}_{i_n}\circ \mathcal{U}_n\circ \mathcal{P}_{i_{n-1}}\circ\cdots\circ \mathcal{U}_1\circ \mathcal{P}_{i_0}(\rho)}{\sum_{i_0,...,i_n}p_{i_0,...,i_n}^2}
\end{equation}
This boosts the probability of the error-free component and quadratically suppresses the noise rate, as long as the error-free channel already has the largest probability among the mixture. 

\begin{figure}[htbp]
\label{fig:vp_n_copies}
\centering
\includegraphics[width=1.0\linewidth]{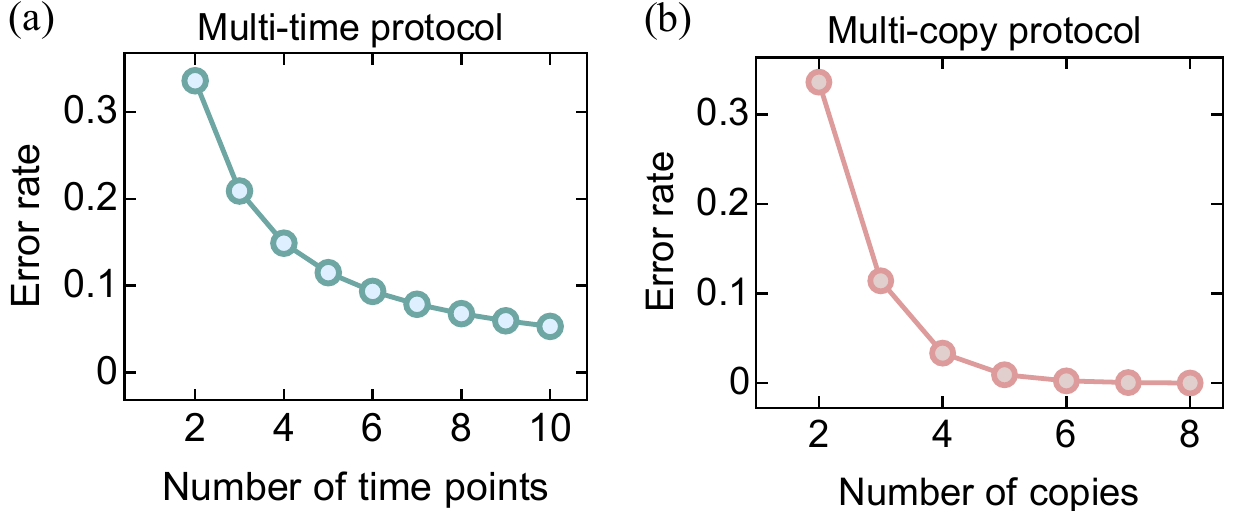}
\caption{The original error rate is set to \( p_e = 0.7 \). (a) An illustrative numerical simulation of the noise-suppression capability of the multi-time protocol versus the number of involved time points. (b) The noise-suppression capability of the multi-copy protocol versus the number of copies.}
\label{fig:sim}
\end{figure}

The advantage brought by this multi-time extension can also be understood from another perspective. Suppose there exists non-Markovian noise with a total error rate of \( p_e \) in a very deep quantum circuit. We can choose the number \( n \) of time-points at which to insert noise-suppression operations throughout the entire circuit. In the weak-noise regime, if only Pauli twirling is applied, there are \( 4^n - 1 \) types of correlated Pauli errors. Although the Pauli errors at different sites are correlated, as a crude estimate, we may assign each site independent probabilities  
$ p_{\!X}=p_{\!Y}=p_{\!Z}=p_{\text{single}}/3 $  
for the three single-qubit Pauli errors.  
Then the total survival probability is roughly  
$ 1-p_e \approx (1-p_{\text{single}})^n $.  
For an error pattern that affects exactly $ n_e $ time points, its probability is approximately  
$ (p_{\text{single}}/3)^{n_e}\,(1-p_{\text{single}})^{n-n_e} $, and there are \( 3^{n_e}C(n,n_e) \) distinct error patterns of this type. Here $C$ denotes the number of combinations, with $C(n,n_e)=n!/[n_e!(n-n_e)!]$. Purification squares the probability of every error pattern (including the error-free one) and re-normalizes; hence, the error rate after purification is obtained as:

\begin{equation} 
\begin{split}
p_{e,\mathrm{eff}}=&\frac{\sum_{n_e=1}^nC(n,n_e)3^{n_e}\left((p_{\text{single}}/3)^{n_e}\,(1-p_{\text{single}})^{n-n_e} \right)^2}{\sum_{n_e=0}^nC(n,n_e)3^{n_e}\left((p_{\text{single}}/3)^{n_e}\,(1-p_{\text{single}})^{n-n_e} \right)^2}\\
=&1-\left( \frac{(1-p_{\text{single}})^2}{(1-p_{\text{single}})^2+p_{\text{single}}^2/3} \right)^n\\
=&1-\left(\frac{(1-p_e)^{2/n}}{(1-p_e)^{2/n}+(1-(1-p_e)^{1/n})^2/3}\right)^n
\end{split}
\end{equation}
A numerical simulation based on this expression is shown in Fig.~\ref{fig:sim}(a). Using a protocol with more time points in a circuit can suppress the error rate to a lower level. The intuitive reason is that the total error probability is distributed among a larger number of error patterns; when each pattern’s probability is squared and then summed, the result becomes smaller as the number of terms increases.

\subsection{Multiple copies scheme}

We can employ more ancillary systems to achieve stronger noise suppression. Consider a two-point non-Markovian noise process: we employ $m$ ancillary systems (so there are $m+1$ copies in total) and replace the CSWAP operations in purification with controlled permutations among the $m+1$ copies, as shown in Fig.~\ref{fig:vp_n_copies}. 

\begin{figure}[htbp]
\label{fig:vp_n_copies}
\centering
\includegraphics[width=0.85\linewidth]{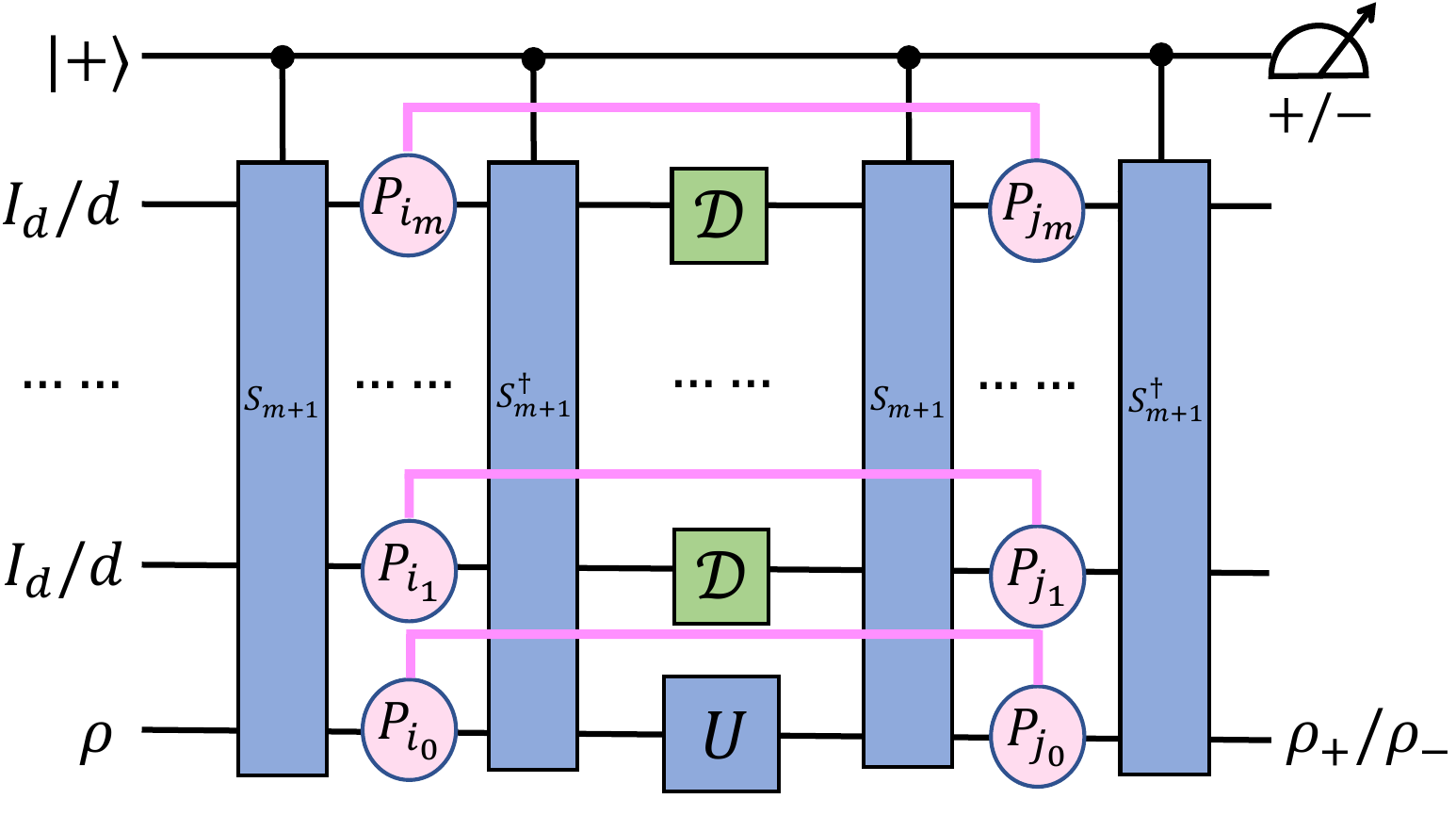}
\caption{In the generalized scheme with $m$ ancillary systems. Pauli twirling is applied at every noise point. Replacing the original CSWAP, purification is now implemented via controlled permutations among the $m{+}1$ systems. $S_{m+1}$ denotes the cyclic permutation of the $m{+}1$ systems, and $S_{m+1}^\dagger$ is its inverse permutation. The joint probability distribution of the two noises on each system is assumed to be the same.}
\end{figure}

This $(m{+}1)$-copy purification is equivalent to taking each component of the Pauli-twirled noise, raising its weight to the $(m{+}1)$-th power, and then re-normalizing to obtain the effective final state:

\begin{equation}
    \rho_{\mathrm{eff}}=\frac{\sum_{ij}(p_{ij})^{m+1}\mathcal{P}_j\circ \mathcal{U} \circ \mathcal{P}_i(\rho)}{\sum_{ij}(p_{ij})^{m+1}}
\end{equation}
In this case, as long as the error-free channel has the largest probability among the original mixture, the $(m{+}1)$-copy scheme achieves $(m{+}1)$-th–power suppression of the noise rate. This shows that as our scheme scales in qubit resources, the noise suppression strengthens exponentially with the resources consumed. 
To illustrate this noise-suppression effect intuitively, we again resort to the simple model used earlier: after Pauli twirling, the probabilities of $X$, $Y$ and $Z$ errors at each time step are all equal to $p_{\text{single}}/3$, and the total error rate satisfies $1 – p_e = (1 – p_{\text{single}})^2$. For a protocol using a total of \(m\) copies, the overall error rate will be suppressed to:

\begin{equation}
\begin{split}   p_{e,\mathrm{eff}}=&\frac{\sum_{n_e=1}^2C(2,n_e)3^{n_e}\left((p_{\text{single}}/3)^{n_e}\,(1-p_{\text{single}})^{2-n_e} \right)^m}{\sum_{n_e=0}^2C(2,n_e)3^{n_e}\left((p_{\text{single}}/3)^{n_e}\,(1-p_{\text{single}})^{2-n_e} \right)^m}\\
=&1-\left(\frac{(1-p_{\text{single}})^m}{(1-p_{\text{single}})^m+3^{1-m}p_{\text{single}}^m}\right)^2\\
=&1-\left(\frac{(1-p_e)^{m/2}}{(1-p_e)^{m/2}+3^{1-m}(1-\sqrt{1-p_e})^m}\right)^2
\end{split}
\end{equation}
The numerical simulation of this expression is shown in Fig.~\ref{fig:sim}(b). A pronounced exponential suppression behavior is observed.

\subsection{Non-identical noises}
\label{app:non_identical}

In practical experiments, it is unrealistic to ensure that the noise on the ancillary register is the same as that on the main register. In fact, as long as the noise rate on the ancillary register is not too large so that the error-free process has the largest probability among the mixture, our scheme can suppress the noise rate.

To generalize the derivation in Appendix~\ref{app:proof_vp} to the case of non-identical noise, we simply no longer assume in Eq.~\eqref{eq:vp_sub} that $p_{ij}$ and $p_{i'j'}$ are the same probability distribution. We rename the probability distributions of the noises on the main register and on the ancillary register as $p$ and $p'$, respectively. Thus, a basic version of our scheme with two-point non-Markovian noise and a single ancillary system yields the effective final state:

\begin{equation}  \rho_{\mathrm{eff}}=\frac{\sum_{ij}p_{ij}p'_{ij}\mathcal{P}_j\circ \mathcal{U} \circ \mathcal{P}_i(\rho)}{\sum_{ij}p_{ij}p'_{ij}}
\end{equation}
$p_{ij}$ $\big(p'_{ij}\big)$ denotes the joint probability that the Pauli errors on the main (ancillary) register are $P_i$ and $P_j$.

When the noisy process on the ancillary register is dominated by the error-free component—i.e., $p'_{00}$ is the largest among all $p'_{ij}$—it is straightforward to see that $\frac{p_{00}p'_{00}}{p_{ij}p'_{ij}}>\frac{p_{00}}{p_{ij}}$ for all $ij\neq 00$, which means that after purification the relative weight of the error-free component on the main register increases.

\section{Experimental Details}\label{app:exp}

The experiments were performed on a five-qubit NMR quantum processor using a Bruker 600 MHz spectrometer at 299K. In this section, we provide the experimental details, including Characterization, Pseudo-pure state preparation, and Measurement.

\subsection{Characterization}
\begin{figure*}[htpb]
\centering
\includegraphics[width=0.8\linewidth]{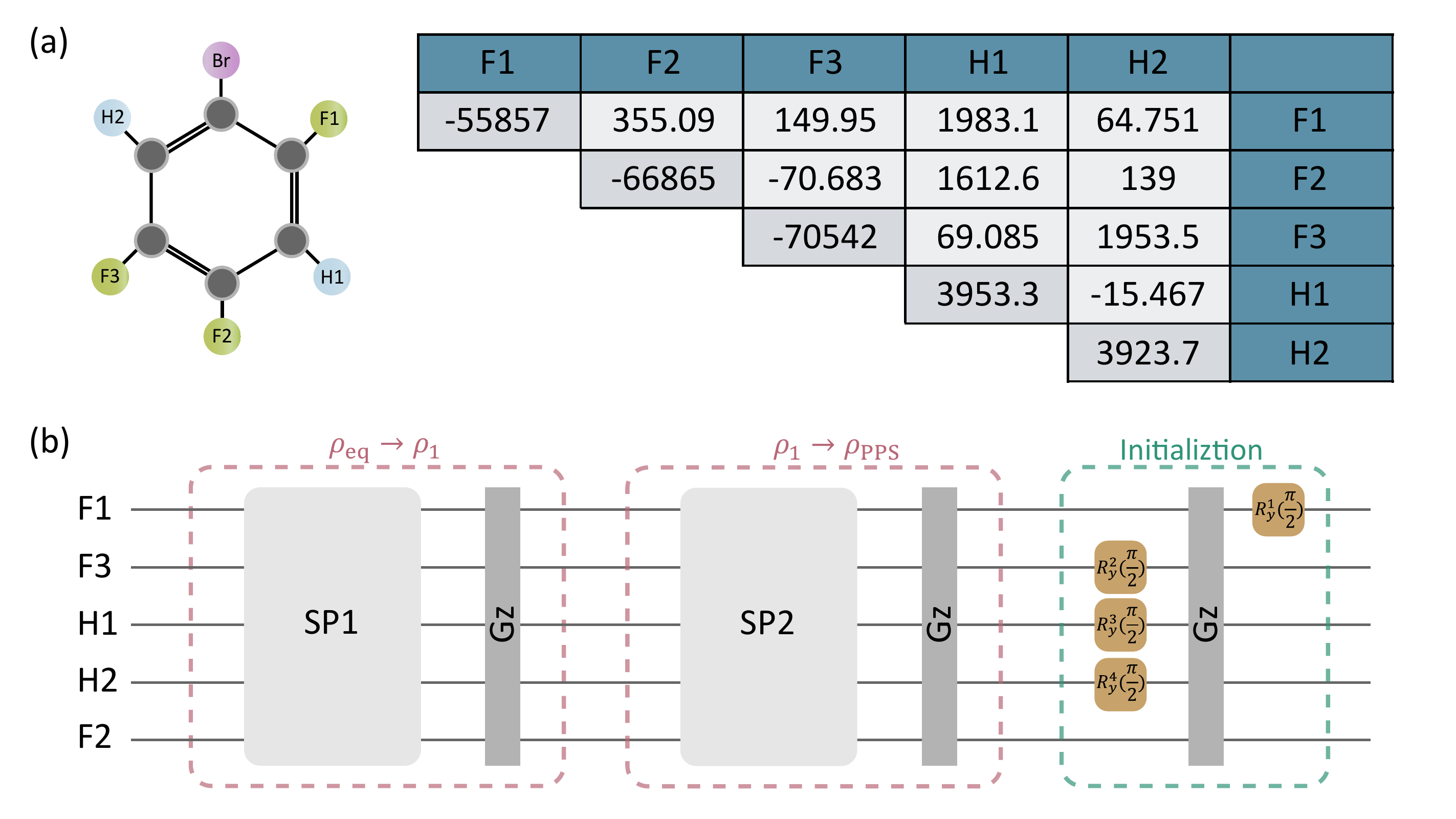}
\caption{(a) The molecular structure of 1-bromo-2, 4, 5-trifluorobenzene. The table on the right side includes the relative parameters at 299K. (b) Quantum circuit for initializing the five-qubit quantum simulator. The operations within the red dashed box generate the pseudo-pure state using the line-selective transition method, while the green dashed box illustrates the preparation of the desired initial state for the noise suppression experiment.}
\label{SM_nMfig:exp1}
\end{figure*}

In our experiments, the quantum register consists of two $^1$H nuclear spins and three $^{19}$F nuclear spins in 1-bromo-2,4,5-trifluorobenzene, dissolved in the liquid crystal solvent N-(4-methoxybenzylidene)-4-butylaniline (MBBA).  Fig.~\ref{SM_nMfig:exp1}(a) shows the molecular structure, and the five nuclear spins (F1, F2, F3, H1, H2) are encoded as qubits 1–5.
Although the sample belongs to a strongly coupled spin system in the liquid-crystal environment, the effective Hamiltonian can still be simplified under high-field and secular approximations. In practice, the transverse components of the dipolar interaction ($X^jX^k$ and $Y^jY^k$) oscillate rapidly in the rotating frame and are strongly suppressed by the finite chemical-shift differences and the reduced dipolar strength due to the orientational order parameter. Consequently, the effective spin–spin couplings can be well approximated by the dipolar interaction $Z^jZ^k$, which provides an accurate description of the experimental dynamics for both homo- and heteronuclear pairs.
The internal Hamiltonian of the NMR system in the rotating frame is denoted as:
\begin{equation}
    \mathcal{H}_{\text{NMR}} = \sum_{j=1}^{5} \pi \nu_j Z^j \;+\; \sum_{j<k} \frac{\pi}{2}\,(J_{jk} + 2D_{jk})\,Z^j Z^k ,
\end{equation}
where the $Z^j$ represents the Pauli-Z operator of the $j$-th spin.
The $\nu_j$ are the chemical shifts, and $J_{jk} + 2D_{jk}$ are the effective scalar plus dipolar coupling constants. The experimentally determined parameters are listed in Fig.~\ref{SM_nMfig:exp1}(a).

\begin{figure*}[htpb]
\centering
\includegraphics[width=1\linewidth]{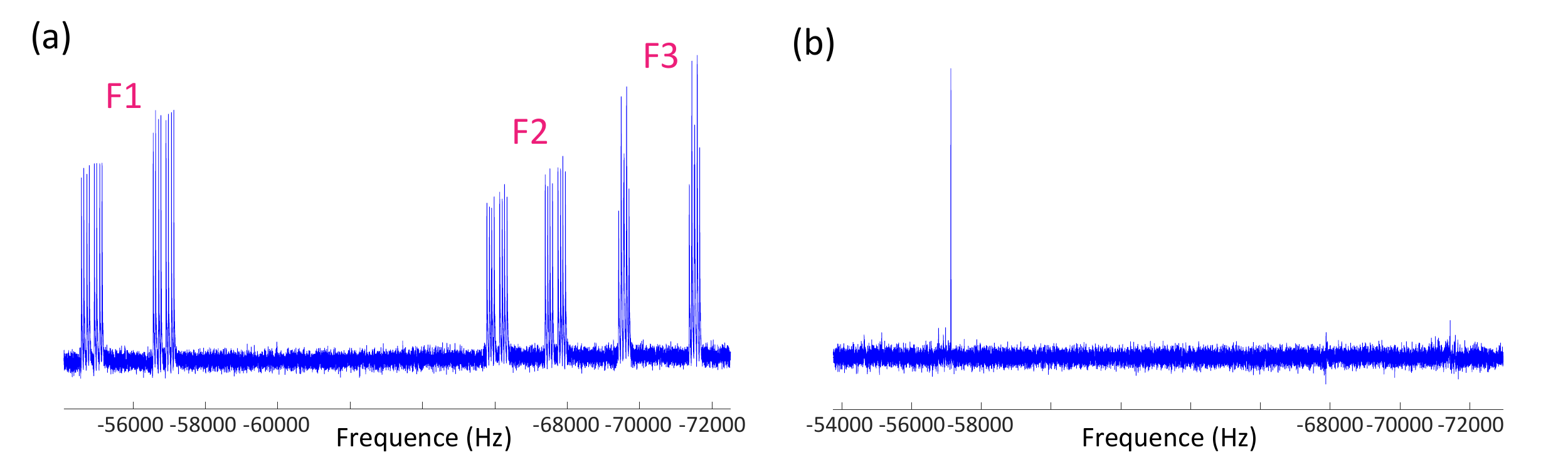}
\caption{(a) Thermal spectrum of the fluorine nuclei. (b) Spectrum of the PPS.}
\label{SM_nMfig:them_pps_spectral}
\end{figure*}

\subsection{Pseudo-pure state preparation}
The thermal equilibrium state of the five-spin ensemble is highly mixed and can be expressed as
$\rho_{\text{eq}} \approx \frac{\mathbb{I}}{2^5} + \epsilon \Delta\rho$,
where $\epsilon \sim 10^{-5}$ denotes the polarization. Since the identity part is invariant under unitary operations and does not contribute to NMR signals, only the deviation density matrix $\Delta\rho$ is experimentally relevant. So the thermal equilibrium state can be approximated as 
$\rho_\text{eq}=\gamma_\text{F}(Z^1+Z^2+Z^3)+\gamma_\text{H}(Z^4+Z^5)$, 
where the $\gamma_\text{F}$ and $\gamma_\text{H}$ are the gyromagnetic ratio of $^{19}$F and $^{1}$H, respectively, and $Z^i$ denotes the Pauli-$Z$ operator acting on the $i$-th qubit.
The corresponding thermal equilibrium spectrum is shown in Fig.~\ref{SM_nMfig:them_pps_spectral}(a).

At room temperature, the thermal equilibrium state of an NMR system is highly mixed and therefore unsuitable as an initialization state for quantum computation. Various techniques have been developed to prepare pseudo-pure states (PPS), including the spatial averaging method~\cite{liu2025neural,Liuhf2025,Xic2024experimental,zheng2025vqb,Huangkyi2024}, the temporal–spatial averaging method, and the cat-state method~\cite{Longxy2022}.
In this work, we prepare the PPS using the line-selective transition method. The pulse sequence consists of two red dashed boxes in Fig.~\ref{SM_nMfig:exp1}(b), each containing a shaped pulse and a gradient-field pulse (Gz).

The first shaped pulse (SP1), followed by a Gz pulse, selectively depletes all population levels except for the two target states $\ket{00000}$ and $\ket{11111}$. This is accomplished by a series of $\pi/2$ line-selective pulses applied between symmetric state pairs, excluding $\ket{00000}$ and $\ket{11111}$.
These line-selective pulses are packed into a large composite pulse, which constitutes the first shaped pulse (SP1) in the sequence. These pulses transform the symmetric states into non–zeroth coherences; for example, $\ketbra{00101}{00101}$ and $\ketbra{11010}{11010}$ are converted into $\ketbra{00101}{11010}$ and $\ketbra{11010}{00101}$. After the subsequent Gz pulse removes these coherences, the resulting density matrix can be expressed as $\rho_1 = \ketbra{00000}{00000} - \ketbra{11111}{11111}$. The populations of $\ket{00000}$ and $\ket{11111}$ are equal in magnitude but opposite in sign, while all other basis states carry zero population.

The second shaped pulse (SP2) redistributes the population on $\ket{11111}$ uniformly among all other computational basis states except $\ket{00000}$. 
This is achieved through a series of line-selective rotations that couple $\ket{11111}$ to each of the other states. The rotation angles are chosen as $2\sin^{-1}(1/31),,2\sin^{-1}(1/30),,\ldots,,2\sin^{-1}(1/2)$, such that each operation sequentially transfers fractions $1/31,,1/30,,\ldots,,1/2$ of the current population of $\ket{11111}$ to the respective states. At the end of this sequence, every state except $\ket{00000}$ carries an equal fraction $1/31$ of the original population of $\ket{11111}$. Because these line-selective rotations are only applied between the all-spin-down state $\ket{11111}$ and states with some spins flipped, no zeroth coherence remains at the end of SP2. A final Gz is then applied to eliminate any residual coherence, yielding the PPS of the five-qubit system.
The spectrum of the PPS is displayed in Fig.~\ref{SM_nMfig:them_pps_spectral}(b).

\subsection{Initialization}
In our noise suppression experiments, we initialize the system to $\ketbra{+} \otimes  I_2/2 \otimes I_2/2 \otimes I_2/2 \otimes \ketbra{0}$. The initial state can be prepared using single-qubit rotation alongside a gradient-field pulse, as shown in the green dashed box of Fig.~\ref{SM_nMfig:exp1}(b). Specifically, single-qubit rotations $R_y(\tfrac{\pi}{2})$ are applied to qubits 2-4 to create $\ket{+}$ state, after which a Gz pulse removes all coherences, leaving each of these qubits in the maximally mixed state $I_2/2$. Finally, an $R_y(\tfrac{\pi}{2})$ rotation is applied to qubit 1 to prepare it in the $\ket{+}$ state, completing the initialization procedure.

\subsection{Measurement}

In an NMR quantum processor, the experimental sample consists not of a single molecule but rather of an ensemble containing a macroscopic number of identical molecules. Consequently, all measurements correspond to ensemble-averaged observables. After the applied operations, the nuclear spins precess around the static magnetic field $B_0$ and gradually relax back toward thermal equilibrium. The precessing transverse magnetization induces an oscillating voltage in the detection coil, meaning that the NMR spectrometer can directly measure only the transverse components of magnetization—i.e., the expectation values of $X$ and $Y$. Therefore, multi-spin observables must be mapped onto single-spin transverse magnetization operators, which can then be read out from the free-induction decay (FID) signals.

In our experimental scheme, for example in the \textit{unitary operation} case, the required observables include $\langle X I I I I\rangle$ and $\langle X I I I Z\rangle$. While $\langle X I I I I\rangle$ can be directly measured, obtaining $\langle X I I I Z\rangle$ is less accurate in practice. In the five-qubit NMR processor, experimental noise reduces the precision of extracting $\langle X I I I Z\rangle$ when it is reconstructed by adding and subtracting the amplitudes of the split spectral peaks of F1. To circumvent this issue, we inserted an additional measurement pulse $e^{-i\frac{\pi}{2}(Z I I I Z/2)}$ to first transform $\langle X I I I Z\rangle$ into $\langle Y I I I I\rangle$. Subsequently, a single-qubit rotation $R_z^1(-\pi/2)$ was applied to map $\langle Y I I I I\rangle$ onto $\langle X I I I I\rangle$, enabling high-fidelity readout. 

\begin{figure*}[htpb]
\centering
\includegraphics[width=1\linewidth]{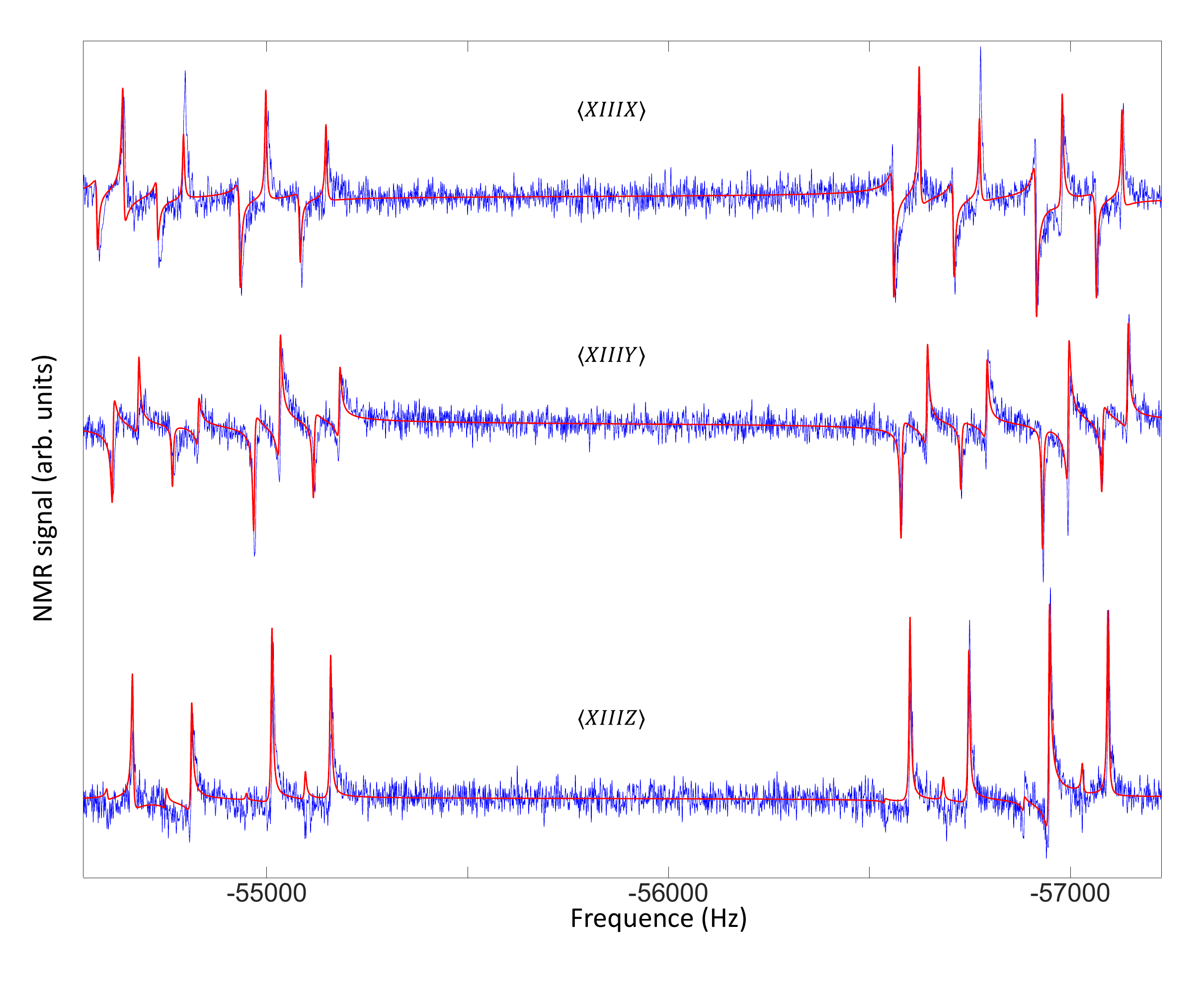}
\caption{The experimental spectra of first qubit (F1) corresponding to the observables $\langle X I I I X\rangle$, $\langle X I I I Y\rangle$, and $\langle X I I I Z\rangle$. The red lines represent the simulation results, while the blue lines show the experimental results.}
\label{SM_nMfig:exp_spectra}
\end{figure*}

For the \textit{non-unitary channel} case, additional observables such as $\langle X I I I X\rangle$ and $\langle X I I I Y\rangle$ are required. These can be accessed analogously by applying single-qubit rotations on the fifth qubit: $R_y^5(-\pi/2)$ maps $\langle X I I I X\rangle$ to $\langle X I I I Z\rangle$, and $R_x^5(\pi/2)$ maps $\langle X I I I Y\rangle$ to $\langle X I I I Z\rangle$. Once transformed, the same procedure described above is applied to obtain reliable measurement outcomes. As an example, Fig.~\ref{SM_nMfig:exp_spectra} shows the experimental spectra of the first qubit (F1) for the observables $\langle X I I I X\rangle$, $\langle X I I I Y\rangle$, and $\langle X I I I Z\rangle$ at $t=0.15$, after applying the measurement pulse to transform them into $\langle X I I I I\rangle$ for readout.

\begin{figure*}[htpb]
\centering
\includegraphics[width=1\linewidth]{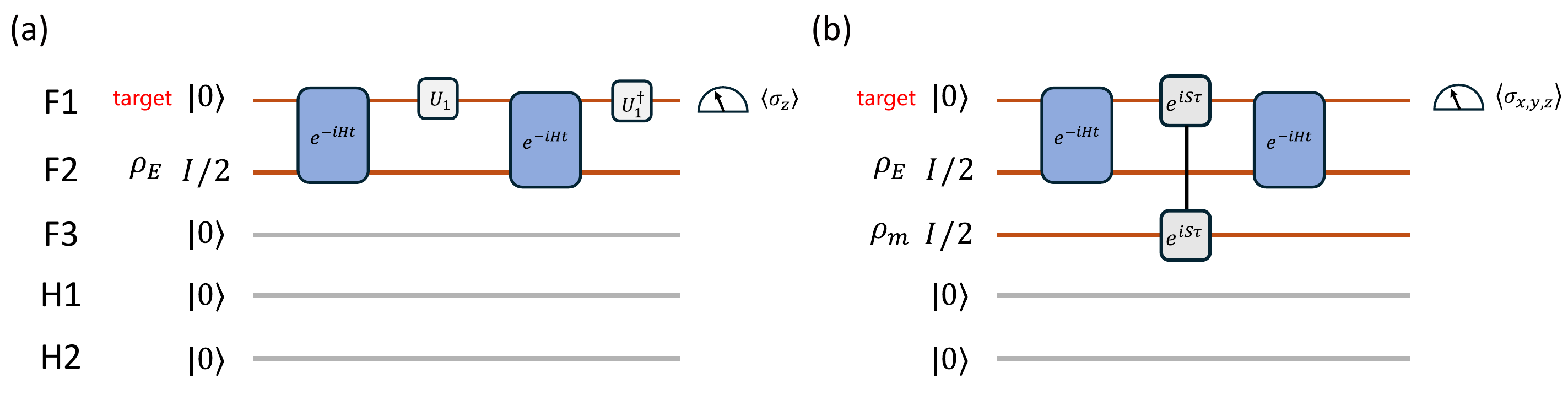}
\caption{Experimental circuits without noise suppression: (a) unitary operation case, and (b) non-unitary channel case.}
\label{SM_nMfig:exp2}
\end{figure*}

\section{Reference Experiments without the Suppression Protocol}\label{app:baselineE}
For comparison, reference experiments without the suppression protocol are carried out for both cases.

In the unitary operation case, only two qubits (F1 and F2) are involved: F1 serves as the target qubit, while F2 acts as the environment qubit $\rho_E$ to introduce non-Markovian noise. The target qubit undergoes the same unitary evolution as in the main experiment, and $\langle Z \rangle$ of F1 is directly measured as a function of $t$. We denote the resulting output as $\bket{Z_{\text{noise}}}$. The experimental circuit is shown in Fig.~\ref{SM_nMfig:exp2}(a).

For the non-unitary channel case, the reference experiment involves three qubits (F1, F2, and F3). The experimental circuit is shown in Fig.~\ref{SM_nMfig:exp2}(b). Here, F1 is the target qubit, F2 is the environment qubit introducing non-Markovian noise, and F3, initialized in the maximally mixed state, undergoes a partial SWAP with the target qubit. For each evolution time $t$, the Pauli expectation values $\langle X \rangle$, $\langle Y \rangle$, and $\langle Z \rangle$ of F1 are measured to reconstruct the target state via quantum state tomography.

\section{Experimental Results of Random Pauli Configurations}\label{app:pauliresult}
\begin{figure*}[htpb]
\centering
\includegraphics[width=1\linewidth]{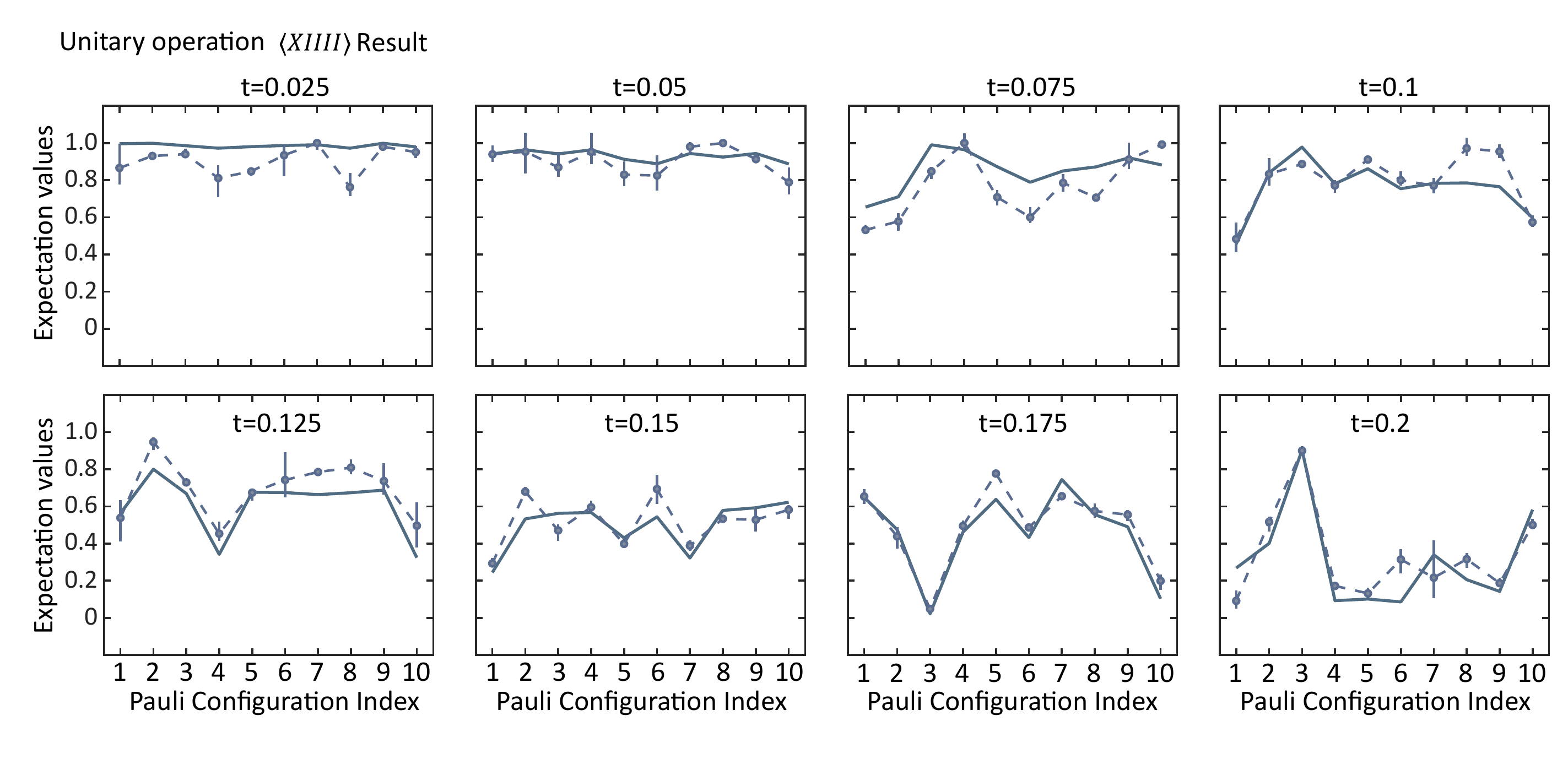}
\caption{Measurement results of $\langle XIIII \rangle$ under 10 distinct Pauli configurations for the unitary operation case. The solid lines indicate theoretical predictions, while the data points with error bars correspond to experimental measurements.}
\label{SM_nMfig:exp31}
\end{figure*}

\begin{figure*}[htpb]
\centering
\includegraphics[width=1\linewidth]{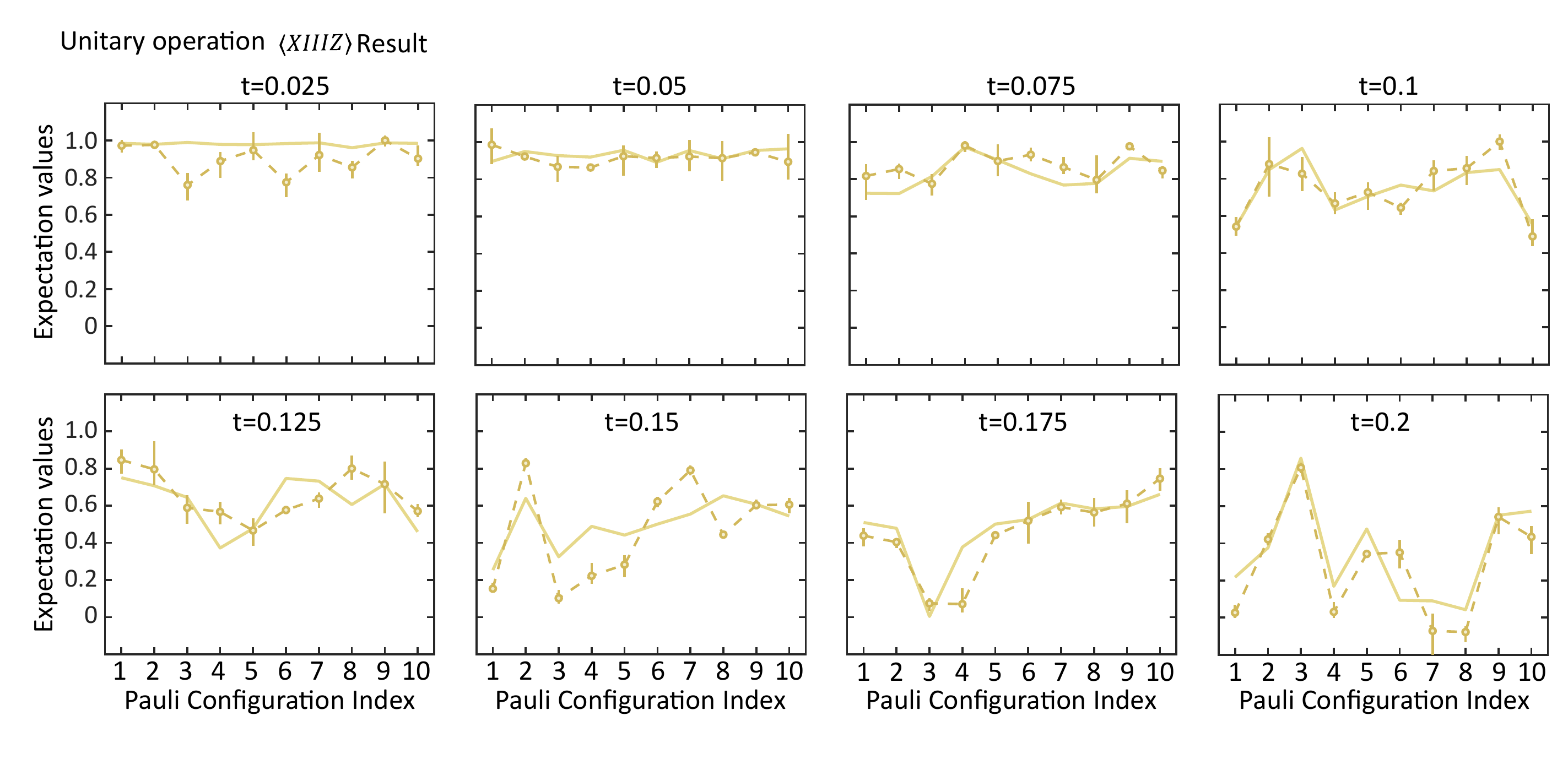}
\caption{Measurement results of $\langle XIIIZ \rangle$ under 10 distinct Pauli configurations for the unitary operation case. The solid lines indicate theoretical predictions, while the data points with error bars correspond to experimental measurements.}
\label{SM_nMfig:exp32}
\end{figure*}

\begin{figure*}[htpb]
\centering
\includegraphics[width=1\linewidth]{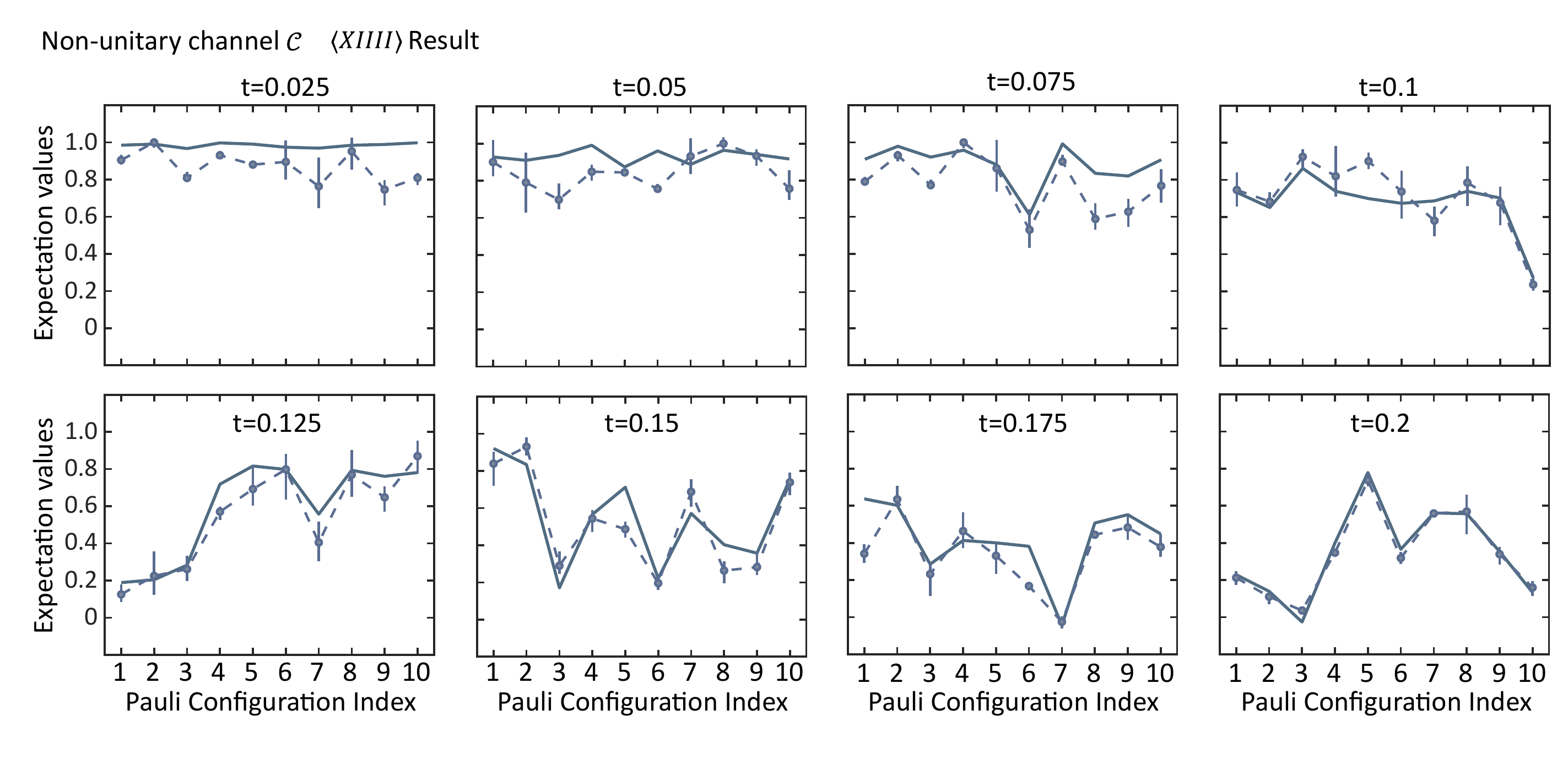}
\caption{Measurement results of $\langle XIIII \rangle$ under 10 distinct Pauli configurations for the non-unitary channel case. The solid lines indicate theoretical predictions, while the data points with error bars correspond to experimental measurements.}
\label{SM_nMfig:exp41}
\end{figure*}

\begin{figure*}[htpb]
\centering
\includegraphics[width=1\linewidth]{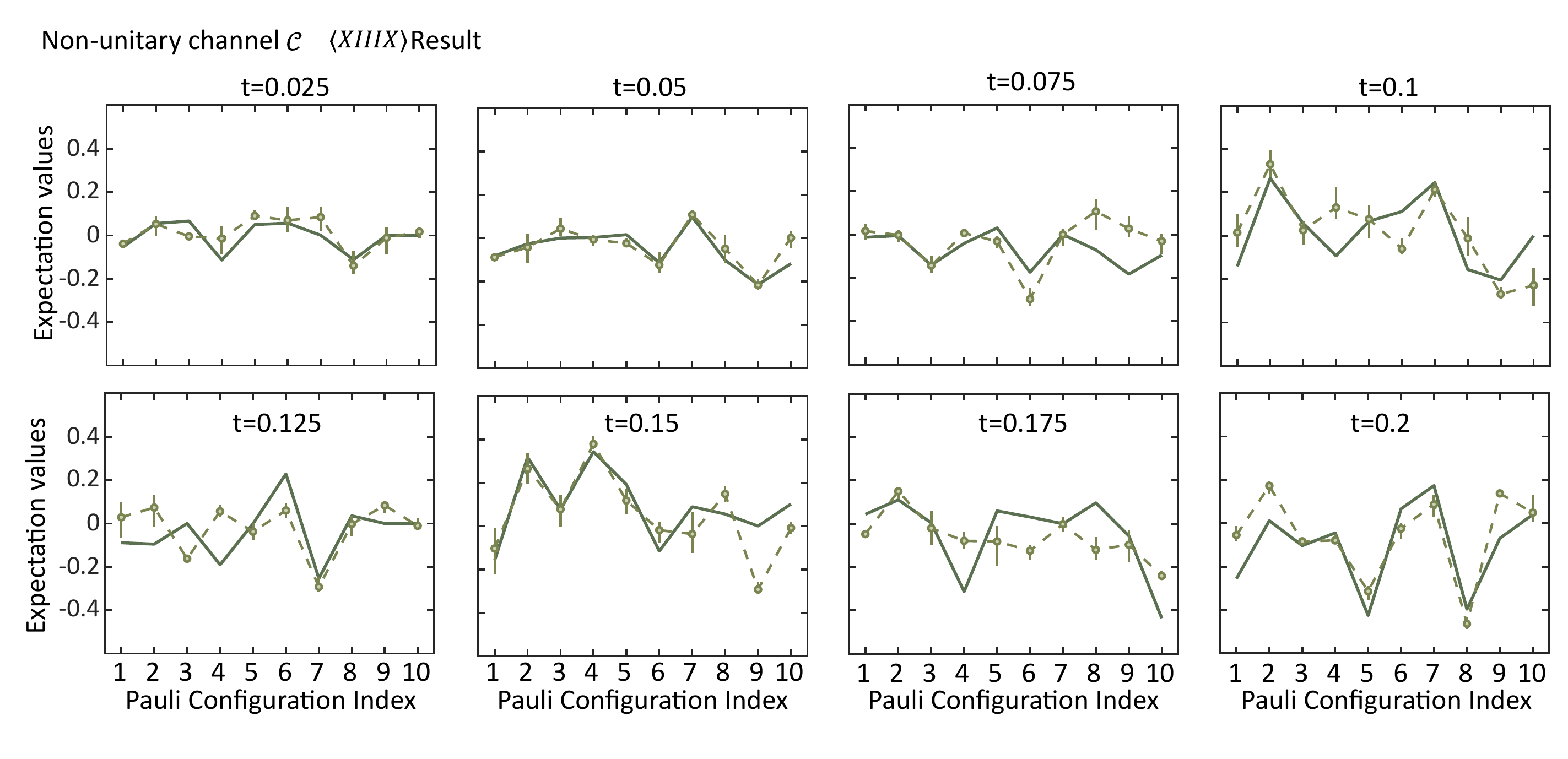}
\caption{Measurement results of $\langle XIIIX \rangle$ under 10 distinct Pauli configurations for the non-unitary channel case. The solid lines indicate theoretical predictions, while the data points with error bars correspond to experimental measurements.}
\label{SM_nMfig:exp42}
\end{figure*}

\begin{figure*}[htpb]
\centering
\includegraphics[width=1\linewidth]{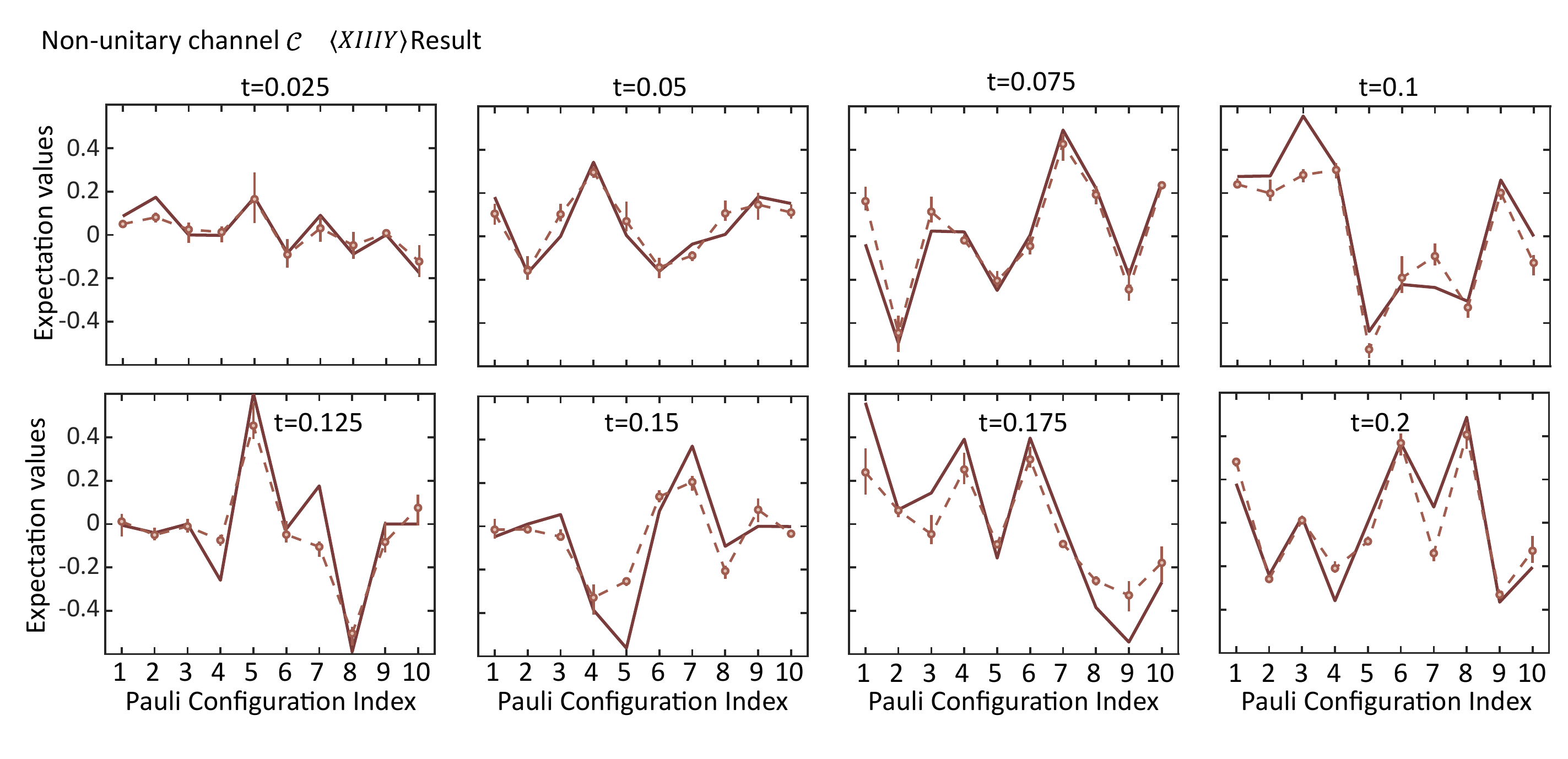}
\caption{Measurement results of $\langle XIIIY \rangle$ under 10 distinct Pauli configurations for the non-unitary channel case. The solid lines indicate theoretical predictions, while the data points with error bars correspond to experimental measurements.}
\label{SM_nMfig:exp43}
\end{figure*}

\begin{figure*}[htpb]
\centering
\includegraphics[width=1\linewidth]{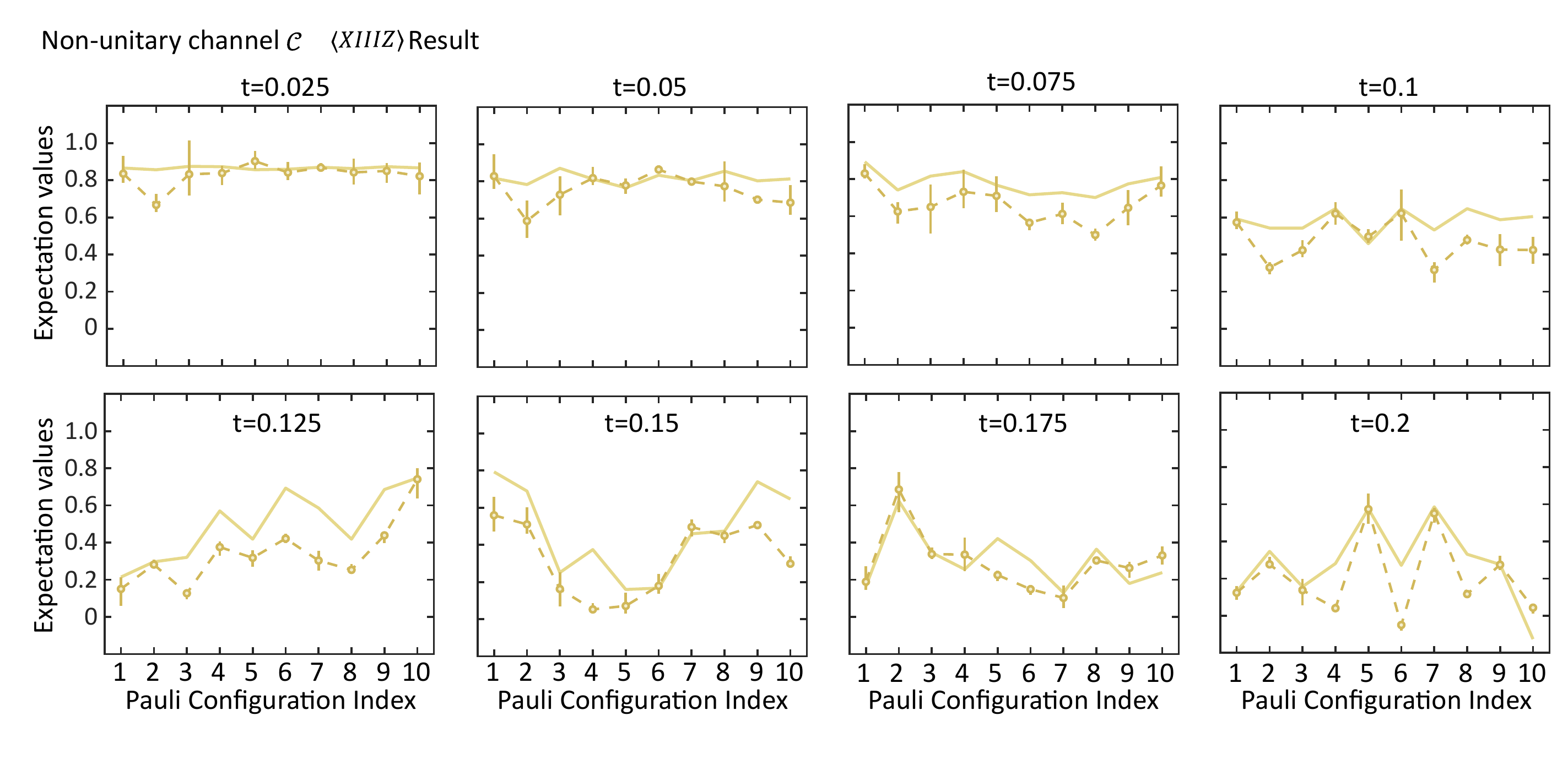}
\caption{Measurement results of $\langle XIIIZ \rangle$ under 10 distinct Pauli configurations for the non-unitary channel case. The solid lines indicate theoretical predictions, while the data points with error bars correspond to experimental measurements.}
\label{SM_nMfig:exp44}
\end{figure*}

For each evolution time, ten random Pauli configurations were sampled, and the target observables were measured in each configuration.
In the unitary operation experiment, only $\langle XIIII \rangle$ and $\langle XIIIZ \rangle$ were measured, as required to evaluate the effective expectation value. The corresponding results are presented in Fig.~\ref{SM_nMfig:exp31} and Fig.~\ref{SM_nMfig:exp32}.
For the non-unitary channel, four observables—$\langle XIIII \rangle$, $\langle XIIIX \rangle$, $\langle XIIIY \rangle$, and $\langle XIIIZ \rangle$—were measured. The measurement results are shown in Figs.~\ref{SM_nMfig:exp41}–\ref{SM_nMfig:exp44}.

\begin{thebibliography}{57}%
\makeatletter
\providecommand \@ifxundefined [1]{%
 \@ifx{#1\undefined}
}%
\providecommand \@ifnum [1]{%
 \ifnum #1\expandafter \@firstoftwo
 \else \expandafter \@secondoftwo
 \fi
}%
\providecommand \@ifx [1]{%
 \ifx #1\expandafter \@firstoftwo
 \else \expandafter \@secondoftwo
 \fi
}%
\providecommand \natexlab [1]{#1}%
\providecommand \enquote  [1]{``#1''}%
\providecommand \bibnamefont  [1]{#1}%
\providecommand \bibfnamefont [1]{#1}%
\providecommand \citenamefont [1]{#1}%
\providecommand \href@noop [0]{\@secondoftwo}%
\providecommand \href [0]{\begingroup \@sanitize@url \@href}%
\providecommand \@href[1]{\@@startlink{#1}\@@href}%
\providecommand \@@href[1]{\endgroup#1\@@endlink}%
\providecommand \@sanitize@url [0]{\catcode `\\12\catcode `\$12\catcode `\&12\catcode `\#12\catcode `\^12\catcode `\_12\catcode `\%12\relax}%
\providecommand \@@startlink[1]{}%
\providecommand \@@endlink[0]{}%
\providecommand \url  [0]{\begingroup\@sanitize@url \@url }%
\providecommand \@url [1]{\endgroup\@href {#1}{\urlprefix }}%
\providecommand \urlprefix  [0]{URL }%
\providecommand \Eprint [0]{\href }%
\providecommand \doibase [0]{https://doi.org/}%
\providecommand \selectlanguage [0]{\@gobble}%
\providecommand \bibinfo  [0]{\@secondoftwo}%
\providecommand \bibfield  [0]{\@secondoftwo}%
\providecommand \translation [1]{[#1]}%
\providecommand \BibitemOpen [0]{}%
\providecommand \bibitemStop [0]{}%
\providecommand \bibitemNoStop [0]{.\EOS\space}%
\providecommand \EOS [0]{\spacefactor3000\relax}%
\providecommand \BibitemShut  [1]{\csname bibitem#1\endcsname}%
\let\auto@bib@innerbib\@empty
\bibitem [{\citenamefont {Breuer}\ and\ \citenamefont {Petruccione}(2006)}]{breuer2006open}%
  \BibitemOpen
  \bibfield  {author} {\bibinfo {author} {\bibfnamefont {H.-P.}\ \bibnamefont {Breuer}}\ and\ \bibinfo {author} {\bibfnamefont {F.}~\bibnamefont {Petruccione}},\ }\href {https://doi.org/10.1093/acprof:oso/9780199213900.001.0001} {\emph {\bibinfo {title} {The Theory of Open Quantum Systems}}}\ (\bibinfo {year} {2006})\BibitemShut {NoStop}%
\bibitem [{\citenamefont {Pollock}\ \emph {et~al.}(2018{\natexlab{a}})\citenamefont {Pollock}, \citenamefont {Rodr\'{\i}guez-Rosario}, \citenamefont {Frauenheim}, \citenamefont {Paternostro},\ and\ \citenamefont {Modi}}]{pollock2018operational}%
  \BibitemOpen
  \bibfield  {author} {\bibinfo {author} {\bibfnamefont {F.~A.}\ \bibnamefont {Pollock}}, \bibinfo {author} {\bibfnamefont {C.}~\bibnamefont {Rodr\'{\i}guez-Rosario}}, \bibinfo {author} {\bibfnamefont {T.}~\bibnamefont {Frauenheim}}, \bibinfo {author} {\bibfnamefont {M.}~\bibnamefont {Paternostro}},\ and\ \bibinfo {author} {\bibfnamefont {K.}~\bibnamefont {Modi}},\ }\bibfield  {title} {\bibinfo {title} {Operational markov condition for quantum processes},\ }\href {https://doi.org/10.1103/PhysRevLett.120.040405} {\bibfield  {journal} {\bibinfo  {journal} {Phys. Rev. Lett.}\ }\textbf {\bibinfo {volume} {120}},\ \bibinfo {pages} {040405} (\bibinfo {year} {2018}{\natexlab{a}})}\BibitemShut {NoStop}%
\bibitem [{\citenamefont {Li}\ \emph {et~al.}(2018)\citenamefont {Li}, \citenamefont {Hall},\ and\ \citenamefont {Wiseman}}]{LI2018concepts}%
  \BibitemOpen
  \bibfield  {author} {\bibinfo {author} {\bibfnamefont {L.}~\bibnamefont {Li}}, \bibinfo {author} {\bibfnamefont {M.~J.}\ \bibnamefont {Hall}},\ and\ \bibinfo {author} {\bibfnamefont {H.~M.}\ \bibnamefont {Wiseman}},\ }\bibfield  {title} {\bibinfo {title} {Concepts of quantum non-markovianity: A hierarchy},\ }\href {https://doi.org/10.1016/j.physrep.2018.07.001} {\bibfield  {journal} {\bibinfo  {journal} {Phys. Rep.}\ }\textbf {\bibinfo {volume} {759}},\ \bibinfo {pages} {1} (\bibinfo {year} {2018})}\BibitemShut {NoStop}%
\bibitem [{\citenamefont {Taranto}\ \emph {et~al.}(2024)\citenamefont {Taranto}, \citenamefont {Quintino}, \citenamefont {Murao},\ and\ \citenamefont {Milz}}]{Taranto2024characterising}%
  \BibitemOpen
  \bibfield  {author} {\bibinfo {author} {\bibfnamefont {P.}~\bibnamefont {Taranto}}, \bibinfo {author} {\bibfnamefont {M.~T.}\ \bibnamefont {Quintino}}, \bibinfo {author} {\bibfnamefont {M.}~\bibnamefont {Murao}},\ and\ \bibinfo {author} {\bibfnamefont {S.}~\bibnamefont {Milz}},\ }\bibfield  {title} {\bibinfo {title} {Characterising the {H}ierarchy of {M}ulti-time {Q}uantum {P}rocesses with {C}lassical {M}emory},\ }\href {https://doi.org/10.22331/q-2024-05-02-1328} {\bibfield  {journal} {\bibinfo  {journal} {{Quantum}}\ }\textbf {\bibinfo {volume} {8}},\ \bibinfo {pages} {1328} (\bibinfo {year} {2024})}\BibitemShut {NoStop}%
\bibitem [{\citenamefont {Terhal}(2015)}]{terhal2015qec}%
  \BibitemOpen
  \bibfield  {author} {\bibinfo {author} {\bibfnamefont {B.~M.}\ \bibnamefont {Terhal}},\ }\bibfield  {title} {\bibinfo {title} {Quantum error correction for quantum memories},\ }\href {https://doi.org/10.1103/RevModPhys.87.307} {\bibfield  {journal} {\bibinfo  {journal} {Rev. Mod. Phys.}\ }\textbf {\bibinfo {volume} {87}},\ \bibinfo {pages} {307} (\bibinfo {year} {2015})}\BibitemShut {NoStop}%
\bibitem [{\citenamefont {Cai}\ \emph {et~al.}(2023)\citenamefont {Cai}, \citenamefont {Babbush}, \citenamefont {Benjamin}, \citenamefont {Endo}, \citenamefont {Huggins}, \citenamefont {Li}, \citenamefont {McClean},\ and\ \citenamefont {O'Brien}}]{cai2023qem}%
  \BibitemOpen
  \bibfield  {author} {\bibinfo {author} {\bibfnamefont {Z.}~\bibnamefont {Cai}}, \bibinfo {author} {\bibfnamefont {R.}~\bibnamefont {Babbush}}, \bibinfo {author} {\bibfnamefont {S.~C.}\ \bibnamefont {Benjamin}}, \bibinfo {author} {\bibfnamefont {S.}~\bibnamefont {Endo}}, \bibinfo {author} {\bibfnamefont {W.~J.}\ \bibnamefont {Huggins}}, \bibinfo {author} {\bibfnamefont {Y.}~\bibnamefont {Li}}, \bibinfo {author} {\bibfnamefont {J.~R.}\ \bibnamefont {McClean}},\ and\ \bibinfo {author} {\bibfnamefont {T.~E.}\ \bibnamefont {O'Brien}},\ }\bibfield  {title} {\bibinfo {title} {Quantum error mitigation},\ }\href {https://doi.org/10.1103/RevModPhys.95.045005} {\bibfield  {journal} {\bibinfo  {journal} {Rev. Mod. Phys.}\ }\textbf {\bibinfo {volume} {95}},\ \bibinfo {pages} {045005} (\bibinfo {year} {2023})}\BibitemShut {NoStop}%
\bibitem [{\citenamefont {Marcos}\ \emph {et~al.}(2011)\citenamefont {Marcos}, \citenamefont {Emary}, \citenamefont {Brandes},\ and\ \citenamefont {Aguado}}]{marcos2011nanostructures}%
  \BibitemOpen
  \bibfield  {author} {\bibinfo {author} {\bibfnamefont {D.}~\bibnamefont {Marcos}}, \bibinfo {author} {\bibfnamefont {C.}~\bibnamefont {Emary}}, \bibinfo {author} {\bibfnamefont {T.}~\bibnamefont {Brandes}},\ and\ \bibinfo {author} {\bibfnamefont {R.}~\bibnamefont {Aguado}},\ }\bibfield  {title} {\bibinfo {title} {Non-markovian effects in the quantum noise of interacting nanostructures},\ }\href {https://doi.org/10.1103/PhysRevB.83.125426} {\bibfield  {journal} {\bibinfo  {journal} {Phys. Rev. B}\ }\textbf {\bibinfo {volume} {83}},\ \bibinfo {pages} {125426} (\bibinfo {year} {2011})}\BibitemShut {NoStop}%
\bibitem [{\citenamefont {Zhang}\ \emph {et~al.}(2022)\citenamefont {Zhang}, \citenamefont {Pokharel}, \citenamefont {Levenson-Falk},\ and\ \citenamefont {Lidar}}]{zhang2022predicting}%
  \BibitemOpen
  \bibfield  {author} {\bibinfo {author} {\bibfnamefont {H.}~\bibnamefont {Zhang}}, \bibinfo {author} {\bibfnamefont {B.}~\bibnamefont {Pokharel}}, \bibinfo {author} {\bibfnamefont {E.}~\bibnamefont {Levenson-Falk}},\ and\ \bibinfo {author} {\bibfnamefont {D.}~\bibnamefont {Lidar}},\ }\bibfield  {title} {\bibinfo {title} {Predicting non-markovian superconducting-qubit dynamics from tomographic reconstruction},\ }\href {https://doi.org/10.1103/PhysRevApplied.17.054018} {\bibfield  {journal} {\bibinfo  {journal} {Phys. Rev. Appl.}\ }\textbf {\bibinfo {volume} {17}},\ \bibinfo {pages} {054018} (\bibinfo {year} {2022})}\BibitemShut {NoStop}%
\bibitem [{\citenamefont {Gul\'acsi}\ and\ \citenamefont {Burkard}(2023)}]{gulacsi2023signitures}%
  \BibitemOpen
  \bibfield  {author} {\bibinfo {author} {\bibfnamefont {B.}~\bibnamefont {Gul\'acsi}}\ and\ \bibinfo {author} {\bibfnamefont {G.}~\bibnamefont {Burkard}},\ }\bibfield  {title} {\bibinfo {title} {Signatures of non-markovianity of a superconducting qubit},\ }\href {https://doi.org/10.1103/PhysRevB.107.174511} {\bibfield  {journal} {\bibinfo  {journal} {Phys. Rev. B}\ }\textbf {\bibinfo {volume} {107}},\ \bibinfo {pages} {174511} (\bibinfo {year} {2023})}\BibitemShut {NoStop}%
\bibitem [{\citenamefont {Agarwal}\ \emph {et~al.}(2024)\citenamefont {Agarwal}, \citenamefont {Lindoy}, \citenamefont {Lall}, \citenamefont {Jamet},\ and\ \citenamefont {Rungger}}]{Agarwal2024modelling}%
  \BibitemOpen
  \bibfield  {author} {\bibinfo {author} {\bibfnamefont {A.}~\bibnamefont {Agarwal}}, \bibinfo {author} {\bibfnamefont {L.~P.}\ \bibnamefont {Lindoy}}, \bibinfo {author} {\bibfnamefont {D.}~\bibnamefont {Lall}}, \bibinfo {author} {\bibfnamefont {F.}~\bibnamefont {Jamet}},\ and\ \bibinfo {author} {\bibfnamefont {I.}~\bibnamefont {Rungger}},\ }\bibfield  {title} {\bibinfo {title} {Modelling non-markovian noise in driven superconducting qubits},\ }\href {https://doi.org/10.1088/2058-9565/ad3d7e} {\bibfield  {journal} {\bibinfo  {journal} {Quantum Sci. Technol.}\ }\textbf {\bibinfo {volume} {9}},\ \bibinfo {pages} {035017} (\bibinfo {year} {2024})}\BibitemShut {NoStop}%
\bibitem [{\citenamefont {Chiribella}\ \emph {et~al.}(2008)\citenamefont {Chiribella}, \citenamefont {D'Ariano},\ and\ \citenamefont {Perinotti}}]{chiribella2008comb}%
  \BibitemOpen
  \bibfield  {author} {\bibinfo {author} {\bibfnamefont {G.}~\bibnamefont {Chiribella}}, \bibinfo {author} {\bibfnamefont {G.~M.}\ \bibnamefont {D'Ariano}},\ and\ \bibinfo {author} {\bibfnamefont {P.}~\bibnamefont {Perinotti}},\ }\bibfield  {title} {\bibinfo {title} {Quantum circuit architecture},\ }\href {https://doi.org/10.1103/PhysRevLett.101.060401} {\bibfield  {journal} {\bibinfo  {journal} {Phys. Rev. Lett.}\ }\textbf {\bibinfo {volume} {101}},\ \bibinfo {pages} {060401} (\bibinfo {year} {2008})}\BibitemShut {NoStop}%
\bibitem [{\citenamefont {Chiribella}\ \emph {et~al.}(2009)\citenamefont {Chiribella}, \citenamefont {D'Ariano},\ and\ \citenamefont {Perinotti}}]{chiribella2009comb}%
  \BibitemOpen
  \bibfield  {author} {\bibinfo {author} {\bibfnamefont {G.}~\bibnamefont {Chiribella}}, \bibinfo {author} {\bibfnamefont {G.~M.}\ \bibnamefont {D'Ariano}},\ and\ \bibinfo {author} {\bibfnamefont {P.}~\bibnamefont {Perinotti}},\ }\bibfield  {title} {\bibinfo {title} {Theoretical framework for quantum networks},\ }\href {https://doi.org/10.1103/PhysRevA.80.022339} {\bibfield  {journal} {\bibinfo  {journal} {Phys. Rev. A}\ }\textbf {\bibinfo {volume} {80}},\ \bibinfo {pages} {022339} (\bibinfo {year} {2009})}\BibitemShut {NoStop}%
\bibitem [{\citenamefont {Rivas}\ \emph {et~al.}(2014)\citenamefont {Rivas}, \citenamefont {Huelga},\ and\ \citenamefont {Plenio}}]{rivas2014quantum}%
  \BibitemOpen
  \bibfield  {author} {\bibinfo {author} {\bibfnamefont {{\'A}.}~\bibnamefont {Rivas}}, \bibinfo {author} {\bibfnamefont {S.~F.}\ \bibnamefont {Huelga}},\ and\ \bibinfo {author} {\bibfnamefont {M.~B.}\ \bibnamefont {Plenio}},\ }\bibfield  {title} {\bibinfo {title} {Quantum non-markovianity: characterization, quantification and detection},\ }\href {https://doi.org/10.1088/0034-4885/77/9/094001} {\bibfield  {journal} {\bibinfo  {journal} {Rep. Prog. Phys.}\ }\textbf {\bibinfo {volume} {77}},\ \bibinfo {pages} {094001} (\bibinfo {year} {2014})}\BibitemShut {NoStop}%
\bibitem [{\citenamefont {Pollock}\ \emph {et~al.}(2018{\natexlab{b}})\citenamefont {Pollock}, \citenamefont {Rodr\'{\i}guez-Rosario}, \citenamefont {Frauenheim}, \citenamefont {Paternostro},\ and\ \citenamefont {Modi}}]{pollock2018processtensor}%
  \BibitemOpen
  \bibfield  {author} {\bibinfo {author} {\bibfnamefont {F.~A.}\ \bibnamefont {Pollock}}, \bibinfo {author} {\bibfnamefont {C.}~\bibnamefont {Rodr\'{\i}guez-Rosario}}, \bibinfo {author} {\bibfnamefont {T.}~\bibnamefont {Frauenheim}}, \bibinfo {author} {\bibfnamefont {M.}~\bibnamefont {Paternostro}},\ and\ \bibinfo {author} {\bibfnamefont {K.}~\bibnamefont {Modi}},\ }\bibfield  {title} {\bibinfo {title} {Non-markovian quantum processes: Complete framework and efficient characterization},\ }\href {https://doi.org/10.1103/PhysRevA.97.012127} {\bibfield  {journal} {\bibinfo  {journal} {Phys. Rev. A}\ }\textbf {\bibinfo {volume} {97}},\ \bibinfo {pages} {012127} (\bibinfo {year} {2018}{\natexlab{b}})}\BibitemShut {NoStop}%
\bibitem [{\citenamefont {Milz}\ and\ \citenamefont {Modi}(2021)}]{milz2021processtensor}%
  \BibitemOpen
  \bibfield  {author} {\bibinfo {author} {\bibfnamefont {S.}~\bibnamefont {Milz}}\ and\ \bibinfo {author} {\bibfnamefont {K.}~\bibnamefont {Modi}},\ }\bibfield  {title} {\bibinfo {title} {Quantum stochastic processes and quantum non-markovian phenomena},\ }\href {https://doi.org/10.1103/PRXQuantum.2.030201} {\bibfield  {journal} {\bibinfo  {journal} {PRX Quantum}\ }\textbf {\bibinfo {volume} {2}},\ \bibinfo {pages} {030201} (\bibinfo {year} {2021})}\BibitemShut {NoStop}%
\bibitem [{\citenamefont {Figueroa-Romero}\ \emph {et~al.}(2021)\citenamefont {Figueroa-Romero}, \citenamefont {Modi}, \citenamefont {Harris}, \citenamefont {Stace},\ and\ \citenamefont {Hsieh}}]{figueroa2021nonmarkovianRB}%
  \BibitemOpen
  \bibfield  {author} {\bibinfo {author} {\bibfnamefont {P.}~\bibnamefont {Figueroa-Romero}}, \bibinfo {author} {\bibfnamefont {K.}~\bibnamefont {Modi}}, \bibinfo {author} {\bibfnamefont {R.~J.}\ \bibnamefont {Harris}}, \bibinfo {author} {\bibfnamefont {T.~M.}\ \bibnamefont {Stace}},\ and\ \bibinfo {author} {\bibfnamefont {M.-H.}\ \bibnamefont {Hsieh}},\ }\bibfield  {title} {\bibinfo {title} {Randomized benchmarking for non-markovian noise},\ }\href {https://doi.org/10.1103/PRXQuantum.2.040351} {\bibfield  {journal} {\bibinfo  {journal} {PRX Quantum}\ }\textbf {\bibinfo {volume} {2}},\ \bibinfo {pages} {040351} (\bibinfo {year} {2021})}\BibitemShut {NoStop}%
\bibitem [{\citenamefont {White}\ \emph {et~al.}(2025)\citenamefont {White}, \citenamefont {Jurcevic}, \citenamefont {Hill},\ and\ \citenamefont {Modi}}]{white2025unifying}%
  \BibitemOpen
  \bibfield  {author} {\bibinfo {author} {\bibfnamefont {G.~A.~L.}\ \bibnamefont {White}}, \bibinfo {author} {\bibfnamefont {P.}~\bibnamefont {Jurcevic}}, \bibinfo {author} {\bibfnamefont {C.~D.}\ \bibnamefont {Hill}},\ and\ \bibinfo {author} {\bibfnamefont {K.}~\bibnamefont {Modi}},\ }\bibfield  {title} {\bibinfo {title} {Unifying non-markovian characterization with an efficient and self-consistent framework},\ }\href {https://doi.org/10.1103/PhysRevX.15.021047} {\bibfield  {journal} {\bibinfo  {journal} {Phys. Rev. X}\ }\textbf {\bibinfo {volume} {15}},\ \bibinfo {pages} {021047} (\bibinfo {year} {2025})}\BibitemShut {NoStop}%
\bibitem [{\citenamefont {Terhal}\ and\ \citenamefont {Burkard}(2005)}]{terhal2005fault}%
  \BibitemOpen
  \bibfield  {author} {\bibinfo {author} {\bibfnamefont {B.~M.}\ \bibnamefont {Terhal}}\ and\ \bibinfo {author} {\bibfnamefont {G.}~\bibnamefont {Burkard}},\ }\bibfield  {title} {\bibinfo {title} {Fault-tolerant quantum computation for local non-markovian noise},\ }\href {https://doi.org/10.1103/PhysRevA.71.012336} {\bibfield  {journal} {\bibinfo  {journal} {Phys. Rev. A}\ }\textbf {\bibinfo {volume} {71}},\ \bibinfo {pages} {012336} (\bibinfo {year} {2005})}\BibitemShut {NoStop}%
\bibitem [{\citenamefont {Aharonov}\ \emph {et~al.}(2006)\citenamefont {Aharonov}, \citenamefont {Kitaev},\ and\ \citenamefont {Preskill}}]{aharonov2006fault}%
  \BibitemOpen
  \bibfield  {author} {\bibinfo {author} {\bibfnamefont {D.}~\bibnamefont {Aharonov}}, \bibinfo {author} {\bibfnamefont {A.}~\bibnamefont {Kitaev}},\ and\ \bibinfo {author} {\bibfnamefont {J.}~\bibnamefont {Preskill}},\ }\bibfield  {title} {\bibinfo {title} {Fault-tolerant quantum computation with long-range correlated noise},\ }\href {https://doi.org/10.1103/PhysRevLett.96.050504} {\bibfield  {journal} {\bibinfo  {journal} {Phys. Rev. Lett.}\ }\textbf {\bibinfo {volume} {96}},\ \bibinfo {pages} {050504} (\bibinfo {year} {2006})}\BibitemShut {NoStop}%
\bibitem [{\citenamefont {F~Kam}\ \emph {et~al.}(2025)\citenamefont {F~Kam}, \citenamefont {Gicev}, \citenamefont {Modi}, \citenamefont {Southwell},\ and\ \citenamefont {Usman}}]{Kam2025detrimental}%
  \BibitemOpen
  \bibfield  {author} {\bibinfo {author} {\bibfnamefont {J.}~\bibnamefont {F~Kam}}, \bibinfo {author} {\bibfnamefont {S.}~\bibnamefont {Gicev}}, \bibinfo {author} {\bibfnamefont {K.}~\bibnamefont {Modi}}, \bibinfo {author} {\bibfnamefont {A.}~\bibnamefont {Southwell}},\ and\ \bibinfo {author} {\bibfnamefont {M.}~\bibnamefont {Usman}},\ }\bibfield  {title} {\bibinfo {title} {Detrimental non-markovian errors for surface code memory},\ }\href {https://doi.org/10.1088/2058-9565/adebab} {\bibfield  {journal} {\bibinfo  {journal} {Quantum Sci. Technol.}\ }\textbf {\bibinfo {volume} {10}},\ \bibinfo {pages} {035060} (\bibinfo {year} {2025})}\BibitemShut {NoStop}%
\bibitem [{\citenamefont {Endo}\ \emph {et~al.}(2018)\citenamefont {Endo}, \citenamefont {Benjamin},\ and\ \citenamefont {Li}}]{endoPracticalQuantumError2018}%
  \BibitemOpen
  \bibfield  {author} {\bibinfo {author} {\bibfnamefont {S.}~\bibnamefont {Endo}}, \bibinfo {author} {\bibfnamefont {S.~C.}\ \bibnamefont {Benjamin}},\ and\ \bibinfo {author} {\bibfnamefont {Y.}~\bibnamefont {Li}},\ }\bibfield  {title} {\bibinfo {title} {Practical {{Quantum Error Mitigation}} for {{Near-Future Applications}}},\ }\href {https://doi.org/10.1103/PhysRevX.8.031027} {\bibfield  {journal} {\bibinfo  {journal} {Phys. Rev. X}\ }\textbf {\bibinfo {volume} {8}},\ \bibinfo {pages} {031027} (\bibinfo {year} {2018})}\BibitemShut {NoStop}%
\bibitem [{\citenamefont {Guo}\ and\ \citenamefont {Yang}(2022)}]{guo2022qem}%
  \BibitemOpen
  \bibfield  {author} {\bibinfo {author} {\bibfnamefont {Y.}~\bibnamefont {Guo}}\ and\ \bibinfo {author} {\bibfnamefont {S.}~\bibnamefont {Yang}},\ }\bibfield  {title} {\bibinfo {title} {Quantum error mitigation via matrix product operators},\ }\href {https://doi.org/10.1103/PRXQuantum.3.040313} {\bibfield  {journal} {\bibinfo  {journal} {PRX Quantum}\ }\textbf {\bibinfo {volume} {3}},\ \bibinfo {pages} {040313} (\bibinfo {year} {2022})}\BibitemShut {NoStop}%
\bibitem [{\citenamefont {Wang}\ and\ \citenamefont {Li}(2025)}]{wang2025non}%
  \BibitemOpen
  \bibfield  {author} {\bibinfo {author} {\bibfnamefont {K.}~\bibnamefont {Wang}}\ and\ \bibinfo {author} {\bibfnamefont {X.}~\bibnamefont {Li}},\ }\bibfield  {title} {\bibinfo {title} {Non-markovian noise mitigation: Practical implementation, error analysis, and the role of spectral properties of the environment},\ }\href {https://doi.org/10.1103/db4q-8ny9} {\bibfield  {journal} {\bibinfo  {journal} {Phys. Rev. A}\ }\textbf {\bibinfo {volume} {112}},\ \bibinfo {pages} {012406} (\bibinfo {year} {2025})}\BibitemShut {NoStop}%
\bibitem [{\citenamefont {Temme}\ \emph {et~al.}(2017)\citenamefont {Temme}, \citenamefont {Bravyi},\ and\ \citenamefont {Gambetta}}]{temme2017mitigation}%
  \BibitemOpen
  \bibfield  {author} {\bibinfo {author} {\bibfnamefont {K.}~\bibnamefont {Temme}}, \bibinfo {author} {\bibfnamefont {S.}~\bibnamefont {Bravyi}},\ and\ \bibinfo {author} {\bibfnamefont {J.~M.}\ \bibnamefont {Gambetta}},\ }\bibfield  {title} {\bibinfo {title} {Error mitigation for short-depth quantum circuits},\ }\href {https://doi.org/10.1103/PhysRevLett.119.180509} {\bibfield  {journal} {\bibinfo  {journal} {Phys. Rev. Lett.}\ }\textbf {\bibinfo {volume} {119}},\ \bibinfo {pages} {180509} (\bibinfo {year} {2017})}\BibitemShut {NoStop}%
\bibitem [{\citenamefont {Kandala}\ \emph {et~al.}(2019)\citenamefont {Kandala}, \citenamefont {Temme}, \citenamefont {C{\'o}rcoles}, \citenamefont {Mezzacapo}, \citenamefont {Chow},\ and\ \citenamefont {Gambetta}}]{Kandala2019ErrorMitigation}%
  \BibitemOpen
  \bibfield  {author} {\bibinfo {author} {\bibfnamefont {A.}~\bibnamefont {Kandala}}, \bibinfo {author} {\bibfnamefont {K.}~\bibnamefont {Temme}}, \bibinfo {author} {\bibfnamefont {A.~D.}\ \bibnamefont {C{\'o}rcoles}}, \bibinfo {author} {\bibfnamefont {A.}~\bibnamefont {Mezzacapo}}, \bibinfo {author} {\bibfnamefont {J.~M.}\ \bibnamefont {Chow}},\ and\ \bibinfo {author} {\bibfnamefont {J.~M.}\ \bibnamefont {Gambetta}},\ }\bibfield  {title} {\bibinfo {title} {Error mitigation extends the computational reach of a noisy quantum processor},\ }\href {https://doi.org/10.1038/s41586-019-1040-7} {\bibfield  {journal} {\bibinfo  {journal} {Nature}\ }\textbf {\bibinfo {volume} {567}},\ \bibinfo {pages} {491} (\bibinfo {year} {2019})}\BibitemShut {NoStop}%
\bibitem [{\citenamefont {Viola}\ and\ \citenamefont {Lloyd}(1998)}]{Viola1998DD}%
  \BibitemOpen
  \bibfield  {author} {\bibinfo {author} {\bibfnamefont {L.}~\bibnamefont {Viola}}\ and\ \bibinfo {author} {\bibfnamefont {S.}~\bibnamefont {Lloyd}},\ }\bibfield  {title} {\bibinfo {title} {Dynamical suppression of decoherence in two-state quantum systems},\ }\href {https://doi.org/10.1103/PhysRevA.58.2733} {\bibfield  {journal} {\bibinfo  {journal} {Phys. Rev. A}\ }\textbf {\bibinfo {volume} {58}},\ \bibinfo {pages} {2733} (\bibinfo {year} {1998})}\BibitemShut {NoStop}%
\bibitem [{\citenamefont {Du}\ \emph {et~al.}(2009)\citenamefont {Du}, \citenamefont {Rong}, \citenamefont {Zhao}, \citenamefont {Wang}, \citenamefont {Yang},\ and\ \citenamefont {Liu}}]{Du2009DD}%
  \BibitemOpen
  \bibfield  {author} {\bibinfo {author} {\bibfnamefont {J.}~\bibnamefont {Du}}, \bibinfo {author} {\bibfnamefont {X.}~\bibnamefont {Rong}}, \bibinfo {author} {\bibfnamefont {N.}~\bibnamefont {Zhao}}, \bibinfo {author} {\bibfnamefont {Y.}~\bibnamefont {Wang}}, \bibinfo {author} {\bibfnamefont {J.}~\bibnamefont {Yang}},\ and\ \bibinfo {author} {\bibfnamefont {R.-B.}\ \bibnamefont {Liu}},\ }\bibfield  {title} {\bibinfo {title} {Preserving electron spin coherence in solids by optimal dynamical decoupling},\ }\href {https://doi.org/10.1038/nature08470} {\bibfield  {journal} {\bibinfo  {journal} {Nature}\ }\textbf {\bibinfo {volume} {461}},\ \bibinfo {pages} {1265} (\bibinfo {year} {2009})}\BibitemShut {NoStop}%
\bibitem [{\citenamefont {Khodjasteh}\ and\ \citenamefont {Lidar}(2005)}]{Khodjasteh2005}%
  \BibitemOpen
  \bibfield  {author} {\bibinfo {author} {\bibfnamefont {K.}~\bibnamefont {Khodjasteh}}\ and\ \bibinfo {author} {\bibfnamefont {D.~A.}\ \bibnamefont {Lidar}},\ }\bibfield  {title} {\bibinfo {title} {Fault-tolerant quantum dynamical decoupling},\ }\href {https://doi.org/10.1103/PhysRevLett.95.180501} {\bibfield  {journal} {\bibinfo  {journal} {Phys. Rev. Lett.}\ }\textbf {\bibinfo {volume} {95}},\ \bibinfo {pages} {180501} (\bibinfo {year} {2005})}\BibitemShut {NoStop}%
\bibitem [{\citenamefont {Khodjasteh}\ and\ \citenamefont {Viola}(2009)}]{Khodjasteh2009}%
  \BibitemOpen
  \bibfield  {author} {\bibinfo {author} {\bibfnamefont {K.}~\bibnamefont {Khodjasteh}}\ and\ \bibinfo {author} {\bibfnamefont {L.}~\bibnamefont {Viola}},\ }\bibfield  {title} {\bibinfo {title} {Dynamically error-corrected gates for universal quantum computation},\ }\href {https://doi.org/10.1103/PhysRevLett.102.080501} {\bibfield  {journal} {\bibinfo  {journal} {Phys. Rev. Lett.}\ }\textbf {\bibinfo {volume} {102}},\ \bibinfo {pages} {080501} (\bibinfo {year} {2009})}\BibitemShut {NoStop}%
\bibitem [{\citenamefont {Zeng}\ \emph {et~al.}(2018)\citenamefont {Zeng}, \citenamefont {Deng}, \citenamefont {Russo},\ and\ \citenamefont {Barnes}}]{Zeng2018DD}%
  \BibitemOpen
  \bibfield  {author} {\bibinfo {author} {\bibfnamefont {J.}~\bibnamefont {Zeng}}, \bibinfo {author} {\bibfnamefont {X.-H.}\ \bibnamefont {Deng}}, \bibinfo {author} {\bibfnamefont {A.}~\bibnamefont {Russo}},\ and\ \bibinfo {author} {\bibfnamefont {E.}~\bibnamefont {Barnes}},\ }\bibfield  {title} {\bibinfo {title} {General solution to inhomogeneous dephasing and smooth pulse dynamical decoupling},\ }\href {https://doi.org/10.1088/1367-2630/aaafe9} {\bibfield  {journal} {\bibinfo  {journal} {New J. Phys.}\ }\textbf {\bibinfo {volume} {20}},\ \bibinfo {pages} {033011} (\bibinfo {year} {2018})}\BibitemShut {NoStop}%
\bibitem [{\citenamefont {Addis}\ \emph {et~al.}(2015)\citenamefont {Addis}, \citenamefont {Bylicka}, \citenamefont {Chru{\'s}ci{\'n}ski},\ and\ \citenamefont {Maniscalco}}]{Addis2015}%
  \BibitemOpen
  \bibfield  {author} {\bibinfo {author} {\bibfnamefont {C.}~\bibnamefont {Addis}}, \bibinfo {author} {\bibfnamefont {B.}~\bibnamefont {Bylicka}}, \bibinfo {author} {\bibfnamefont {D.}~\bibnamefont {Chru{\'s}ci{\'n}ski}},\ and\ \bibinfo {author} {\bibfnamefont {S.}~\bibnamefont {Maniscalco}},\ }\bibfield  {title} {\bibinfo {title} {Dynamical decoupling efficiency versus quantum non-markovianity},\ }\href {https://doi.org/10.1088/1367-2630/17/12/123004} {\bibfield  {journal} {\bibinfo  {journal} {New J. Phys.}\ }\textbf {\bibinfo {volume} {17}},\ \bibinfo {pages} {123004} (\bibinfo {year} {2015})}\BibitemShut {NoStop}%
\bibitem [{\citenamefont {Liu}\ \emph {et~al.}(2024)\citenamefont {Liu}, \citenamefont {Xiao},\ and\ \citenamefont {Cai}}]{liu2024non}%
  \BibitemOpen
  \bibfield  {author} {\bibinfo {author} {\bibfnamefont {Z.}~\bibnamefont {Liu}}, \bibinfo {author} {\bibfnamefont {Y.}~\bibnamefont {Xiao}},\ and\ \bibinfo {author} {\bibfnamefont {Z.}~\bibnamefont {Cai}},\ }\bibfield  {title} {\bibinfo {title} {Non-markovian noise suppression simplified through channel representation},\ }\href {https://arxiv.org/abs/2412.11220} {\bibfield  {journal} {\bibinfo  {journal} {arXiv:2412.11220}\ } (\bibinfo {year} {2024})}\BibitemShut {NoStop}%
\bibitem [{\citenamefont {Cirac}\ \emph {et~al.}(1999)\citenamefont {Cirac}, \citenamefont {Ekert},\ and\ \citenamefont {Macchiavello}}]{cirac1999optimal}%
  \BibitemOpen
  \bibfield  {author} {\bibinfo {author} {\bibfnamefont {J.~I.}\ \bibnamefont {Cirac}}, \bibinfo {author} {\bibfnamefont {A.~K.}\ \bibnamefont {Ekert}},\ and\ \bibinfo {author} {\bibfnamefont {C.}~\bibnamefont {Macchiavello}},\ }\bibfield  {title} {\bibinfo {title} {Optimal purification of single qubits},\ }\href {https://doi.org/10.1103/PhysRevLett.82.4344} {\bibfield  {journal} {\bibinfo  {journal} {Phys. Rev. Lett.}\ }\textbf {\bibinfo {volume} {82}},\ \bibinfo {pages} {4344} (\bibinfo {year} {1999})}\BibitemShut {NoStop}%
\bibitem [{\citenamefont {Cotler}\ \emph {et~al.}(2019)\citenamefont {Cotler}, \citenamefont {Choi}, \citenamefont {Lukin}, \citenamefont {Gharibyan}, \citenamefont {Grover}, \citenamefont {Tai}, \citenamefont {Rispoli}, \citenamefont {Schittko}, \citenamefont {Preiss}, \citenamefont {Kaufman}, \citenamefont {Greiner}, \citenamefont {Pichler},\ and\ \citenamefont {Hayden}}]{cotler2019cooling}%
  \BibitemOpen
  \bibfield  {author} {\bibinfo {author} {\bibfnamefont {J.}~\bibnamefont {Cotler}}, \bibinfo {author} {\bibfnamefont {S.}~\bibnamefont {Choi}}, \bibinfo {author} {\bibfnamefont {A.}~\bibnamefont {Lukin}}, \bibinfo {author} {\bibfnamefont {H.}~\bibnamefont {Gharibyan}}, \bibinfo {author} {\bibfnamefont {T.}~\bibnamefont {Grover}}, \bibinfo {author} {\bibfnamefont {M.~E.}\ \bibnamefont {Tai}}, \bibinfo {author} {\bibfnamefont {M.}~\bibnamefont {Rispoli}}, \bibinfo {author} {\bibfnamefont {R.}~\bibnamefont {Schittko}}, \bibinfo {author} {\bibfnamefont {P.~M.}\ \bibnamefont {Preiss}}, \bibinfo {author} {\bibfnamefont {A.~M.}\ \bibnamefont {Kaufman}}, \bibinfo {author} {\bibfnamefont {M.}~\bibnamefont {Greiner}}, \bibinfo {author} {\bibfnamefont {H.}~\bibnamefont {Pichler}},\ and\ \bibinfo {author} {\bibfnamefont {P.}~\bibnamefont {Hayden}},\ }\bibfield  {title} {\bibinfo {title} {Quantum virtual cooling},\ }\href {https://doi.org/10.1103/PhysRevX.9.031013} {\bibfield  {journal} {\bibinfo  {journal} {Phys. Rev.
  X}\ }\textbf {\bibinfo {volume} {9}},\ \bibinfo {pages} {031013} (\bibinfo {year} {2019})}\BibitemShut {NoStop}%
\bibitem [{\citenamefont {Huggins}\ \emph {et~al.}(2021)\citenamefont {Huggins}, \citenamefont {McArdle}, \citenamefont {O'Brien}, \citenamefont {Lee}, \citenamefont {Rubin}, \citenamefont {Boixo}, \citenamefont {Whaley}, \citenamefont {Babbush},\ and\ \citenamefont {McClean}}]{hugginsVirtualDistillationQuantum2021}%
  \BibitemOpen
  \bibfield  {author} {\bibinfo {author} {\bibfnamefont {W.~J.}\ \bibnamefont {Huggins}}, \bibinfo {author} {\bibfnamefont {S.}~\bibnamefont {McArdle}}, \bibinfo {author} {\bibfnamefont {T.~E.}\ \bibnamefont {O'Brien}}, \bibinfo {author} {\bibfnamefont {J.}~\bibnamefont {Lee}}, \bibinfo {author} {\bibfnamefont {N.~C.}\ \bibnamefont {Rubin}}, \bibinfo {author} {\bibfnamefont {S.}~\bibnamefont {Boixo}}, \bibinfo {author} {\bibfnamefont {K.~B.}\ \bibnamefont {Whaley}}, \bibinfo {author} {\bibfnamefont {R.}~\bibnamefont {Babbush}},\ and\ \bibinfo {author} {\bibfnamefont {J.~R.}\ \bibnamefont {McClean}},\ }\bibfield  {title} {\bibinfo {title} {Virtual {{distillation}} for {{quantum error mitigation}}},\ }\href {https://doi.org/10.1103/PhysRevX.11.041036} {\bibfield  {journal} {\bibinfo  {journal} {Phys. Rev. X}\ }\textbf {\bibinfo {volume} {11}},\ \bibinfo {pages} {041036} (\bibinfo {year} {2021})}\BibitemShut {NoStop}%
\bibitem [{\citenamefont {Koczor}(2021)}]{koczorExponentialErrorSuppression2021}%
  \BibitemOpen
  \bibfield  {author} {\bibinfo {author} {\bibfnamefont {B.}~\bibnamefont {Koczor}},\ }\bibfield  {title} {\bibinfo {title} {Exponential error suppression for near-term quantum devices},\ }\href {https://doi.org/10.1103/PhysRevX.11.031057} {\bibfield  {journal} {\bibinfo  {journal} {Phys. Rev. X}\ }\textbf {\bibinfo {volume} {11}},\ \bibinfo {pages} {031057} (\bibinfo {year} {2021})}\BibitemShut {NoStop}%
\bibitem [{\citenamefont {Childs}\ \emph {et~al.}(2025)\citenamefont {Childs}, \citenamefont {Fu}, \citenamefont {Leung}, \citenamefont {Li}, \citenamefont {Ozols},\ and\ \citenamefont {Vyas}}]{Childs2025streamingquantum}%
  \BibitemOpen
  \bibfield  {author} {\bibinfo {author} {\bibfnamefont {A.~M.}\ \bibnamefont {Childs}}, \bibinfo {author} {\bibfnamefont {H.}~\bibnamefont {Fu}}, \bibinfo {author} {\bibfnamefont {D.}~\bibnamefont {Leung}}, \bibinfo {author} {\bibfnamefont {Z.}~\bibnamefont {Li}}, \bibinfo {author} {\bibfnamefont {M.}~\bibnamefont {Ozols}},\ and\ \bibinfo {author} {\bibfnamefont {V.}~\bibnamefont {Vyas}},\ }\bibfield  {title} {\bibinfo {title} {Streaming quantum state purification},\ }\href {https://doi.org/10.22331/q-2025-01-21-1603} {\bibfield  {journal} {\bibinfo  {journal} {{Quantum}}\ }\textbf {\bibinfo {volume} {9}},\ \bibinfo {pages} {1603} (\bibinfo {year} {2025})}\BibitemShut {NoStop}%
\bibitem [{\citenamefont {Liu}\ \emph {et~al.}(2025{\natexlab{a}})\citenamefont {Liu}, \citenamefont {Zhang}, \citenamefont {Fei},\ and\ \citenamefont {Cai}}]{liu2025virtual}%
  \BibitemOpen
  \bibfield  {author} {\bibinfo {author} {\bibfnamefont {Z.}~\bibnamefont {Liu}}, \bibinfo {author} {\bibfnamefont {X.}~\bibnamefont {Zhang}}, \bibinfo {author} {\bibfnamefont {Y.-Y.}\ \bibnamefont {Fei}},\ and\ \bibinfo {author} {\bibfnamefont {Z.}~\bibnamefont {Cai}},\ }\bibfield  {title} {\bibinfo {title} {Virtual channel purification},\ }\href {https://journals.aps.org/prxquantum/abstract/10.1103/PRXQuantum.6.020325} {\bibfield  {journal} {\bibinfo  {journal} {PRX Quantum}\ }\textbf {\bibinfo {volume} {6}},\ \bibinfo {pages} {020325} (\bibinfo {year} {2025}{\natexlab{a}})}\BibitemShut {NoStop}%
\bibitem [{\citenamefont {Wallman}\ and\ \citenamefont {Emerson}(2016)}]{wallmanNoiseTailoringScalable2016}%
  \BibitemOpen
  \bibfield  {author} {\bibinfo {author} {\bibfnamefont {J.~J.}\ \bibnamefont {Wallman}}\ and\ \bibinfo {author} {\bibfnamefont {J.}~\bibnamefont {Emerson}},\ }\bibfield  {title} {\bibinfo {title} {Noise tailoring for scalable quantum computation via randomized compiling},\ }\href {https://doi.org/10.1103/PhysRevA.94.052325} {\bibfield  {journal} {\bibinfo  {journal} {Phys. Rev. A}\ }\textbf {\bibinfo {volume} {94}},\ \bibinfo {pages} {052325} (\bibinfo {year} {2016})}\BibitemShut {NoStop}%
\bibitem [{\citenamefont {Cai}\ \emph {et~al.}(2020)\citenamefont {Cai}, \citenamefont {Xu},\ and\ \citenamefont {Benjamin}}]{Cai2020Mitigating}%
  \BibitemOpen
  \bibfield  {author} {\bibinfo {author} {\bibfnamefont {Z.}~\bibnamefont {Cai}}, \bibinfo {author} {\bibfnamefont {X.}~\bibnamefont {Xu}},\ and\ \bibinfo {author} {\bibfnamefont {S.~C.}\ \bibnamefont {Benjamin}},\ }\bibfield  {title} {\bibinfo {title} {Mitigating coherent noise using pauli conjugation},\ }\href {https://doi.org/10.1038/s41534-019-0233-0} {\bibfield  {journal} {\bibinfo  {journal} {npj Quantum Inf.}\ }\textbf {\bibinfo {volume} {6}},\ \bibinfo {pages} {17} (\bibinfo {year} {2020})}\BibitemShut {NoStop}%
\bibitem [{\citenamefont {Wallman}(2018)}]{Wallman2018randomized}%
  \BibitemOpen
  \bibfield  {author} {\bibinfo {author} {\bibfnamefont {J.~J.}\ \bibnamefont {Wallman}},\ }\bibfield  {title} {\bibinfo {title} {Randomized benchmarking with gate-dependent noise},\ }\href {https://doi.org/10.22331/q-2018-01-29-47} {\bibfield  {journal} {\bibinfo  {journal} {{Quantum}}\ }\textbf {\bibinfo {volume} {2}},\ \bibinfo {pages} {47} (\bibinfo {year} {2018})}\BibitemShut {NoStop}%
\bibitem [{\citenamefont {Winick}\ \emph {et~al.}(2022)\citenamefont {Winick}, \citenamefont {Wallman}, \citenamefont {Dahlen}, \citenamefont {Hincks}, \citenamefont {Ospadov},\ and\ \citenamefont {Emerson}}]{winickConceptsConditionsError2022}%
  \BibitemOpen
  \bibfield  {author} {\bibinfo {author} {\bibfnamefont {A.}~\bibnamefont {Winick}}, \bibinfo {author} {\bibfnamefont {J.~J.}\ \bibnamefont {Wallman}}, \bibinfo {author} {\bibfnamefont {D.}~\bibnamefont {Dahlen}}, \bibinfo {author} {\bibfnamefont {I.}~\bibnamefont {Hincks}}, \bibinfo {author} {\bibfnamefont {E.}~\bibnamefont {Ospadov}},\ and\ \bibinfo {author} {\bibfnamefont {J.}~\bibnamefont {Emerson}},\ }\bibfield  {title} {\bibinfo {title} {Concepts and conditions for error suppression through randomized compiling},\ }\href {http://arxiv.org/abs/2212.07500} {\bibfield  {journal} {\bibinfo  {journal} {arXiv:2212.07500}\ } (\bibinfo {year} {2022})}\BibitemShut {NoStop}%
\bibitem [{\citenamefont {{Figueroa-Romero}}\ \emph {et~al.}(2024)\citenamefont {{Figueroa-Romero}}, \citenamefont {Papi{\v c}}, \citenamefont {Auer}, \citenamefont {Hsieh}, \citenamefont {Modi},\ and\ \citenamefont {de~Vega}}]{figueroa-romeroOperationalMarkovianizationRandomized2024}%
  \BibitemOpen
  \bibfield  {author} {\bibinfo {author} {\bibfnamefont {P.}~\bibnamefont {{Figueroa-Romero}}}, \bibinfo {author} {\bibfnamefont {M.}~\bibnamefont {Papi{\v c}}}, \bibinfo {author} {\bibfnamefont {A.}~\bibnamefont {Auer}}, \bibinfo {author} {\bibfnamefont {M.-H.}\ \bibnamefont {Hsieh}}, \bibinfo {author} {\bibfnamefont {K.}~\bibnamefont {Modi}},\ and\ \bibinfo {author} {\bibfnamefont {I.}~\bibnamefont {de~Vega}},\ }\bibfield  {title} {\bibinfo {title} {Operational markovianization in randomized benchmarking},\ }\href {https://doi.org/10.1088/2058-9565/ad3f44} {\bibfield  {journal} {\bibinfo  {journal} {Quantum Science and Technology}\ }\textbf {\bibinfo {volume} {9}},\ \bibinfo {pages} {35020} (\bibinfo {year} {2024})}\BibitemShut {NoStop}%
\bibitem [{\citenamefont {Smolin}\ and\ \citenamefont {DiVincenzo}(1996)}]{smolin1996cswap}%
  \BibitemOpen
  \bibfield  {author} {\bibinfo {author} {\bibfnamefont {J.~A.}\ \bibnamefont {Smolin}}\ and\ \bibinfo {author} {\bibfnamefont {D.~P.}\ \bibnamefont {DiVincenzo}},\ }\bibfield  {title} {\bibinfo {title} {Five two-bit quantum gates are sufficient to implement the quantum fredkin gate},\ }\href {https://doi.org/10.1103/PhysRevA.53.2855} {\bibfield  {journal} {\bibinfo  {journal} {Phys. Rev. A}\ }\textbf {\bibinfo {volume} {53}},\ \bibinfo {pages} {2855} (\bibinfo {year} {1996})}\BibitemShut {NoStop}%
\bibitem [{\citenamefont {Yu}\ \emph {et~al.}(2013)\citenamefont {Yu}, \citenamefont {Duan},\ and\ \citenamefont {Ying}}]{yu2013cswap}%
  \BibitemOpen
  \bibfield  {author} {\bibinfo {author} {\bibfnamefont {N.}~\bibnamefont {Yu}}, \bibinfo {author} {\bibfnamefont {R.}~\bibnamefont {Duan}},\ and\ \bibinfo {author} {\bibfnamefont {M.}~\bibnamefont {Ying}},\ }\bibfield  {title} {\bibinfo {title} {Five two-qubit gates are necessary for implementing the toffoli gate},\ }\href {https://doi.org/10.1103/PhysRevA.88.010304} {\bibfield  {journal} {\bibinfo  {journal} {Phys. Rev. A}\ }\textbf {\bibinfo {volume} {88}},\ \bibinfo {pages} {010304} (\bibinfo {year} {2013})}\BibitemShut {NoStop}%
\bibitem [{\citenamefont {Nie}\ \emph {et~al.}(2024)\citenamefont {Nie}, \citenamefont {Zhu}, \citenamefont {Fan}, \citenamefont {Long}, \citenamefont {Liu}, \citenamefont {Huang}, \citenamefont {Xi}, \citenamefont {Che}, \citenamefont {Zheng}, \citenamefont {Feng}, \citenamefont {Yang},\ and\ \citenamefont {Lu}}]{Niexf2024}%
  \BibitemOpen
  \bibfield  {author} {\bibinfo {author} {\bibfnamefont {X.}~\bibnamefont {Nie}}, \bibinfo {author} {\bibfnamefont {X.}~\bibnamefont {Zhu}}, \bibinfo {author} {\bibfnamefont {Y.-a.}\ \bibnamefont {Fan}}, \bibinfo {author} {\bibfnamefont {X.}~\bibnamefont {Long}}, \bibinfo {author} {\bibfnamefont {H.}~\bibnamefont {Liu}}, \bibinfo {author} {\bibfnamefont {K.}~\bibnamefont {Huang}}, \bibinfo {author} {\bibfnamefont {C.}~\bibnamefont {Xi}}, \bibinfo {author} {\bibfnamefont {L.}~\bibnamefont {Che}}, \bibinfo {author} {\bibfnamefont {Y.}~\bibnamefont {Zheng}}, \bibinfo {author} {\bibfnamefont {Y.}~\bibnamefont {Feng}}, \bibinfo {author} {\bibfnamefont {X.}~\bibnamefont {Yang}},\ and\ \bibinfo {author} {\bibfnamefont {D.}~\bibnamefont {Lu}},\ }\bibfield  {title} {\bibinfo {title} {Self-consistent determination of single-impurity anderson model using hybrid quantum-classical approach on a spin quantum simulator},\ }\href {https://doi.org/10.1103/PhysRevLett.133.140602} {\bibfield  {journal} {\bibinfo  {journal}
  {Phys. Rev. Lett.}\ }\textbf {\bibinfo {volume} {133}},\ \bibinfo {pages} {140602} (\bibinfo {year} {2024})}\BibitemShut {NoStop}%
\bibitem [{\citenamefont {Li}\ \emph {et~al.}(2014)\citenamefont {Li}, \citenamefont {Zhou}, \citenamefont {Ju}, \citenamefont {Chen}, \citenamefont {Zheng}, \citenamefont {Lu}, \citenamefont {Rong}, \citenamefont {Duan}, \citenamefont {Peng},\ and\ \citenamefont {Du}}]{Lizhaokai2014}%
  \BibitemOpen
  \bibfield  {author} {\bibinfo {author} {\bibfnamefont {Z.}~\bibnamefont {Li}}, \bibinfo {author} {\bibfnamefont {H.}~\bibnamefont {Zhou}}, \bibinfo {author} {\bibfnamefont {C.}~\bibnamefont {Ju}}, \bibinfo {author} {\bibfnamefont {H.}~\bibnamefont {Chen}}, \bibinfo {author} {\bibfnamefont {W.}~\bibnamefont {Zheng}}, \bibinfo {author} {\bibfnamefont {D.}~\bibnamefont {Lu}}, \bibinfo {author} {\bibfnamefont {X.}~\bibnamefont {Rong}}, \bibinfo {author} {\bibfnamefont {C.}~\bibnamefont {Duan}}, \bibinfo {author} {\bibfnamefont {X.}~\bibnamefont {Peng}},\ and\ \bibinfo {author} {\bibfnamefont {J.}~\bibnamefont {Du}},\ }\bibfield  {title} {\bibinfo {title} {Experimental realization of a compressed quantum simulation of a 32-spin ising chain},\ }\href {https://doi.org/10.1103/PhysRevLett.112.220501} {\bibfield  {journal} {\bibinfo  {journal} {Phys. Rev. Lett.}\ }\textbf {\bibinfo {volume} {112}},\ \bibinfo {pages} {220501} (\bibinfo {year} {2014})}\BibitemShut {NoStop}%
\bibitem [{\citenamefont {Chen}\ \emph {et~al.}(2021)\citenamefont {Chen}, \citenamefont {Cheng}, \citenamefont {Li}, \citenamefont {Nie}, \citenamefont {Yu}, \citenamefont {Yung},\ and\ \citenamefont {Peng}}]{CHEN202123}%
  \BibitemOpen
  \bibfield  {author} {\bibinfo {author} {\bibfnamefont {X.}~\bibnamefont {Chen}}, \bibinfo {author} {\bibfnamefont {B.}~\bibnamefont {Cheng}}, \bibinfo {author} {\bibfnamefont {Z.}~\bibnamefont {Li}}, \bibinfo {author} {\bibfnamefont {X.}~\bibnamefont {Nie}}, \bibinfo {author} {\bibfnamefont {N.}~\bibnamefont {Yu}}, \bibinfo {author} {\bibfnamefont {M.-H.}\ \bibnamefont {Yung}},\ and\ \bibinfo {author} {\bibfnamefont {X.}~\bibnamefont {Peng}},\ }\bibfield  {title} {\bibinfo {title} {Experimental cryptographic verification for near-term quantum cloud computing},\ }\href {https://doi.org/https://doi.org/10.1016/j.scib.2020.08.013} {\bibfield  {journal} {\bibinfo  {journal} {Sci Bull.}\ }\textbf {\bibinfo {volume} {66}},\ \bibinfo {pages} {23} (\bibinfo {year} {2021})}\BibitemShut {NoStop}%
\bibitem [{\citenamefont {Khaneja}\ \emph {et~al.}(2005)\citenamefont {Khaneja}, \citenamefont {Reiss}, \citenamefont {Kehlet}, \citenamefont {Schulte-Herbr{\"u}ggen},\ and\ \citenamefont {Glaser}}]{khaneja2005optimal}%
  \BibitemOpen
  \bibfield  {author} {\bibinfo {author} {\bibfnamefont {N.}~\bibnamefont {Khaneja}}, \bibinfo {author} {\bibfnamefont {T.}~\bibnamefont {Reiss}}, \bibinfo {author} {\bibfnamefont {C.}~\bibnamefont {Kehlet}}, \bibinfo {author} {\bibfnamefont {T.}~\bibnamefont {Schulte-Herbr{\"u}ggen}},\ and\ \bibinfo {author} {\bibfnamefont {S.~J.}\ \bibnamefont {Glaser}},\ }\bibfield  {title} {\bibinfo {title} {Optimal control of coupled spin dynamics: design of nmr pulse sequences by gradient ascent algorithms},\ }\href {https://doi.org/10.1016/j.jmr.2004.11.004} {\bibfield  {journal} {\bibinfo  {journal} {J. Magn. Reson.}\ }\textbf {\bibinfo {volume} {172}},\ \bibinfo {pages} {296} (\bibinfo {year} {2005})}\BibitemShut {NoStop}%
\bibitem [{PEN(2001)}]{PENG2001509}%
  \BibitemOpen
  \bibfield  {title} {\bibinfo {title} {Preparation of pseudo-pure states by line-selective pulses in nuclear magnetic resonance},\ }\href {https://doi.org/https://doi.org/10.1016/S0009-2614(01)00421-3} {\bibfield  {journal} {\bibinfo  {journal} {Chem. Phys. Lett.}\ }\textbf {\bibinfo {volume} {340}},\ \bibinfo {pages} {509} (\bibinfo {year} {2001})}\BibitemShut {NoStop}%
\bibitem [{\citenamefont {Seltman}(2012)}]{seltman2012approximations}%
  \BibitemOpen
  \bibfield  {author} {\bibinfo {author} {\bibfnamefont {H.}~\bibnamefont {Seltman}},\ }\bibfield  {title} {\bibinfo {title} {Approximations for mean and variance of a ratio},\ }\href {https://www.stat.cmu.edu/~hseltman/files/ratio.pdf} {\bibfield  {journal} {\bibinfo  {journal} {unpublished note}\ } (\bibinfo {year} {2012})}\BibitemShut {NoStop}%
\bibitem [{\citenamefont {Liu}\ \emph {et~al.}(2025{\natexlab{b}})\citenamefont {Liu}, \citenamefont {Hur}, \citenamefont {Zhang}, \citenamefont {Che}, \citenamefont {Long}, \citenamefont {Wang}, \citenamefont {Huang}, \citenamefont {Fan}, \citenamefont {Zheng}, \citenamefont {Feng}, \citenamefont {Zhou}, \citenamefont {Ng}, \citenamefont {Nie}, \citenamefont {Park},\ and\ \citenamefont {Lu}}]{liu2025neural}%
  \BibitemOpen
  \bibfield  {author} {\bibinfo {author} {\bibfnamefont {H.}~\bibnamefont {Liu}}, \bibinfo {author} {\bibfnamefont {T.}~\bibnamefont {Hur}}, \bibinfo {author} {\bibfnamefont {S.}~\bibnamefont {Zhang}}, \bibinfo {author} {\bibfnamefont {L.}~\bibnamefont {Che}}, \bibinfo {author} {\bibfnamefont {X.}~\bibnamefont {Long}}, \bibinfo {author} {\bibfnamefont {X.}~\bibnamefont {Wang}}, \bibinfo {author} {\bibfnamefont {K.}~\bibnamefont {Huang}}, \bibinfo {author} {\bibfnamefont {Y.-a.}\ \bibnamefont {Fan}}, \bibinfo {author} {\bibfnamefont {Y.}~\bibnamefont {Zheng}}, \bibinfo {author} {\bibfnamefont {Y.}~\bibnamefont {Feng}}, \bibinfo {author} {\bibfnamefont {Y.}~\bibnamefont {Zhou}}, \bibinfo {author} {\bibfnamefont {J.}~\bibnamefont {Ng}}, \bibinfo {author} {\bibfnamefont {X.}~\bibnamefont {Nie}}, \bibinfo {author} {\bibfnamefont {D.~K.}\ \bibnamefont {Park}},\ and\ \bibinfo {author} {\bibfnamefont {D.}~\bibnamefont {Lu}},\ }\bibfield  {title} {\bibinfo {title} {Neural quantum embedding via deterministic quantum
  computation with one qubit},\ }\href {https://doi.org/10.1103/y8wr-yml4} {\bibfield  {journal} {\bibinfo  {journal} {Phys. Rev. Lett.}\ }\textbf {\bibinfo {volume} {135}},\ \bibinfo {pages} {080603} (\bibinfo {year} {2025}{\natexlab{b}})}\BibitemShut {NoStop}%
\bibitem [{\citenamefont {Liu}\ \emph {et~al.}(2025{\natexlab{c}})\citenamefont {Liu}, \citenamefont {Liu}, \citenamefont {Chen}, \citenamefont {Nie}, \citenamefont {Liu},\ and\ \citenamefont {Lu}}]{Liuhf2025}%
  \BibitemOpen
  \bibfield  {author} {\bibinfo {author} {\bibfnamefont {H.}~\bibnamefont {Liu}}, \bibinfo {author} {\bibfnamefont {Z.}~\bibnamefont {Liu}}, \bibinfo {author} {\bibfnamefont {S.}~\bibnamefont {Chen}}, \bibinfo {author} {\bibfnamefont {X.}~\bibnamefont {Nie}}, \bibinfo {author} {\bibfnamefont {X.}~\bibnamefont {Liu}},\ and\ \bibinfo {author} {\bibfnamefont {D.}~\bibnamefont {Lu}},\ }\bibfield  {title} {\bibinfo {title} {Certifying quantum temporal correlation via randomized measurements: Theory and experiment},\ }\href {https://doi.org/10.1103/PhysRevLett.134.040201} {\bibfield  {journal} {\bibinfo  {journal} {Phys. Rev. Lett.}\ }\textbf {\bibinfo {volume} {134}},\ \bibinfo {pages} {040201} (\bibinfo {year} {2025}{\natexlab{c}})}\BibitemShut {NoStop}%
\bibitem [{\citenamefont {Xi}\ \emph {et~al.}(2024)\citenamefont {Xi}, \citenamefont {Liu}, \citenamefont {Liu}, \citenamefont {Huang}, \citenamefont {Long}, \citenamefont {Ebler}, \citenamefont {Nie}, \citenamefont {Dahlsten},\ and\ \citenamefont {Lu}}]{Xic2024experimental}%
  \BibitemOpen
  \bibfield  {author} {\bibinfo {author} {\bibfnamefont {C.}~\bibnamefont {Xi}}, \bibinfo {author} {\bibfnamefont {X.}~\bibnamefont {Liu}}, \bibinfo {author} {\bibfnamefont {H.}~\bibnamefont {Liu}}, \bibinfo {author} {\bibfnamefont {K.}~\bibnamefont {Huang}}, \bibinfo {author} {\bibfnamefont {X.}~\bibnamefont {Long}}, \bibinfo {author} {\bibfnamefont {D.}~\bibnamefont {Ebler}}, \bibinfo {author} {\bibfnamefont {X.}~\bibnamefont {Nie}}, \bibinfo {author} {\bibfnamefont {O.}~\bibnamefont {Dahlsten}},\ and\ \bibinfo {author} {\bibfnamefont {D.}~\bibnamefont {Lu}},\ }\bibfield  {title} {\bibinfo {title} {Experimental validation of enhanced information capacity by quantum switch in accordance with thermodynamic laws},\ }\href {https://doi.org/10.1103/PhysRevLett.133.040401} {\bibfield  {journal} {\bibinfo  {journal} {Phys. Rev. Lett.}\ }\textbf {\bibinfo {volume} {133}},\ \bibinfo {pages} {040401} (\bibinfo {year} {2024})}\BibitemShut {NoStop}%
\bibitem [{\citenamefont {Zheng}\ \emph {et~al.}(2025)\citenamefont {Zheng}, \citenamefont {Nie}, \citenamefont {Liu}, \citenamefont {Luo}, \citenamefont {Lu},\ and\ \citenamefont {Liu}}]{zheng2025vqb}%
  \BibitemOpen
  \bibfield  {author} {\bibinfo {author} {\bibfnamefont {Y.}~\bibnamefont {Zheng}}, \bibinfo {author} {\bibfnamefont {X.}~\bibnamefont {Nie}}, \bibinfo {author} {\bibfnamefont {H.}~\bibnamefont {Liu}}, \bibinfo {author} {\bibfnamefont {Y.}~\bibnamefont {Luo}}, \bibinfo {author} {\bibfnamefont {D.}~\bibnamefont {Lu}},\ and\ \bibinfo {author} {\bibfnamefont {X.}~\bibnamefont {Liu}},\ }\bibfield  {title} {\bibinfo {title} {Experimental virtual quantum broadcasting},\ }\href {https://doi.org/10.1103/8vrg-tvsd} {\bibfield  {journal} {\bibinfo  {journal} {Phys. Rev. A}\ }\textbf {\bibinfo {volume} {111}},\ \bibinfo {pages} {L060402} (\bibinfo {year} {2025})}\BibitemShut {NoStop}%
\bibitem [{\citenamefont {Huang}\ \emph {et~al.}(2024)\citenamefont {Huang}, \citenamefont {Xi}, \citenamefont {Long}, \citenamefont {Liu}, \citenamefont {Fan}, \citenamefont {Wang}, \citenamefont {Zheng}, \citenamefont {Feng}, \citenamefont {Nie},\ and\ \citenamefont {Lu}}]{Huangkyi2024}%
  \BibitemOpen
  \bibfield  {author} {\bibinfo {author} {\bibfnamefont {K.}~\bibnamefont {Huang}}, \bibinfo {author} {\bibfnamefont {C.}~\bibnamefont {Xi}}, \bibinfo {author} {\bibfnamefont {X.}~\bibnamefont {Long}}, \bibinfo {author} {\bibfnamefont {H.}~\bibnamefont {Liu}}, \bibinfo {author} {\bibfnamefont {Y.-a.}\ \bibnamefont {Fan}}, \bibinfo {author} {\bibfnamefont {X.}~\bibnamefont {Wang}}, \bibinfo {author} {\bibfnamefont {Y.}~\bibnamefont {Zheng}}, \bibinfo {author} {\bibfnamefont {Y.}~\bibnamefont {Feng}}, \bibinfo {author} {\bibfnamefont {X.}~\bibnamefont {Nie}},\ and\ \bibinfo {author} {\bibfnamefont {D.}~\bibnamefont {Lu}},\ }\bibfield  {title} {\bibinfo {title} {Experimental realization of self-contained quantum refrigeration},\ }\href {https://doi.org/10.1103/PhysRevLett.132.210403} {\bibfield  {journal} {\bibinfo  {journal} {Phys. Rev. Lett.}\ }\textbf {\bibinfo {volume} {132}},\ \bibinfo {pages} {210403} (\bibinfo {year} {2024})}\BibitemShut {NoStop}%
\bibitem [{\citenamefont {Long}\ \emph {et~al.}(2022)\citenamefont {Long}, \citenamefont {He}, \citenamefont {Zhang}, \citenamefont {Tang}, \citenamefont {Lin}, \citenamefont {Liu}, \citenamefont {Nie}, \citenamefont {Feng}, \citenamefont {Li}, \citenamefont {Xin}, \citenamefont {Ai},\ and\ \citenamefont {Lu}}]{Longxy2022}%
  \BibitemOpen
  \bibfield  {author} {\bibinfo {author} {\bibfnamefont {X.}~\bibnamefont {Long}}, \bibinfo {author} {\bibfnamefont {W.-T.}\ \bibnamefont {He}}, \bibinfo {author} {\bibfnamefont {N.-N.}\ \bibnamefont {Zhang}}, \bibinfo {author} {\bibfnamefont {K.}~\bibnamefont {Tang}}, \bibinfo {author} {\bibfnamefont {Z.}~\bibnamefont {Lin}}, \bibinfo {author} {\bibfnamefont {H.}~\bibnamefont {Liu}}, \bibinfo {author} {\bibfnamefont {X.}~\bibnamefont {Nie}}, \bibinfo {author} {\bibfnamefont {G.}~\bibnamefont {Feng}}, \bibinfo {author} {\bibfnamefont {J.}~\bibnamefont {Li}}, \bibinfo {author} {\bibfnamefont {T.}~\bibnamefont {Xin}}, \bibinfo {author} {\bibfnamefont {Q.}~\bibnamefont {Ai}},\ and\ \bibinfo {author} {\bibfnamefont {D.}~\bibnamefont {Lu}},\ }\bibfield  {title} {\bibinfo {title} {Entanglement-enhanced quantum metrology in colored noise by quantum zeno effect},\ }\href {https://doi.org/10.1103/PhysRevLett.129.070502} {\bibfield  {journal} {\bibinfo  {journal} {Phys. Rev. Lett.}\ }\textbf {\bibinfo {volume} {129}},\
  \bibinfo {pages} {070502} (\bibinfo {year} {2022})}\BibitemShut {NoStop}%
\end{thebibliography}
\end{document}